\documentclass[twocolumn]{aastex631} %
\usepackage{comment} %
\usepackage[maxfloats=256]{morefloats} %

\shorttitle{M-dwarf Compositions in the Subaru-IRD Survey} %
\shortauthors{Ishikawa et al.}
\graphicspath{{./}{figures/}}

\begin{document}

\title{Elemental abundances of nearby M dwarfs
based on high-resolution near-infrared spectra \\
obtained by the Subaru/IRD survey:
Proof of concept
}  %

\email{hiroyuki.ishikawa@nao.ac.jp} %

\author[0000-0001-6309-4380]{Hiroyuki Tako Ishikawa}
\affiliation{Astrobiology Center, 2-21-1 Osawa, Mitaka, Tokyo 181-8588, Japan}
\affiliation{National Astronomical Observatory of Japan, 2-21-1 Osawa, Mitaka, Tokyo 181-8588, Japan}

\author[0000-0002-8975-6829]{Wako Aoki}
\affiliation{National Astronomical Observatory of Japan, 2-21-1 Osawa, Mitaka, Tokyo 181-8588, Japan}
\affiliation{Department of Astronomical Science, The Graduate University for Advanced Studies, SOKENDAI, 2-21-1 Osawa, Mitaka, Tokyo 181-8588, Japan} %

\author[0000-0003-3618-7535]{Teruyuki Hirano}
\affiliation{Astrobiology Center, 2-21-1 Osawa, Mitaka, Tokyo 181-8588, Japan}
\affiliation{National Astronomical Observatory of Japan, 2-21-1 Osawa, Mitaka, Tokyo 181-8588, Japan}

\author[0000-0001-6181-3142]{Takayuki Kotani}
\affiliation{Astrobiology Center, 2-21-1 Osawa, Mitaka, Tokyo 181-8588, Japan}
\affiliation{National Astronomical Observatory of Japan, 2-21-1 Osawa, Mitaka, Tokyo 181-8588, Japan}
\affiliation{Department of Astronomical Science, The Graduate University for Advanced Studies, SOKENDAI, 2-21-1 Osawa, Mitaka, Tokyo 181-8588, Japan}

\author[0000-0002-4677-9182]{Masayuki Kuzuhara}
\affiliation{Astrobiology Center, 2-21-1 Osawa, Mitaka, Tokyo 181-8588, Japan}
\affiliation{National Astronomical Observatory of Japan, 2-21-1 Osawa, Mitaka, Tokyo 181-8588, Japan}

\author[0000-0002-5051-6027]{Masashi Omiya}
\affiliation{Astrobiology Center, 2-21-1 Osawa, Mitaka, Tokyo 181-8588, Japan}
\affiliation{National Astronomical Observatory of Japan, 2-21-1 Osawa, Mitaka, Tokyo 181-8588, Japan}

\author[0000-0003-4676-0251]{Yasunori Hori}
\affiliation{Astrobiology Center, 2-21-1 Osawa, Mitaka, Tokyo 181-8588, Japan}
\affiliation{National Astronomical Observatory of Japan, 2-21-1 Osawa, Mitaka, Tokyo 181-8588, Japan}

\author[0000-0002-5486-7828]{Eiichiro Kokubo}
\affiliation{National Astronomical Observatory of Japan, 2-21-1 Osawa, Mitaka, Tokyo 181-8588, Japan}

\author[0000-0002-9294-1793]{Tomoyuki Kudo}
\affiliation{Subaru Telescope, National Astronomical Observatory of Japan, 650 North Aohoku Place, Hilo, HI 96720, USA}

\author[0000-0001-9986-4797]{Takashi Kurokawa}
\affiliation{Tokyo University of Agriculture and Technology, Koganei, Tokyo 184-8588, Japan}
\affiliation{Astrobiology Center, 2-21-1 Osawa, Mitaka, Tokyo 181-8588, Japan}

\author[0000-0001-9194-1268]{Nobuhiko Kusakabe}
\affiliation{Astrobiology Center, 2-21-1 Osawa, Mitaka, Tokyo 181-8588, Japan}
\affiliation{National Astronomical Observatory of Japan, 2-21-1 Osawa, Mitaka, Tokyo 181-8588, Japan}

\author[0000-0001-8511-2981]{Norio Narita}
\affiliation{Komaba Institute for Science, The University of Tokyo, 3-8-1 Komaba, Meguro, Tokyo 153-8902, Japan}
\affiliation{JST, PRESTO, 3-8-1 Komaba, Meguro, Tokyo 153-8902, Japan}
\affiliation{Astrobiology Center, 2-21-1 Osawa, Mitaka, Tokyo 181-8588, Japan}
\affiliation{Instituto de Astrof\'{i}sica de Canarias (IAC), 38205 La Laguna, Tenerife, Spain}

\author[0000-0001-9326-8134]{Jun Nishikawa}
\affiliation{National Astronomical Observatory of Japan, 2-21-1 Osawa, Mitaka, Tokyo 181-8588, Japan}
\affiliation{Department of Astronomical Science, The Graduate University for Advanced Studies, SOKENDAI, 2-21-1 Osawa, Mitaka, Tokyo 181-8588, Japan}
\affiliation{Astrobiology Center, 2-21-1 Osawa, Mitaka, Tokyo 181-8588, Japan}

\author[0000-0002-8300-7990]{Masahiro Ogihara}
\affiliation{National Astronomical Observatory of Japan, 2-21-1 Osawa, Mitaka, Tokyo 181-8588, Japan}
\affiliation{Tokyo Institute of Technology, Meguro, Tokyo 152-8550, Japan}

\author{Akitoshi Ueda} %
\affiliation{National Astronomical Observatory of Japan, 2-21-1 Osawa, Mitaka, Tokyo 181-8588, Japan}

\author[0000-0002-7405-3119]{Thayne Currie}
\affiliation{Subaru Telescope, National Astronomical Observatory of Japan, 650 North Aohoku Place, Hilo, HI 96720, USA}
\affiliation{NASA-Ames Research Center, Moffett Field, CA, USA.}
\affiliation{Eureka Scientific, Oakland, 10 CA, USA.}

\author[0000-0002-1493-300X]{Thomas Henning} %
\affiliation{Max-Planck-Institut f\"ur Astronomie, K\"onigstuhl 17, 69117 Heidelberg, Germany}

\author[0000-0002-8607-358X]{Yui Kasagi}
\affiliation{Department of Astronomical Science, The Graduate University for Advanced Studies, SOKENDAI, 2-21-1 Osawa, Mitaka, Tokyo 181-8588, Japan}
\affiliation{National Astronomical Observatory of Japan, 2-21-1 Osawa, Mitaka, Tokyo 181-8588, Japan}

\author[0000-0002-6267-0849]{Jared R. Kolecki}
\affiliation{Department of Astronomy, The Ohio State University, Columbus, Ohio 43210, USA}

\author[0000-0003-2815-7774]{Jungmi Kwon}
\affiliation{Department of Astronomy, Graduate School of Science, The University of Tokyo, 7-3-1 Hongo, Bunkyo-ku, Tokyo 113-0033, Japan}

\author[0000-0002-0963-0872]{Masahiro N. Machida}
\affiliation{Department of Earth and Planetary Sciences, Faculty of Sciences Kyushu University, Fukuoka 819-0395, Japan}

\author[0000-0003-0241-8956]{Michael W. McElwain}
\affiliation{Exoplanets and Stellar Astrophysics Laboratory, NASA Goddard Space Flight Center, Greenbelt, MD 20771, USA}

\author[0000-0002-6660-9375]{Takao Nakagawa}
\affiliation{Institute of Space and Astronautical Science, Japan Aerospace Exploration Agency, 3-1-1 Yoshinodai, Chuo-ku, Sagamihara, Kanagawa 252-5210, Japan}

\author[0000-0003-4018-2569]{Sebastien Vievard} %
\affiliation{Astrobiology Center, 2-21-1 Osawa, Mitaka, Tokyo 181-8588, Japan}
\affiliation{Subaru Telescope, National Astronomical Observatory of Japan, 650 North Aohoku Place, Hilo, HI 96720, USA}

\author[0000-0002-4361-8885]{Ji Wang}
\affiliation{Department of Astronomy, The Ohio State University, Columbus, Ohio 43210, USA}

\author[0000-0002-6510-0681]{Motohide Tamura}
\affiliation{Department of Astronomy, Graduate School of Science, The University of Tokyo, 7-3-1 Hongo, Bunkyo-ku, Tokyo 113-0033, Japan}
\affiliation{Astrobiology Center, 2-21-1 Osawa, Mitaka, Tokyo 181-8588, Japan}
\affiliation{National Astronomical Observatory of Japan, 2-21-1 Osawa, Mitaka, Tokyo 181-8588, Japan}

\author[0000-0001-7505-2487]{Bun'ei Sato}
\affiliation{Tokyo Institute of Technology, Meguro, Tokyo 152-8550, Japan}

\begin{abstract} %
  Detailed chemical analyses of M dwarfs are scarce but necessary to constrain the formation environment and internal structure of planets being found around them. %
  We present elemental abundances of 13 M dwarfs (2900 $< T_{\mathrm{eff}} <$ 3500 K)
  \added{%
  observed in the Subaru/IRD planet search project.
  }%
  They are mid-to-late M dwarfs whose abundance of individual elements has not been well studied. %
  We use the
  \added{%
  high-resolution ($\sim$70,000) near-infrared (970--1750 nm)
  }%
  spectra %
  to measure the abundances of Na, Mg, Si, K, Ca, Ti, V, Cr, Mn, Fe, and Sr by the line-by-line analysis based on model atmospheres, %
   with typical errors ranging from 0.2 dex for [Fe/H] to 0.3--0.4 dex for other [X/H].
  We measure radial velocities from the spectra and combine them with Gaia astrometry to calculate the Galactocentric space velocities $UVW$. %
  The resulting [Fe/H] values agree with previous estimates based on medium-resolution $K$-band spectroscopy, %
  showing a wide distribution of metallicity ($-$0.6 $<$ [Fe/H] $<$ $+$0.4).
  The abundance ratios of individual elements [X/Fe] are generally aligned with the solar values in all targets. %
  While the [X/Fe] distributions are comparable to those of nearby FGK stars, most of which belong to the thin disk population,  %
  the most metal-poor object, GJ 699, could be a thick disk star. %
  The $UVW$ velocities also support this.
  The results raise the prospect that near-infrared spectra of M dwarfs obtained in the planet search projects can be used to grasp the trend of elemental abundances and Galactic stellar population of nearby M dwarfs. %

\end{abstract}


\section{Introduction} \label{sec:IRD_intro}

M dwarfs are the most ubiquitous stars in the solar neighborhood.
Due to their small masses, small radii, and low luminosities, they are also promising targets for recent and upcoming projects seeking to identify and characterize rocky, habitable zone planets through radial-velocity, transits, and direct imaging~\citep[e.g.,][]{2010Sci...327..977B, 2018SPIE10702E..11K, 2019BAAS...51c.162L}. %

Elemental abundances of planet-host stars give crucial information on the formation, characteristics, or habitability of the orbiting planets.
Especially, stellar abundances of refractory elements are a proxy of the composition of planetary building blocks.
They are essential to constrain the planetary internal structures that are the keys to habitability, such as core size and mantle composition (e.g., \citealt{2017A&A...597A..37D}, \citealt{2017ApJ...845...61U}, \citealt{2021Sci...374..330A}). %
\citet{2020A&A...633A..10B} calculated the chemical composition of planetary building blocks, taking into account the stellar abundance variation of some key elements.
Their results show that different elemental abundances can significantly change the amount of rock-forming materials.

Another observational indication from elemental abundances on planet formation scenario is the correlation between the stellar elemental abundances and the planet occurrence rate.
There is a well-known positive correlation between the stellar metallicity and the occurrence rate of giant planets~\citep[e.g.,][]{2005ApJ...622.1102F}.
The correlation has also been investigated for individual refractory elements such as Si, Mg, and Ti for FGK-type stars, reporting stronger correlations than the case of overall metallicity~\citep[e.g.,][]{2011ApJ...738...97B, 2012A&A...543A..89A}.
Around M dwarfs, gas giants are relatively rare, and small planets dominate.
The correlations between the occurrence of planets and the metallicity of host stars are not as clear as for FGK samples.
Some previous studies have suggested that gas giants show a positive correlation also around M dwarfs, and small planets do not have a significant correlation or even a weak anti-correlation with metallicity~\citep[e.g.,][]{2013A&A...551A..36N, 2016MNRAS.457.2877G, 2018RMxAA..54...65H}. %
Detailed abundance analysis of M dwarfs could help to settle these debates, along with the increase of the validation of planets around M dwarfs.  %

M dwarfs are also important in the context of Galactic history because of their ubiquity and longer main-sequence lifetimes than
the current age of the universe, which means they are the ideal tracer of the Galactic chemical evolution.
The abundance ratio of individual elements is one of the essential diagnostics to explore the chemical evolution of the Galaxy. %
Several studies have identified the chemically separated populations by probing the elemental abundance trends (especially [$\alpha$/Fe] ratio) for stars of individual populations~\citep[e.g.,][]{2006MNRAS.367.1329R, 2014A&A...562A..71B, 2015AJ....150..148H}. %
However, most of the previous knowledge of the detailed chemical composition has been limited to the more massive members than M dwarfs. %
Recently, similar trends have been reported also for M dwarfs using [$\alpha$/Fe]~\citep{2020AJ....159...30H} or [Ti/Fe]~\citep{2020MNRAS.494.2718W}.

The occurrence rates of planets may also depend on the Galactic population.
\citet{2020A&A...643A.106B} found the fact that
the occurrence rates of low-mass close-in planets are significantly larger at high-$\alpha$ FGK stars in the thick disk than low-$\alpha$ FGK stars in the thin disk
in the range of [Fe/H] $<$ 0. %
Also around the M dwarfs in the thick disk and possibly in the halo, some planets have been confirmed recently~\citep[e.g.,][]{2020AJ....159..160G, 2021ApJ...909..115C, 2021arXiv210408306P}.
Abundance measurements of individual elements of nearby M dwarfs enable the extension of these previous insights to the dominant type of stars in the Galaxy. %

Overall metallicity or the iron abundance has been estimated by empirically calibrated methods~\citep[e.g.,][]{2013AJ....145...52M, 2014AJ....147..160M} for many M dwarfs.
Abundance ratios of individual elements are not determined for many M dwarfs, in particular for late-M dwarfs due to the dense forest of molecular lines, but should be important to constrain stellar populations and the impact on planet formation.
Moreover, abundance ratios of individual elements are required to obtain reliable metallicity as demonstrated by \citet{2020PASJ...72..102I} (hereafter Ish20).
They reported that spectral lines of M dwarfs are sensitive to changes in the abundances of not only the elements responsible for the lines but also other elements, especially dominant electron donors such as Na and Ca.

A standard method to determine the elemental abundances of FGK-type stars is the comparison of line profiles or equivalent widths (EW) of atomic lines between the observed high-resolution spectra and the calculated synthetic spectra.
In recent years, the standard analysis has also been applied to M dwarfs using spectral bandpasses that are relatively unaffected by molecular absorption, mainly from 0.7 to 1.7 \micron. %
An increasing number of previous studies have estimated the overall metallicity or the iron abundance [Fe/H] of M dwarfs by such model-based analyses of individual spectral lines~\citep[e.g.,][]{2012A&A...542A..33O, 2017A&A...604A..97L, 2018A&A...620A.180R,
2019A&A...627A.161P, 2021A&A...649A.147S, 2021ApJ...917...11S, 2021arXiv211007329M}. %
In addition to [Fe/H], some studies have estimated [Ti/H] or [$\alpha$/Fe] by using mainly Ti lines, which are prominent in the near-infrared spectra of M dwarfs~\citep[e.g.,][]{2017ApJ...851...26V, 2020AJ....159...30H, 2020MNRAS.494.2718W}. %
The abundance determination of individual elements other than iron or titanium is still limited to a few studies described below. %

Meanwhile, the model-based abundance analysis of major refractory elements such as Na or Mg hitherto has been reported only for seven M dwarfs.
\citet{2017ApJ...835..239S} determined the stellar parameters and abundances of thirteen individual elements on two planet-hosting early-M dwarfs (effective temperature $T_{\mathrm{eff}} \sim 3900$ K) by spectrum synthesis fitting to $H$-band spectra. %
\citet{2018ApJ...860L..15S} applied the same method to later M dwarfs ($T_{\mathrm{eff}} \sim 3200$ K) to derive abundances for eight elements.
Ish20 determined the abundances of eight elements for five M dwarfs ($T_{\mathrm{eff}} \sim \;$3200--3800 K), which have G- or K-type binary companions with known elemental abundances, and confirmed the abundance agreement between each binary pair.
The model-based analysis on the individual spectral lines of the major refractory elements in stars cooler than 3200 K has been unprecedented.

As for the other elements,
\citet{2014PASJ...66...98T}, \citet{2015PASJ...67...26T}, \citet{2016PASJ...68...13T}, and \citet{2016PASJ...68...84T} determined the carbon and oxygen abundances and their isotopic ratios based on CO and $\mathrm{H_{2}O}$ lines in the $K$-band. %
\citet{2020ApJ...890..133S} also estimated the C and O abundances of M dwarfs for the purpose of $T_{\mathrm{eff}}$ estimation. %
\citet{2020A&A...642A.227A} derived the abundances of neutron-capture elements Rb, Sr, and Zr in 57 early-M dwarfs by spectral synthesis fits to the atomic lines in red optical and near-infrared spectra to investigate the chemical enrichment history of the Galaxy.
\added{%
More recently, \citet{2021A&A...654A.118S} determined the vanadium abundances of 135 early M dwarfs observed with CARMENES~\citep{2014SPIE.9147E..1FQ, 2020SPIE11447E..3CQ} and found a tight correlation between [V/H] and [Fe/H]. %
}%

On the other hand, data-driven approaches, which do not necessarily employ model atmospheres of M dwarfs, are also beginning to be used for spectroscopic studies of M dwarfs~\citep[e.g.,][]{2018MNRAS.476.1120S, 2020MNRAS.491.2280S, 2020ApJ...892...31B, 2020A&A...636A...9A, 2020A&A...642A..22P, 2021ApJS..253...45L}.
For individual elements, \citet{2020A&A...644A..68M} developed the methodology using the principal component analysis and sparse Bayesian method
with 19 binary pairs of M and FGK dwarfs as a training dataset
to determine the abundances of 15 elements of M dwarfs with $T_{\mathrm{eff}} >$ 3200 K.
The progress in the application of data science is promising, but in parallel, abundance analyses with the non-empirical line-by-line method based on model atmospheres are essential to interpret the physical meaning of obtained elemental abundances.

In this paper, we present the elemental abundances obtained by analysis of near-infrared spectra using model atmospheres for 13 mid-to-late M dwarfs (2900 $< T_{\mathrm{eff}} <$ 3500 K) observed in a planet search project using the Subaru Telescope. %
The IRD-SSP project and our target selection are introduced in Section \ref{sec:IRD_targets_data}.
We describe the analysis on the effective temperatures, elemental abundances, and kinematics of our targets in Section \ref{sec:IRD_analysis}, and their results in Section \ref{sec:IRD_results}.
Based on the results, we discuss the position of our M dwarfs in the Galactic context and its indication to the planet searches in Section \ref{sec:IRD_discussion}.
Finally, a summary is provided in Section \ref{sec:IRD_summary}.

\section{Targets and data} \label{sec:IRD_targets_data}
To constrain the chemical properties of nearby M dwarfs, we conducted the abundance analysis for 13 M dwarfs as a pilot sample from the targets of a planet search project using the InfraRed Doppler instrument \citep[IRD;][]{2012SPIE.8446E..1TT, 2018SPIE10702E..11K} mounted on the Subaru Telescope.
The IRD is a fiber-fed echelle high-resolution near-infrared spectrometer.
It covers the $Y$-, $J$-, and $H$-bands (970--1750 nm) simultaneously with a maximum spectral resolution of $\sim$70,000 and has a laser frequency comb \citep[LFC;][]{2016OExpr..24.8120K, 2016SPIE.9912E..1RK} for the precise wavelength calibration, which enables high-precision radial velocity (RV) measurements. %
The main purpose of the instrument is the planet search around nearby mid-M to late-M dwarfs (spectral type later than M4V).
The large aperture (8.2 m) of the Subaru Telescope and the wide coverage of the IRD over the near-infrared wavelength, where the flux peak of M dwarfs is located, enable the systematic survey of fainter, thus late-type, M dwarfs than ever before.
The Doppler survey project is running since February 2019 in the framework of the Subaru Strategic Program (IRD-SSP).

The high-S/N spectra obtained in the IRD-SSP are also suitable data to investigate the elemental abundances of nearby M dwarfs.
Besides, the knowledge of elemental abundances of the IRD-SSP targets will also help to interpret the characteristics of planets that will be detected in the project. %

During a five-year survey period of IRD-SSP, each object is observed at multiple epochs in order to precisely investigate the RV variations. %
We use the IRD stellar template spectrum made for each object as a reference for the relative RV at each epoch \citep{2020PASJ...72...93H}. %
The template spectra are generated by deconvolving the instrumental broadening profiles (IP), removing telluric absorptions, and combining multiple frames.
IRD uses another fiber to obtain spectra of %
the wavelength calibration sources generated with an LFC.
The IP is derived for each frame and each spectral segment of $\sim$1 nm width, based on the simultaneously observed spectra of LFC.
The removal of telluric lines is performed by fitting the theoretical telluric spectra synthesized with Line-By-Line Radiative Transfer Model \citep{1992JGR....9715761C}, %
or by utilizing the spectra of telluric standard stars (mostly A-type stars), observed on the same nights.
The imperfection of the synthesis might leave some residual telluric features.
These residuals can be mostly removed by taking the median of the spectra obtained over seasons %
because the telluric features move against the intrinsic stellar spectra due to the barycentric motion of the Earth. %

The IRD-SSP is planned to monitor 60 carefully selected M dwarfs taking 175 observing nights over the five years duration of the project.
Through the pre-selection, 150 objects are included in the IRD input catalog.
This selection is based on the following six criteria to guarantee the accuracy and efficiency of the RV measurements~\citep{2018SSPproposal..S}:
(1) $J$-band magnitudes $J_\mathrm{mag} <$ 11.5,
(2) stellar mass $M_{\star} <$ 0.25 M$_{\odot}$  ($T_{\mathrm{eff}} <$ $\sim$3400 K),
(3) rotational velocities $v \sin{i} <$ 5 km s$^{-1}$,
(4) no significant H$\alpha$ (656.28 nm) emission (EW$_{\mathrm{H}\alpha} < -$0.75~\AA),
(5) rotational periods $P_\mathrm{rot} > $ 70 days, and
(6) expected to be a single star.
The information of stellar properties used in the criteria is collected from the literature and confirmed with pre-selection observations using medium-resolution optical spectroscopy~\citep{2021PASJ...73..154K}.

To ensure the quality of the template spectra of the pilot sample, we selected 13 M dwarfs which were already observed with IRD more than eight times on well-separated nights. %
The original data were obtained between February 2019 and August 2020.
In addition to the survey targets, two other objects observed as RV standards, GJ 436 and GJ 699, are included in our sample.
Table \ref{tab:IRD_targets} shows the basic information of the targets.
At this time, among our targets, GJ 436 and GJ 699 have been reported to have planets \citep{2004ApJ...617..580B, 2018Natur.563..365R}.

\begin{deluxetable*}{lcccccccc} %
  \tablecaption{Basic information of the target M dwarfs}
  \label{tab:IRD_targets}
    \tablehead{
      \multicolumn{1}{c}{Name}  &  R.A.\,(J2000.0)  &  Decl.\,(J2000.0)  &  $V_\mathrm{mag}$  &  $J_\mathrm{mag}$  &  $K_\mathrm{mag}$  &  $T_{\mathrm{eff}}$ (K)  &  $\log{g}$  &  Luminosity (L$_\odot$)    %
    }
    \decimals
    \startdata
      GJ 436     &  11:42:11.09  &  $+$26:42:23.7  &  10.7  &   6.90  &   6.07  &  3456$\pm$157  &  4.804$\pm$0.005  &  0.0233$\pm$0.0056  \\
      GJ 699     &  17:57:48.50  &  $+$04:41:36.1  &   9.5  &   5.24  &   4.52  &  3259$\pm$157  &  5.076$\pm$0.028  &  0.0038$\pm$0.0010  \\
      LSPM J1306+3050     &  13:06:50.25  &  $+$30:50:54.9  &  15.5  &  10.23  &   9.31  &  2970$\pm$157  &  5.127$\pm$0.037  &  0.0020$\pm$0.0005  \\
      LSPM J1523+1727     &  15:23:51.13  &  $+$17:27:57.0  &  14.2  &   9.11  &   8.28  &  3151$\pm$157  &  5.050$\pm$0.023  &  0.0038$\pm$0.0010  \\
      LSPM J1652+6304     &  16:52:49.46  &  $+$63:04:38.9  &  14.4  &   9.59  &   8.76  &  3115$\pm$157  &  5.024$\pm$0.019  &  0.0043$\pm$0.0011  \\
      LSPM J1703+5124     &  17:03:23.85  &  $+$51:24:21.9  &  13.6  &   8.77  &   7.92  &  3103$\pm$157  &  5.037$\pm$0.021  &  0.0039$\pm$0.0010  \\
      LSPM J1802+3731     &  18:02:46.25  &  $+$37:31:04.9  &  15.3  &   9.72  &   8.89  &  2992$\pm$157  &  5.110$\pm$0.034  &  0.0022$\pm$0.0006  \\
      LSPM J1816+0452     &  18:16:31.54  &  $+$04:52:45.6  &  15.0  &   9.80  &   8.83  &  3072$\pm$157  &  5.045$\pm$0.022  &  0.0036$\pm$0.0009  \\
      LSPM J1909+1740     &  19:09:50.98  &  $+$17:40:07.5  &  13.6  &   8.82  &   7.90  &  3125$\pm$157  &  5.024$\pm$0.019  &  0.0043$\pm$0.0011  \\
      LSPM J2026+5834     &  20:26:05.29  &  $+$58:34:22.5  &  14.0  &   9.03  &   8.10  &  3025$\pm$157  &  5.072$\pm$0.027  &  0.0029$\pm$0.0008  \\
      LSPM J2043+0445     &  20:43:23.88  &  $+$04:45:55.3  &  15.4  &  10.08  &   9.14  &  2979$\pm$157  &  5.073$\pm$0.027  &  0.0027$\pm$0.0007  \\
      LSPM J2053+1037     &  20:53:33.04  &  $+$10:37:02.0  &  13.9  &   9.35  &   8.48  &  3168$\pm$157  &  5.037$\pm$0.021  &  0.0042$\pm$0.0011  \\
      LSPM J2343+3632     &  23:43:06.29  &  $+$36:32:13.2  &  12.5  &   8.11  &   7.23  &  3175$\pm$157  &  4.993$\pm$0.015  &  0.0055$\pm$0.0014  \\
    \enddata
  \tablecomments{
  The stellar parameters are adopted from the TESS Input Catalog (TIC) Version 8 \citep{2019AJ....158..138S}. %
  }
\end{deluxetable*}

Magnetically active stars are intended to be excluded by the above selection criteria.
Thereby, we do not consider the influence of the magnetic field and related activities on the analysis.
It should be noted, however, that almost half or more objects in the volume-limited sample of mid-M to late-M dwarfs are magnetically active~\citep[e.g.,][]{2018A&A...614A..76J}, so we should be aware of a potential selection bias that this selection may impose on our results.

\section{Analysis} \label{sec:IRD_analysis}
We determined the abundances of individual elements for the 13 targets by the analysis method presented in Ish20.
Based on the points made by Ish20, we determine the abundance of each element from the absorption lines of the individual elements in a way that is consistent with each other.
We investigated their kinematics as a hint of Galactic stellar population. %

An overview of the analysis procedures is given here. %
We measured EW of an absorption line of interest from the normalized one-dimensional spectra and compared it with EW from synthetic spectra calculated with the radiative transfer code, iteratively modifying the abundance of the responsible element assumed in the calculation until observed and calculated EWs agree. %
The radiative transfer is calculated based on the same assumptions as in the model atmosphere program of \citet{1978A&A....62...29T}, i.e., local thermodynamical equilibrium (LTE), plane-parallel geometry, radiative and hydrostatic equilibrium, and chemical equilibrium.
The atmospheric layer structure is derived from the interpolated grid of MARCS~\citep{2008A&A...486..951G} and the spectral line data are taken from the Vienna Atomic Line Database (VALD; \citealt{1999A&AS..138..119K}, \citealt{2015PhyS...90e4005R}).
Dust is not considered as an opacity source since dust formation in the atmosphere is negligible at $T_{\mathrm{eff}} > $ 2800 K (e.g., \citealt{2013MmSAI..84.1053A}). %
For each element (X), the abundance ratio ([X/H]) was determined by averaging the results from individual absorption lines measured in this study. %
Abundances of elements other than those of interest were assumed to be solar values in the first step.
In the next step, however, the abundance results of the previous step are adopted to perform the analysis with the same procedure.
This is iterated until the difference between the assumed values and the results converge to $\pm$0.005 for all elements consistently. %
We estimated the error $\sigma_\mathrm{Total}$ by taking the quadrature sum of
\added{%
four types of errors:
}%
(1) the standard deviation of the results derived from individual lines divided by square root of the number of lines used ($\sigma / \sqrt{N}$) ($\sigma_\mathrm{SEM}$), %
(2) errors propagated from uncertainties of stellar parameters ($\sigma_{T_{\mathrm{eff}}}$, $\sigma_{\log{g}}$, and $\sigma_{\xi}$),
\added{%
(3) changes in the abundance result of the element of interest when the abundances of all other elements are modified to be 0.2 dex higher or lower than the final results ($\sigma_\mathrm{OE}$), and
}%
(4) errors due to the uncertainty of continuum levels ($\sigma_\mathrm{cont}$).
More detailed procedures to estimate the individual types of errors are given in Ish20.

There are two major modifications in the analysis procedure from that in Ish20.
One is that we adopted $T_{\mathrm{eff}}$ and $\log{g}$ values from TESS Input Catalog (TIC) Version 8~\citep{2019AJ....158..138S}.
The other is the procedure of EW measurement. %
\added{%
The details of the two are described in Sections \ref{sec:IRD_stellar_parameters} and \ref{sec:IRD_equivalent_width_measurement} respectively. %
}%

\subsection{Stellar parameters} \label{sec:IRD_stellar_parameters}

\added{%
The first modification from the analysis procedure of Ish20 is the selection of stellar parameters.
We here adopt $T_{\mathrm{eff}}$ and $\log{g}$ values stored in the TIC for homogeneity in the 13 targets, whereas Ish20 used these parameters from various sources in the literature.
}%

The $T_{\mathrm{eff}}$ of M dwarfs in TIC ($T_{\mathrm{eff\mathchar`-TIC}}$) is based on the empirical relation between the $T_{\mathrm{eff}}$ and the $G_\mathrm{BP} - G_\mathrm{RP}$ color from the Gaia photometry.
The empirical relation was calibrated by the sample of \citet[][]{2015ApJ...804...64M} (hereafter Man15) and their $T_{\mathrm{eff}}$ estimates.
\added{%
The uncertainty of the $T_{\mathrm{eff}}$ is also taken from the TIC.
It is dominated by the root mean square scatter of the fitting to derive the empirical relation and is uniformly given as 157 K for M dwarfs in the catalog. %
We adopted it to estimate the $\sigma_{T_{\mathrm{eff}}}$.
}%

\begin{figure}
  \plotone{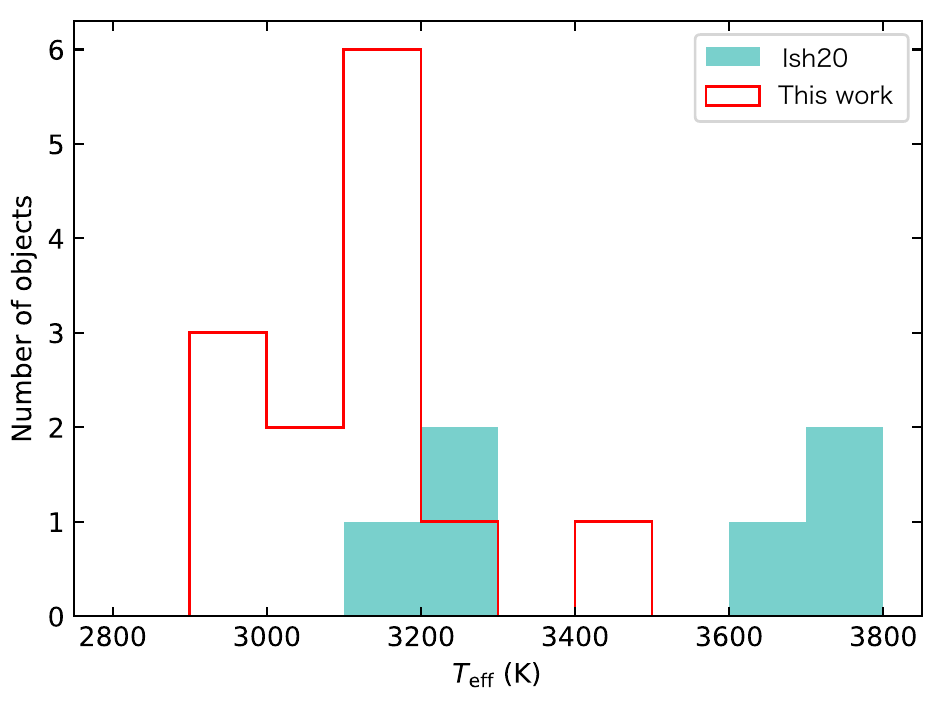} %
  \caption{
  The distribution of $T_{\mathrm{eff}}$ from TIC of the M dwarfs.
  The green and red histograms show the samples analyzed in Ish20 and this work, respectively.
  The sample intentionally chooses M-dwarfs with $T_{\mathrm{eff}}$ $>$ 2800 such that dust formation does not affect the spectra.
  } \label{fig:hist_Teff}
\end{figure}

Figure \ref{fig:hist_Teff} shows the $T_{\mathrm{eff\mathchar`-TIC}}$ distribution of our target M dwarfs compared with the samples used to verify the analysis method by Ish20.
Note that the sample of this study includes objects with 100--200 K lower temperatures than those in Ish20 but the difference would not have a significant impact on the modeling of the atmospheric structure in this temperature range. %
Once the $T_{\mathrm{eff}}$ becomes lower than about 2800 K, dust formation may start to affect the spectra~\citep[e.g.][]{1996A&A...305L...1T}.
It is important to confirm the reliability of the abundance analysis for M dwarfs with such lower temperatures with binary systems in future work. %

\begin{figure}
  \plotone{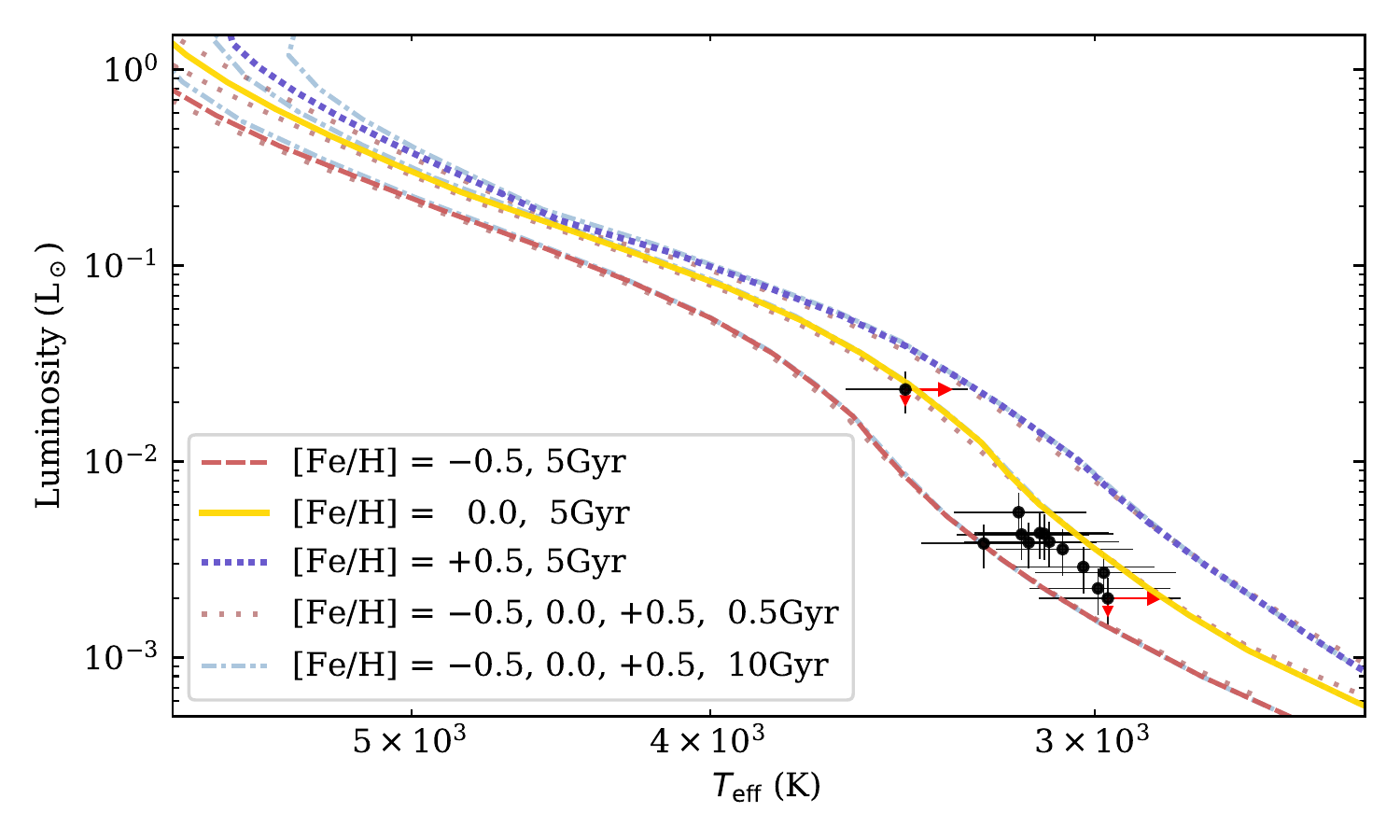}
  \caption{
  HR diagram, i.e., the Luminosity--$T_{\mathrm{eff}}$ relations, for the 13 target M dwarfs superimposed on the 5-Gyr theoretical models for [Fe/H] = $-$0.5, 0.0, $+$0.5 from the PARSEC library.
  \added{%
  Models with different ages (0.5 Gyr and 10 Gyr) are also plotted in the background for reference, indicating that changes within a reasonable age range do not make a difference in the figure. %
  }%
  The black dots and error bars represent the temperature and luminosity recorded in the TIC for the 13 objects.
  The red arrows on two representative points illustrate the changes that occur when the adopted $T_{\mathrm{eff}}$ is adjusted downward by 100 K
  (see Section \ref{sec:Teff_shift} for discussion). %
  } \label{fig:HR-diagram}
\end{figure}

We plot our target sample on the Hertzsprung–Russell (HR) diagram in Figure \ref{fig:HR-diagram} to show that our target sample covers the mid-M to late-M dwarf stars.
The superposed lines show the predictions from the PARSEC stellar evolutionary models version 2.1~\citep{2012MNRAS.427..127B, 2014MNRAS.444.2525C} for different metallicities
\added{%
at ages of 0.5, 5, and 10 Gyr (including the pre-main sequence lifetime).
}%

The $\log{g}$ in TIC is calculated from the mass and radius, which are estimated from the mass-$M_K$ relation in \citet{2019ApJ...871...63M} and the radius-$M_K$ relation in Man15, respectively.
The typical error in the mass-$M_K$ relation is reported to be 2--3\,\%
and that in the radius-$M_K$ relation is 3--4\,\%. %
\added{%
The propagation of these errors results in a $\log{g}$ uncertainty of $\sim$0.02--0.03 dex.
We used the uncertainties reported for the individual objects in TIC to calculate the $\sigma_{\log{g}}$.
}%

\added{%
For simplicity, $\xi$ was fixed to 0.5 km s$^{-1}$ for all objects, and $\pm$0.5 km s$^{-1}$ was adopted as the uncertainty used for the $\sigma_{\xi}$ estimation.
}%
It is confirmed by Ish20 that this simplification is realistic and adds negligible effects to the results.
All the adopted stellar parameters are summarized in Table \ref{tab:IRD_targets}.

\subsection{Effective temperature from FeH lines} \label{sec:IRD_FeH_Teff}

We also investigated the possibility to determine $T_{\mathrm{eff}}$ using FeH molecular lines, which are very sensitive to temperature. %
Our approach is to find $T_{\mathrm{eff}}$ at which the synthetic spectra best reproduce the strengths of FeH lines. %
An advantage of this approach is that
we can determine the $T_{\mathrm{eff}}$ of all the IRD samples homogeneously with the same method, which is suitable for comparing the abundances between individual objects.

The Wing-Ford band of FeH lines at $\sim$990--1020 nm \citep{1969PASP...81..527W} is one of the most prominent molecular features observed in the near-infrared spectra of M dwarfs.
We select 47 well-isolated FeH lines in the band to measure EW as indicators of the $T_{\mathrm{eff}}$.
The line data are based on the list edited by Plez \citep{2012A&A...542A..33O}.
We also tried analyzing the FeH lines in the $H$-bands with the line list of \citet{2010AJ....140..919H} but the continuum level around them is so uncertain due to numerous weak lines of $\mathrm{H_{2}O}$ that we decide not to include the analysis in this work.
The selected line list is provided in Table \ref{tab:IRD_FeH_linelist}.
Since there is no information on the van der Waals damping parameters for these lines, we calculated the pressure broadening by applying the hydrogenic approximation of \citet{1955psmb.book.....U}. %
Synthetic spectra based on the line list well reproduce the individual lines in the band.
\startlongtable
\begin{deluxetable}{lccc} %
  \tablecaption{FeH molecular line list}
  \label{tab:IRD_FeH_linelist}
    \tablehead{
      Species &
      $\lambda$ (nm)\tablenotemark{$*$} &
      $E_\mathrm{low}$ (eV)\tablenotemark{$\dag$}
      & $\log gf$\tablenotemark{$\ddag$}
      }
      \decimals
      \startdata
      FeH  &   990.0489  &    0.225  &   $-$0.663 \\
      FeH  &   990.4987  &    0.134  &   $-$0.770 \\
      FeH  &   990.5869  &    0.177  &   $-$0.712 \\
      FeH  &   991.1999  &    0.153  &   $-$0.791 \\
      FeH  &   991.4706  &    0.086  &   $-$0.954 \\
      FeH  &   991.6441  &    0.061  &   $-$1.075 \\
      FeH  &   992.4870  &    0.084  &   $-$1.111 \\
      FeH  &   992.8899  &    0.137  &   $-$0.891 \\
      FeH  &   993.3279  &    0.176  &   $-$0.760 \\
      FeH  &   993.5206  &    0.056  &   $-$1.695 \\
      FeH  &   993.5985  &    0.066  &   $-$0.928 \\
      FeH  &   994.1592  &    0.199  &   $-$0.732 \\
      FeH  &   994.4518  &    0.139  &   $-$0.891 \\
      FeH  &   994.5785  &    0.173  &   $-$0.816 \\
      FeH  &   994.7128  &    0.053  &   $-$0.983 \\
      FeH  &   995.0866  &    0.407  &   $-$0.552 \\
      FeH  &   995.3064  &    0.156  &   $-$0.852 \\
      FeH  &   995.6641  &    0.376  &   $-$0.568 \\
      FeH  &   995.7314  &    0.194  &   $-$0.783 \\
      FeH  &   996.2861  &    0.175  &   $-$0.816 \\
      FeH  &   996.5166  &    0.442  &   $-$0.537 \\
      FeH  &   997.0814  &    0.216  &   $-$0.754 \\
      FeH  &   997.3037  &    0.410  &   $-$0.552 \\
      FeH  &   997.3805  &    0.196  &   $-$0.783 \\
      FeH  &   997.4475  &    0.108  &   $-$1.162 \\
      FeH  &   999.4838  &    0.051  &   $-$1.588 \\
      FeH  &  1000.3440  &    0.266  &   $-$0.700 \\
      FeH  &  1001.9530  &    0.066  &   $-$1.446 \\
      FeH  &  1003.3060  &    0.427  &   $-$0.565 \\
      FeH  &  1004.3970  &    0.322  &   $-$0.654 \\
      FeH  &  1005.2320  &    0.066  &   $-$1.544 \\
      FeH  &  1005.5340  &    0.097  &   $-$1.640 \\
      FeH  &  1005.7510  &    0.239  &   $-$0.778 \\
      FeH  &  1005.8290  &    0.051  &   $-$1.262 \\
      FeH  &  1006.1630  &    0.115  &   $-$1.684 \\
      FeH  &  1006.5960  &    0.084  &   $-$1.198 \\
      FeH  &  1007.1900  &    0.262  &   $-$0.749 \\
      FeH  &  1009.4640  &    0.073  &   $-$1.080 \\
      FeH  &  1009.5940  &    0.288  &   $-$0.722 \\
      FeH  &  1009.9350  &    0.468  &   $-$0.550 \\
      FeH  &  1009.9620  &    0.108  &   $-$1.254 \\
      FeH  &  1010.7970  &    0.106  &   $-$1.050 \\
      FeH  &  1011.0850  &    0.538  &   $-$0.523 \\
      FeH  &  1011.5000  &    0.086  &   $-$1.015 \\
      FeH  &  1011.7960  &    0.315  &   $-$0.697 \\
      FeH  &  1011.9700  &    0.416  &   $-$0.597 \\
      FeH  &  1012.0330  &    0.108  &   $-$1.050 \\
    \enddata
  \tablenotetext{$*$}{Wavelength in vacuum}
  \tablenotetext{$\dag$}{Lower excitation potential}
  \tablenotetext{$\ddag$}{Oscillator strength}
\end{deluxetable}

\begin{figure}
    \plotone{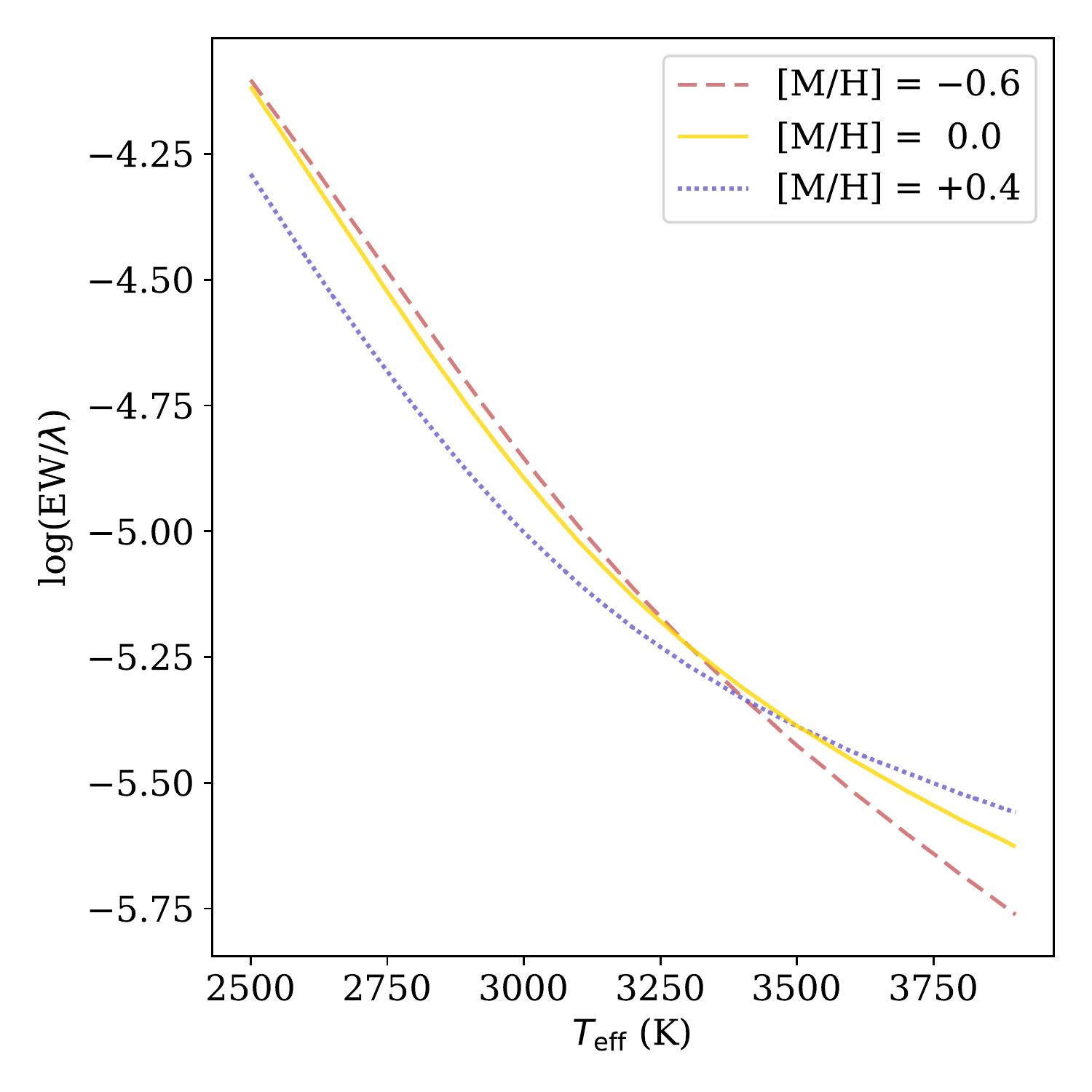}
    \caption{Sensitivity of EWs of
    \added{%
    a representative FeH line
    }%
    in model spectra to the adopted $T_{\mathrm{eff}}$. %
    The vertical axis shows the logarithmic EW normalized by the wavelength. %
    \added{%
    Different colors or line styles represent the cases with different metallicities adopted in EW calculation.
    }%
    } \label{fig:FeH_Teff_sensi}
\end{figure}

We check that the FeH lines are exclusively sensitive to $T_{\mathrm{eff}}$ rather than [Fe/H] or $\log{g}$. %
\added{%
Figure \ref{fig:FeH_Teff_sensi} shows the EW change of a representative FeH line calculated using model atmospheres %
as a function of $T_{\mathrm{eff}}$.
Different colors represent the cases with different metallicities adopted in EW calculation. %
The other parameters of $\log{g}$ = 5.0 and $\xi$ = 0.5 km s$^{-1}$ are adopted for all the cases. %
We confirmed that
}%
the change in $\log{g}$ of 0.1 dex does not affect the EWs by more than 0.05 dex.
\added{%
Focusing on the $T_{\mathrm{eff}}$ range around our targets, the variation in the EWs along with the change in the overall metallicity [M/H] of $-$0.6 or $+$0.4 dex is less than 0.15 dex for most of the lines. %
An EW change of 0.15 dex roughly corresponds to the result of a 100 K change in $T_{\mathrm{eff}}$.
Thus, the effect of metallicity on the $T_{\mathrm{eff}}$ estimation from the FeH lines is not significant here and further details will be dealt with in a future paper.
}%
The EWs are insensitive to metallicity partly due to the change in the continuous opacity of negative hydrogen ions (H$^-$), as in the case of atomic lines reported by Ish20. %
The other cause can be the temperature increase due to the line-blanketing effect of increased metal, which dissociates the FeH molecules.
In the $T_{\mathrm{eff}}$ range studied here, these effects that reduce the EWs of FeH lines are balanced by the effect of the increase of iron abundance, resulting in the FeH line being insensitive to the overall metallicity.
\added{%
The effect of reducing EWs is larger at $T_{\mathrm{eff}}$ lower than $\sim$3300 K, whereas that of increasing EWs is larger at higher $T_{\mathrm{eff}}$, as found in Figure \ref{fig:FeH_Teff_sensi}. %
}%

We measured the EWs of the FeH lines by fitting the Gaussian profiles to observational data.
The Gaussian profiles can fit the line profiles satisfactorily because the damping wings of the FeH lines are not significant.
\added{%
We determined $T_{\mathrm{eff}}$ from each FeH line by demanding that the observed EW be reproduced by the calculation. %
}%
The $\log{g}$ = 5.0, $\xi$ = 0.5 km s$^{-1}$, and [M/H] = 0.0 are assumed throughout the $T_{\mathrm{eff}}$ estimation for the simplicity. %
The average of all the values is taken as the $T_{\mathrm{eff\mathchar`-FeH}}$ for a certain object.
The $T_{\mathrm{eff\mathchar`-FeH}}$ of each target is listed in Table \ref{tab:IRD_Teff_FeH}.
The errors given in the table are the standard deviation ($\sigma$) of estimates based on individual FeH lines divided by the square root of the number of lines ($\sigma / \sqrt{N}$) and do not include any possible systematic errors.
\begin{deluxetable}{lc} %
  \tablecaption{$T_{\mathrm{eff\mathchar`-FeH}}$ of the target M dwarfs}
  \label{tab:IRD_Teff_FeH}
    \tablehead{
      \multicolumn{1}{c}{Objects}  &  $T_{\mathrm{eff\mathchar`-FeH}}$ (K)
    }
    \decimals
    \startdata
      GJ 436     &  3478$\pm$9  \\
      GJ 699     &  3319$\pm$8  \\
      LSPM J1306+3050     &  3077$\pm$6  \\
      LSPM J1523+1727     &  3239$\pm$7  \\
      LSPM J1652+6304     &  3210$\pm$7  \\
      LSPM J1703+5124     &  3196$\pm$6  \\
      LSPM J1802+3731     &  3145$\pm$6  \\
      LSPM J1816+0452     &  3169$\pm$7  \\
      LSPM J1909+1740     &  3209$\pm$6  \\
      LSPM J2026+5834     &  3151$\pm$6  \\
      LSPM J2043+0445     &  3131$\pm$7  \\
      LSPM J2053+1037     &  3203$\pm$6  \\
      LSPM J2343+3632     &  3186$\pm$7  \\
    \enddata
  \tablecomments{
  \added{%
  The errors given here do not include any possible systematic errors but only the standard deviation ($\sigma$) of estimates based on individual FeH lines divided by the square root of the number of lines ($\sigma / \sqrt{N}$). %
  }%
  }
\end{deluxetable}

\begin{figure}
  \plotone{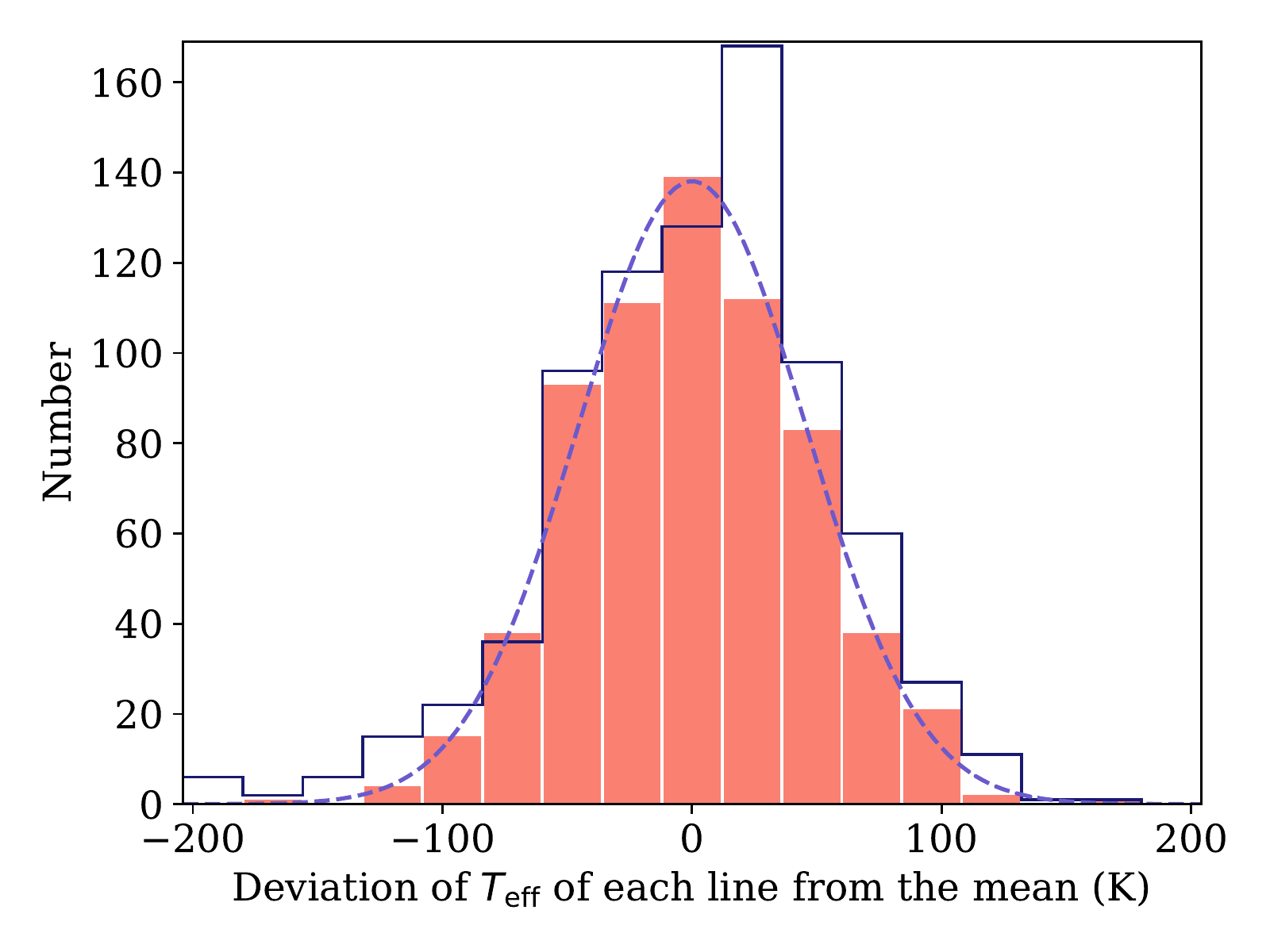}
    \caption{A histogram of the deviation of $T_{\mathrm{eff}}$ derived by each FeH line from the mean of $T_{\mathrm{eff}}$ derived by all the lines of the corresponding object.
    The data of all objects are stacked.
    The red-filled histogram is the final case, where we selected 47 FeH lines, and the dashed curve shows a Gaussian fit to it.
    Also shown as a reference with the histogram outlined in black is the case with 57 FeH lines before the final selection, which deviates from the normal distribution.
    } \label{fig:hist_teff_dev_stacked}
\end{figure}

We confirmed that the scatter among the estimated $T_{\mathrm{eff}}$ from individual FeH lines is dominated by statistical errors that follow a normal distribution.
Figure \ref{fig:hist_teff_dev_stacked} shows a histogram of the deviation of individual-line results from the mean for each object.
The Shapiro-Wilk test results in a p-value of 0.4, which means that it is a normal distribution. %
During the final line selection process,
we excluded FeH lines that often lead to outlier $T_{\mathrm{eff}}$, which should be due to unidentified contamination and make a $T_{\mathrm{eff}}$-distribution asymmetric.
The histogram before the selection is also shown in Figure \ref{fig:hist_teff_dev_stacked} as a reference.

\begin{figure}
   \plotone{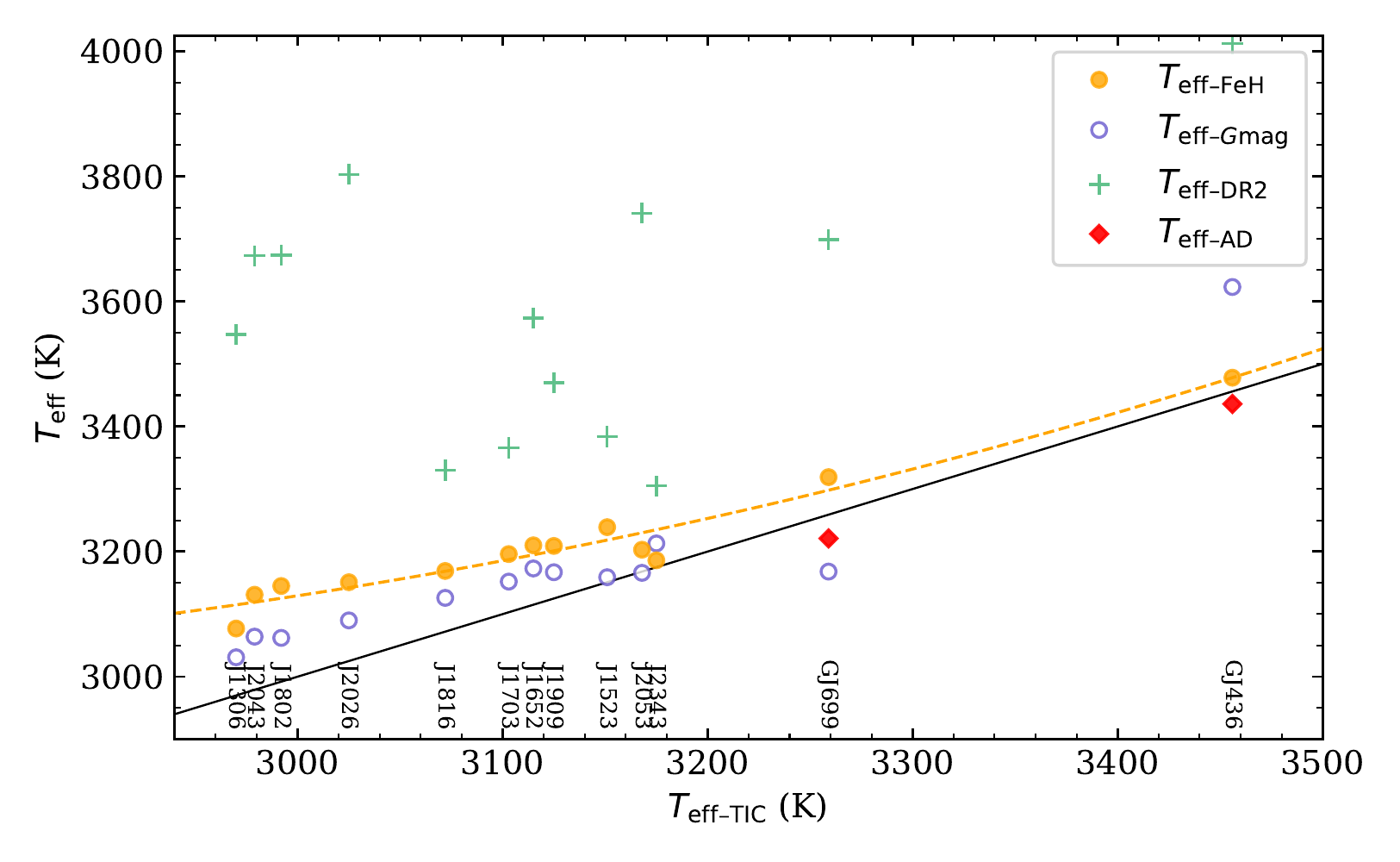} %
    \caption{
    Comparison of different estimates of $T_{\mathrm{eff}}$ as a function of the $T_{\mathrm{eff\mathchar`-TIC}}$, which we adopt for the abundance analysis.
    The orange filled circles present our estimates from the FeH lines in the IRD spectra ($T_{\mathrm{eff\mathchar`-FeH}}$).
    The other $T_{\mathrm{eff}}$ shown by the purple open circles ($T_{\mathrm{eff\mathchar`-}G\mathrm{mag}}$), green plus signs ($T_{\mathrm{eff\mathchar`-DR2}}$), and red diamonds ($T_{\mathrm{eff\mathchar`-AD}}$) show the values of
    empirical estimates from absolute $G$-band magnitudes, those reported in Gaia DR2, and those calculated from angular diameters measured with interferometry, respectively.
    The black solid line shows the one-to-one relation. %
    The orange dashed curve shows a quadratic fitting to the $T_{\mathrm{eff\mathchar`-FeH}}$.
    } \label{fig:Teff_feh_vs_Teff_color}
\end{figure}

We compare the $T_{\mathrm{eff}}$ values estimated with various methods in Figure \ref{fig:Teff_feh_vs_Teff_color}.
The $T_{\mathrm{eff\mathchar`-FeH}}$ that we derived from FeH lines, indicated by the orange filled circles in the figure, lies higher than $T_{\mathrm{eff\mathchar`-TIC}}$. %
The purple circles are the estimates from the absolute $G$-band magnitude based on the empirical relation calibrated by \citet{2019MNRAS.484.2674R}.
The absolute $G$-band magnitude is calculated from $G$-band photometry and parallaxes measured by Gaia.
The green plus signs are the $T_{\mathrm{eff}}$ estimated with Apsis software~\citep{2018A&A...616A...8A} as part of Gaia DR2~\citep{2018A&A...616A...1G}.

The $T_{\mathrm{eff\mathchar`-TIC}}$, $T_{\mathrm{eff\mathchar`-FeH}}$, and $T_{\mathrm{eff\mathchar`-}G\mathrm{mag}}$ share a common trend among our targets. %
This implies that, although there are offsets between them, statistical errors in these $T_{\mathrm{eff}}$ estimates are small, which is thanks to the substantial number of FeH lines, the high photometric accuracy of Gaia, and the strong $T_{\mathrm{eff}}$ dependence of the FeH lines and Gaia photometry. %
The difference between $T_{\mathrm{eff\mathchar`-FeH}}$ and $T_{\mathrm{eff\mathchar`-TIC}}$ tends to be larger at lower temperatures, as shown by the quadratic fit (orange dashed line in the figure).
This indicates that the departure of $T_{\mathrm{eff\mathchar`-FeH}}$ and/or $T_{\mathrm{eff\mathchar`-TIC}}$ from the true value are larger for cooler stars. %
We examined the case of a cooler object, TRAPPIST-1 ($T_{\mathrm{eff}} \sim$ 2400--2600 K ~\citep[e.g.,][]{2017Natur.542..456G, 2018ApJ...853...30V}), using the data observed by \citet{2020ApJ...890L..27H}.
We found that the FeH lines in our synthetic spectra have wider wings than observed ones, unless arbitrary calibrations were made to the approximation formula of \citet{1955psmb.book.....U}.
This suggests that the $T_{\mathrm{eff\mathchar`-FeH}}$ might be overestimated in cooler stars. %
Besides, the $T_{\mathrm{eff\mathchar`-FeH}}$, $T_{\mathrm{eff\mathchar`-TIC}}$, and $T_{\mathrm{eff\mathchar`-AD}}$ of the hottest object GJ 436 agree with each other within 50 K. %
This reinforces the fact that the $T_{\mathrm{eff}}$ discrepancy between these methods is due to problems specific to low temperatures.

However, there is also an uncertainty in the empirical estimation of $T_{\mathrm{eff\mathchar`-TIC}}$ at lower temperatures.
The empirical relation based on the $G_\mathrm{BP} - G_\mathrm{RP}$ color %
is calibrated by the targets of Man15, where $T_{\mathrm{eff}}$ was estimated based on fitting of the BT-Settl-model spectra to the observed medium-resolution spectra. %
This fitting method in Man15 was confirmed by \citet{2013ApJ...779..188M} using the objects with angular diameter measurements, but their samples do not include objects cooler than 3238 K. %
Hence, the $T_{\mathrm{eff\mathchar`-TIC}}$ for cooler stars is not observationally calibrated. %

$T_{\mathrm{eff}}$ can be determined from the stellar angular diameter ($T_{\mathrm{eff\mathchar`-AD}}$) independently of any models based on the relation $F_{\mathrm{bol}} \propto L/d^2 \propto (R/d)^2 {T_{\mathrm{eff}}}^4 \propto \theta^2 {T_{\mathrm{eff}}}^4$, where $L$, $d$, $R$, and $\theta$ are the mean luminosity, distance, radius, and angular diameter of the star, respectively. %
However, the faintness and small angular diameters of late-M dwarfs limit its application with sufficient accuracy to only a few objects very close to the Sun.
The coolest example is GJ 406 ($T_{\mathrm{eff\mathchar`-AD}}$ = 2657 $\pm$ 20 K) reported by \citet{2019MNRAS.484.2674R}.
Although the temperature of this object is not available in TIC unfortunately, %
a high-resolution near-infrared spectrum of this object is available in the CARMENES GTO Data Archive~\citep{2018AaA...612A..49R}.
We determined $T_{\mathrm{eff\mathchar`-FeH}}$ = 2871 K from the spectrum. %
Considering the possibility that the $T_{\mathrm{eff\mathchar`-FeH}}$ is more overestimated for cooler objects, it may be possible to reproduce the $T_{\mathrm{eff\mathchar`-AD}}$ by calibrating the $T_{\mathrm{eff\mathchar`-FeH}}$ appropriately in future work.
The extension of the reach of interferometric measurements to cooler M dwarfs is expected to drive the further understanding of the accurate $T_{\mathrm{eff}}$ in this regime.

It should be noted that GJ 406 is magnetically active.
\citet{2018AaA...612A..49R} reported the significant emission of H$\alpha$ (EW$_{\mathrm{H}\alpha} \sim -$9.25~\AA). %
We confirmed that the FeH lines that are sensitive to the Zeeman effect~\citep{2006ApJ...644..497R} are broader than other lines. %
The FeH lines which are clearly broader than the bulk of other lines are excluded by visual inspection from the $T_{\mathrm{eff\mathchar`-FeH}}$ estimation. %
The IRD-SSP sample does not include magnetically active M dwarfs.
If the analysis is further applied to the magnetically active stars, we need to select the lines taking the Land\'{e} g-factors into consideration. %

The investigation above suggests that the equivalent widths of FeH lines can be used to estimate $T_{\mathrm{eff}}$ of mid-to-late M dwarfs if an appropriate correction is applied. %
We decide not to adopt them in the abundance analysis in this work because the sample size is too small to calibrate the $T_{\mathrm{eff\mathchar`-FeH}}$ scale.
\added{%
Instead
}%
we adopt the $T_{\mathrm{eff\mathchar`-TIC}}$ and its uncertainty recorded in the TIC.
However, the comparison with $T_{\mathrm{eff\mathchar`-FeH}}$ in Figure \ref{fig:Teff_feh_vs_Teff_color} suggests that this uncertainty may be overestimated.

\subsection{Equivalent width measurement} \label{sec:IRD_equivalent_width_measurement} %
\added{%
The second modification from Ish20 is the procedure of EW measurement.
The details in the spectral line selection and the EW measurement methods are described below.
}%

The spectral lines used in the analysis are selected following the same criteria as Ish20.
We investigated the behavior of the absorption lines using synthetic spectra and confirmed that the sensitivity of most lines to the metallicity and stellar parameters is comparable even at the lower temperature range (2900 $< T_{\mathrm{eff}} <$ 3200 K) with that at $\sim$3200 K.

The exception is the three K I lines, which are much more sensitive to $T_{\mathrm{eff}}$ than [K/H] at the low temperature.
We analyze the K I lines only for the hottest target GJ 436.
On the other hand, we added some lines which were unavailable in Ish20 due to the detector gaps of CARMENES, thanks to the high continuity of the wavelength coverage of IRD.
The additional lines were selected by applying the same considerations that were noted in Ish20, such as contamination and sensitivity. %
One of the notable lines is the Sr II line at 1091.79 nm.
We analyzed it for all the objects to determine [Sr/H].
Another is the Si I line at 1198.75 nm, which is a unique Si I line detectable in the wavelength range of IRD for late-M dwarfs, but we found that it is not sensitive to Si abundance enough at the lower $T_{\mathrm{eff}}$. %
We used the line only for GJ 436 together with two other Si I lines which are not detectable for the other objects.   %
We also added a V I line at 1225.28 nm only for GJ 436 due to the relatively high abundance sensitivity at higher $T_{\mathrm{eff}}$.
\startlongtable
\begin{deluxetable}{lcccc} %
  \tablecaption{Atomic line list}
  \label{tab:IRD_linelist}
    \tablehead{
      Species &
      $\lambda$ (nm)\tablenotemark{$*$} &
      $E_\mathrm{low}$ (eV)\tablenotemark{$\dag$} &
      $\log gf$\tablenotemark{$\ddag$} &
      vdW $\,$\tablenotemark{${\S}$}
      }
      \decimals
      \startdata
      Na I  &  1074.9393  &    3.191  &   $-$1.294 &    .... \\
      Na I  &  1083.7814  &    3.617  &   $-$0.503 &    .... \\
      Na I  &  1268.2639  &    3.617  &   $-$0.043 &   $-$6.653  \\
      Mg I  &  1183.1409  &    4.346  &   $-$0.333 &   $-$7.192  \\
      Si I  &  1078.9804  &    4.930  &   $-$0.303 &   $-$7.272  \\
      Si I  &  1083.0054  &    4.954  &    0.302 &   $-$7.266  \\
      Si I  &  1198.7478  &    4.930  &    0.239 &   $-$7.298  \\
      K I  &  1177.2861  &    1.617  &   $-$0.450 &   $-$7.326  \\
      K I  &  1243.5675  &    1.610  &   $-$0.439 &   $-$7.022  \\
      Ca I  &  1083.6349  &    4.877  &   $-$0.244 &   $-$7.590  \\
      Ca I  &  1195.9227  &    4.131  &   $-$0.849 &   $-$7.300  \\
      Ca I  &  1195.9228  &    4.131  &   $-$0.849 &   $-$7.300  \\
      Ca I  &  1281.9551  &    3.910  &   $-$0.765 &   $-$7.520  \\
      Ca I  &  1282.7375  &    3.910  &   $-$0.997 &   $-$7.520  \\
      Ca I  &  1283.0568  &    3.910  &   $-$1.478 &   $-$7.520  \\
      Ca I  &  1291.2601  &    4.430  &   $-$0.224 &   $-$7.710  \\
      Ca I  &  1303.7119  &    4.441  &   $-$0.064 &   $-$7.710  \\
      Ca I  &  1306.1457  &    4.441  &   $-$1.092 &   $-$7.710  \\
      Ca I  &  1313.8534  &    4.451  &    0.085 &   $-$7.710  \\
      Ti I  &   967.8197  &    0.836  &   $-$0.804 &   $-$7.800  \\
      Ti I  &   972.1625  &    1.503  &   $-$1.181 &   $-$7.780  \\
      Ti I  &   972.1626  &    1.503  &   $-$1.181 &   $-$7.780  \\
      Ti I  &   973.1075  &    0.818  &   $-$1.206 &   $-$7.800  \\
      Ti I  &   974.6277  &    0.813  &   $-$1.306 &   $-$7.800  \\
      Ti I  &   974.6278  &    0.813  &   $-$1.306 &   $-$7.800  \\
      Ti I  &   977.2980  &    0.848  &   $-$1.581 &   $-$7.800  \\
      Ti I  &   979.0371  &    0.826  &   $-$1.444 &   $-$7.800  \\
      Ti I  &   979.0372  &    0.826  &   $-$1.444 &   $-$7.800  \\
      Ti I  &   983.4836  &    1.887  &   $-$1.130 &   $-$7.634  \\
      Ti I  &  1000.5831  &    2.160  &   $-$1.210 &   $-$7.780  \\
      Ti I  &  1005.1583  &    1.443  &   $-$1.930 &   $-$7.780  \\
      Ti I  &  1039.9651  &    0.848  &   $-$1.539 &   $-$7.810  \\
      Ti I  &  1058.7533  &    0.826  &   $-$1.775 &   $-$7.810  \\
      Ti I  &  1061.0624  &    0.848  &   $-$2.697 &   $-$7.810  \\
      Ti I  &  1066.4544  &    0.818  &   $-$1.915 &   $-$7.810  \\
      Ti I  &  1073.5804  &    0.826  &   $-$2.515 &   $-$7.810  \\
      Ti I  &  1077.7818  &    0.818  &   $-$2.666 &   $-$7.810  \\
      Ti I  &  1083.0863  &    0.836  &   $-$3.910 &   $-$7.810  \\
      Ti I  &  1085.0605  &    0.826  &   $-$3.922 &   $-$7.810  \\
      Ti I  &  1180.0415  &    1.430  &   $-$2.280 &   $-$7.790  \\
      Ti I  &  1180.0416  &    1.430  &   $-$2.280 &   $-$7.790  \\
      Ti I  &  1189.6132  &    1.430  &   $-$1.730 &   $-$7.790  \\
      Ti I  &  1281.4983  &    2.160  &   $-$1.390 &   $-$7.750  \\
      Ti I  &  1282.5179  &    1.460  &   $-$1.190 &   $-$7.790  \\
      Ti I  &  1282.5180  &    1.460  &   $-$1.190 &   $-$7.790  \\
      Ti I  &  1292.3433  &    2.154  &   $-$1.560 &   $-$7.750  \\
      Ti I  &  1292.3433  &    2.153  &   $-$1.560 &   $-$7.750  \\
      V I  &  1225.2755  &    2.359  &   $-$0.999 &   $-$7.780  \\
      Cr I  &  1065.0557  &    3.011  &   $-$1.582 &   $-$7.770  \\
      Cr I  &  1080.4319  &    3.011  &   $-$1.562 &   $-$7.780  \\
      Cr I  &  1081.9873  &    3.013  &   $-$1.894 &   $-$7.780  \\
      Cr I  &  1082.4625  &    3.013  &   $-$1.520 &   $-$7.780  \\
      Cr I  &  1090.8697  &    3.438  &   $-$0.561 &   $-$7.530  \\
      Cr I  &  1093.2864  &    3.010  &   $-$1.858 &   $-$7.780  \\
      Cr I  &  1139.3869  &    3.322  &   $-$0.423 &   $-$7.640  \\
      Cr I  &  1161.3739  &    3.321  &    0.055 &   $-$7.640  \\
      Cr I  &  1291.3622  &    2.708  &   $-$1.779 &   $-$7.800  \\
      Cr I  &  1294.0559  &    2.710  &   $-$1.896 &   $-$7.800  \\
      Mn I  &  1290.3289  &    2.114  &   $-$1.070 &    .... \\
      Mn I  &  1297.9459  &    2.888  &   $-$1.090 &    .... \\
      Fe I  &  1038.1843  &    2.223  &   $-$4.148 &   $-$7.800  \\
      Fe I  &  1039.8643  &    2.176  &   $-$3.393 &   $-$7.800  \\
      Fe I  &  1042.5884  &    2.692  &   $-$3.616 &   $-$7.810  \\
      Fe I  &  1061.9630  &    3.267  &   $-$3.127 &   $-$7.780  \\
      Fe I  &  1078.6004  &    3.111  &   $-$2.567 &   $-$7.790  \\
      Fe I  &  1082.1238  &    3.960  &   $-$1.948 &   $-$7.820  \\
      Fe I  &  1088.4739  &    2.845  &   $-$3.604 &   $-$7.810  \\
      Fe I  &  1088.7244  &    3.929  &   $-$1.925 &   $-$7.820  \\
      Fe I  &  1089.9284  &    3.071  &   $-$2.694 &   $-$7.790  \\
      Fe I  &  1288.3289  &    2.279  &   $-$3.458 &   $-$7.820  \\
      Sr II  &  1091.7877  &    1.805  &   $-$0.526 &   $-$7.641  \\
    \enddata
  \tablenotetext{$*$}{Wavelength in vacuum}
  \tablenotetext{$\dag$}{Lower excitation potential}
  \tablenotetext{$\ddag$}{Oscillator strength}
  \tablenotetext{${\S}$}{Van der Waals damping parameter}
\end{deluxetable}

\begin{figure*}\begin{center}
  \includegraphics[width=80mm]{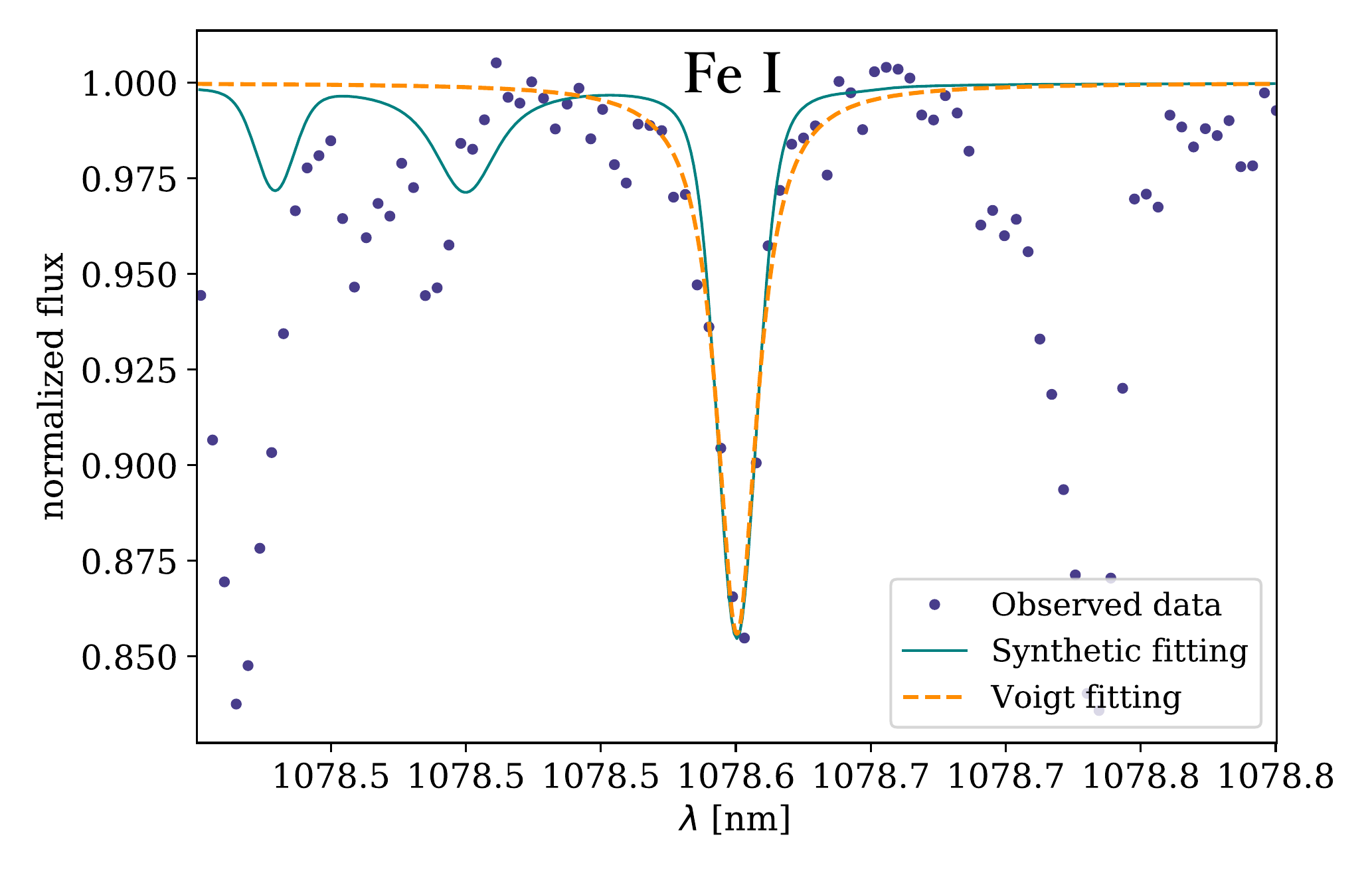}
  \includegraphics[width=80mm]{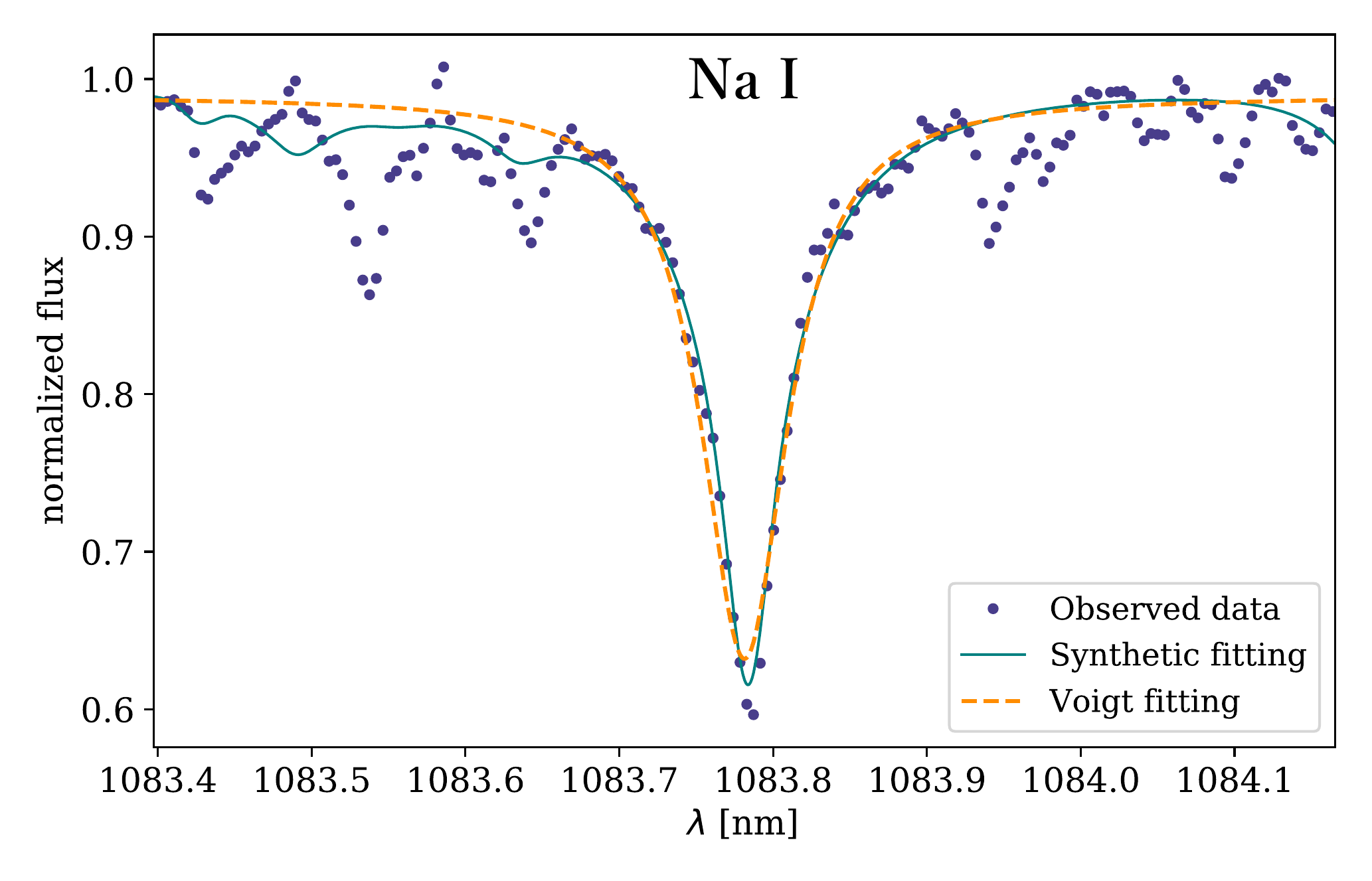}
  \caption{
  Two examples of atomic lines to show the small difference between the fitting by synthetic spectra and that by Voigt profiles.
  The data shown by black dots are the observed spectra of $\rho$ Cnc B, the coolest ($T_{\mathrm{eff}} \sim 3166$ K) target of Ish20. The green solid line and orange dashed line are the best fit by synthetic spectra and Voigt profiles, respectively.
  Note that the two panels have different scales for both axes.
  } \label{fig:compare_EW_splot_vs_fit}
\end{center}\end{figure*}

\added{%
To determine the EW of each of these selected lines, we fitted synthetic spectra to the narrow wavelength region (0.04--0.2 nm depending on lines) around the individual line in the observed spectra.
This is different from Ish20, where the EWs were measured by fitting the Gaussian or Voigt profiles except for Na I and Mn I lines.
}%
This modification is mainly for the automation and homogenization of the analysis procedure.
Another benefit is that for broad absorption lines, synthetic spectra can reproduce their profiles more accurately than the Gaussian and Voigt profiles.
We initially fixed for all objects the selected line list and the wavelength range to fit for each line.
The wavelength range for fitting was selected to cover the line profile from the wings to the core, while avoiding wavelengths where contamination is suspected. %
This allows us to measure EW automatically and consistently for all objects.
When there is contamination of other absorption lines for which reliable data are available, we fitted the synthetic spectra including the blending lines in the calculation. %
The EW of the lines of interest is then calculated by employing the best-fit abundance of the corresponding element.
For confirmation, we applied this modification to the M dwarfs in binary systems analyzed by Ish20, and found that the change in the results is within the error range and the agreement with the primary stars is maintained.
Figure \ref{fig:compare_EW_splot_vs_fit} shows
\added{%
examples
}%
 of atomic lines overplotted by the best-fit synthetic spectra and Voigt profiles. %
The differences are negligible within the accuracy of our analysis, but the stronger the line, the less able the Voigt profile is to simultaneously reproduce its broad wings and relatively narrow core.

Subsequently, we visually check the agreement between the best-fit synthetic spectrum and the observed one for each line to exclude lines that show unrealistic fitting. %
\added{%
Many of the candidate Ca I lines were excluded because their core width could not be reproduced in the model spectra. %
This could be due to the effect of non-LTE as pointed out by \citet{2021A&A...649A.103O}. %
Among the three Na I lines, the weakest line at 1074.94 nm was excluded for GJ 699, and the strongest line at 1268.26 nm was excluded for GJ436.
The same set of absorption lines was homogeneously used for all 11 objects with 2900 $< T_{\mathrm{eff}} <$ 3200 because they show similar spectra.
The Fe I line at 1088.47 nm and the Ti I line at 1083.09 nm were excluded for all 11 objects because they appear wider in the observed data of some objects than in the models.
}%

\subsection{Kinematics} \label{sec:IRD_kinematics}
For the kinematic classification of the Galactic populations, we calculated the Galactocentric space velocities of each M dwarf.

We first measured the radial velocity ($v_\mathrm{rad}$) using the IRD spectra in a relatively telluric-free bandpass, i.e. 1030--1330 nm.
In these wavelengths, we took the cross-correlation between the observed spectra and synthetic spectra calculated from the VALD line list and the MARCS model atmosphere with $T_{\mathrm{eff}}$ = 3000 K, $\log{g}$ = 5.0, and [Fe/H] = 0.0 to obtain the relative velocity.
The barycentric correction is made for the obtained relative velocity based on the observation dates and location. %
We shifted these velocities by $-$0.1 km s$^{-1}$ for all our M dwarfs so as the velocity of GJ 699 measured by the same procedure matches the value from \citet{2018AaA...612A..49R}.
This small correction does not affect the results.

Second, the space velocities $UVW$ were calculated by combining the $v_\mathrm{rad}$ with the astrometric measurements (i.e. coordinates, proper motions, and parallaxes) in Gaia EDR3 \citep{2021A&A...649A...1G}.
The $U$, $V$, and $W$ are the radial, azimuthal, and vertical components of Galactocentric space velocities with respect to the local standard of rest (LSR), respectively.
They are calculated adopting the Sun-Galactic center distance of 8.2 kpc \citep{2017MNRAS.465...76M}, the height of the Sun above the Galactic midplane of 0.025 kpc \citep{2008ApJ...673..864J}, and the solar motion relative to the LSR of ($U_\odot$, $V_\odot$, $W_\odot$) = (11.10, 12.24, 7.25) km s$^{-1}$ \citep{2010MNRAS.403.1829S}.
The coordinate transformation to the Galactocentric Cartesian system was conducted with the astropy.coordinates package. %

Table \ref{tab:IRD_kinematics} lists the radial velocities $v_\mathrm{rad}$ and the space velocities $UVW$ of our M dwarfs with their distances and proper motion measured by Gaia.
\added{%
The distances were simply calculated as the inverse of the parallax, but they do not differ by more than 0.1\,\%
from the photogeometric estimates by \citet{2021yCat.1352....0B}.
}%
The table also includes the kinematics of the six M dwarfs analyzed by Ish20 for which we calculate $UVW$ with the same procedure adopting the $v_\mathrm{rad}$ from \citet{2018AaA...612A..49R}.
An exception is G 102-4, whose astrometric data and $v_\mathrm{rad}$ are adopted from \citet{2018MNRAS.479.1332M} and \citet{2015ApJ...802L..10T}, respectively.
Note that its distance is not the parallactic distance but the spectro-photometric estimate based on the $M_J$--spectral type relation of \citet{2017A&A...597A..47C} assuming solar metallicity so must be handled with care.

\begin{deluxetable*}{lccccccc}
  \tablecaption{Kinematics of the target M dwarfs}
  \label{tab:IRD_kinematics}
    \tablehead{ %
      Object & distance & PM$_\mathrm{R.A.}$ & PM$_\mathrm{Decl.}$ & $v_\mathrm{rad}$ & $U$ & $V$ & $W$ \\
      {} & (pc) & (arcsec/yr) & (arcsec/yr) & (km s$^{-1}$)    & (km s$^{-1}$) & (km s$^{-1}$) & (km s$^{-1}$) %
    }
    \decimals
    \startdata
      GJ 436              &   9.8 &   0.8951 &  $-$0.8135 &     9.5 &    61.4 &    $-$6.3 &    26.3 \\
      GJ 699             &   1.8 &  $-$0.8016 &  10.3624 &  $-$110.6 &  $-$130.2 &    17.0 &    26.1 \\
      LSPM J1306+3050    &  14.1 &   0.1822 &  $-$0.4748 &     9.8 &    39.4 &    $-$6.0 &    18.3 \\
      LSPM J1523+1727    &  11.2 &  $-$0.3919 &  $-$1.2593 &    42.2 &    68.5 &   $-$37.8 &    37.5 \\
      LSPM J1652+6304    &  15.2 &   0.1474 &   0.1529 &   $-$21.1 &     3.3 &     2.9 &   $-$15.7 \\
      LSPM J1703+5124    &   9.9 &   0.1244 &   0.6101 &    37.3 &    $-$9.9 &    51.2 &    24.5 \\
      LSPM J1802+3731    &  12.2 &   0.1712 &  $-$1.1417 &     4.1 &    72.7 &    $-$0.5 &   $-$16.2 \\
      LSPM J1909+1740    &  10.3 &  $-$0.6406 &  $-$0.4241 &   $-$13.8 &    27.0 &   $-$20.0 &    24.5 \\
      LSPM J2026+5834    &   9.6 &   0.2610 &   0.5429 &   $-$59.4 &   $-$12.1 &   $-$48.2 &    $-$1.0 \\
      LSPM J1816+0452    &  14.7 &  $-$0.1554 &   0.4071 &   $-$53.2 &   $-$47.0 &    $-$1.5 &    20.2 \\
      LSPM J2043+0445    &  15.0 &   0.4463 &  $-$0.1455 &   $-$48.1 &   $-$30.8 &   $-$26.8 &    $-$4.6 \\
      LSPM J2053+1037    &  12.8 &  $-$0.4925 &  $-$0.4472 &   $-$20.2 &    36.2 &   $-$21.9 &    22.9 \\
      LSPM J2343+3632    &   8.4 &   0.9418 &  $-$0.1513 &    $-$2.9 &   $-$18.5 &    $-$7.1 &    $-$6.5 \\
      HD 233153          &  12.3 &   0.0039 &  $-$0.5160 &     2.0\tablenotemark{$*$} &    $-$3.2 &   $-$10.2 &    $-$6.7 \\
      HD 154363 B          &  10.5 &  $-$0.9172 &  $-$1.1322 &    34.7\tablenotemark{$*$} &    58.2 &   $-$49.2 &    28.2 \\
      BX Cet             &   7.2 &   1.8017 &   1.4505 &    70.4\tablenotemark{$*$} &   $-$94.5 &    21.5 &     7.1 \\
      G 102-4             &  19.5\tablenotemark{$\dag$} &   0.0951\tablenotemark{$\dag$} &  $-$0.2152\tablenotemark{$\dag$} &    17.3\tablenotemark{$\ddag$} &    $-$0.4 &   $-$12.2 &     0.8 \\
      $\rho$ Cnc B          &  12.6 &  $-$0.4812 &  $-$0.2445 &    73.3\tablenotemark{$*$} &   $-$60.5 &   $-$17.1 &    27.5 \\
      BD-02 2198         &  14.2 &   0.0741 &  $-$0.2931 &   $-$17.1\tablenotemark{$*$} &    36.3 &     7.9 &    $-$0.2 \\
    \enddata
  \tablenotetext{$*$}{From \citet{2018AaA...612A..49R}.}
  \tablenotetext{$\dag$}{From \citet{2018MNRAS.479.1332M}.}
  \tablenotetext{$\ddag$}{From \citet{2015ApJ...802L..10T}.}
\end{deluxetable*}

\section{Results} \label{sec:IRD_results}

\subsection{Elemental abundance} \label{sec:IRD_elemental_abundance}
The results and corresponding errors of elemental abundance determination from the IRD-SSP data are tabulated in Table \ref{tab:IRD_results} and plotted in Figure \ref{fig:res_IRD_SSP_targets}.
We found that the abundance ratios of individual elements [X/H] are aligned with those of iron [Fe/H] in all M dwarfs, with no stars showing a peculiar abundance pattern.
We also calculated an error-weighted average of abundances of all the elements for each object ([M$_\mathrm{ave}$/H]; the horizontal dashed lines in Figure \ref{fig:res_IRD_SSP_targets}) and confirmed that none of the elements deviate from it in all objects beyond the measurement errors. %

\startlongtable
\begin{deluxetable*}{llccccccccc} %
  \tablecaption{Abundance results with individual error sources}
  \label{tab:IRD_results}
    \tablehead{
      \multicolumn{1}{c}{Object} &
      Element &
      [X/H]  &
      $N_\mathrm{line}$  &
      $\sigma_\mathrm{SEM}$  &
      $\sigma_{T_{\mathrm{eff}}}$  &
      $\sigma_{\log{g}}$  &
      $\sigma_{\xi}$  &
      $\sigma_\mathrm{OE}$  &
      $\sigma_\mathrm{cont}$  &
      $\sigma_\mathrm{Total}$
      }
    \decimals
    \startdata
      GJ 436
      & Na &  0.04 &     2 &  0.12 &  0.12 &  0.00 &  0.01 &  0.07 &  0.04 &  0.19 \\
      & Mg &  0.26 &     1 &  0.18 &  0.06 &  0.01 &  0.01 &  0.14 &  0.05 &  0.24 \\
      & Si &  0.09 &     3 &  0.10 &  0.36 &  0.00 &  0.01 &  0.18 &  0.05 &  0.41 \\
      & K & $-$0.05 &     2 &  0.12 &  0.21 &  0.01 &  0.03 &  0.10 &  0.03 &  0.27 \\
      & Ca &  0.07 &     9 &  0.05 &  0.05 &  0.00 &  0.02 &  0.11 &  0.02 &  0.13 \\
      & Ti &  0.29 &    22 &  0.02 &  0.04 &  0.01 &  0.04 &  0.22 &  0.02 &  0.23 \\
      & V &  0.21 &     1 &  0.18 &  0.02 &  0.00 &  0.01 &  0.09 &  0.09 &  0.22 \\
      & Cr &  0.12 &    10 &  0.04 &  0.02 &  0.00 &  0.02 &  0.11 &  0.03 &  0.13 \\
      & Mn &  0.19 &     2 &  0.12 &  0.04 &  0.01 &  0.03 &  0.15 &  0.04 &  0.21 \\
      & Fe &  0.08 &    10 &  0.06 &  0.02 &  0.00 &  0.03 &  0.10 &  0.03 &  0.13 \\
      & Sr &  0.12 &     1 &  0.18 &  0.02 &  0.00 &  0.05 &  0.14 &  0.10 &  0.25 \\
      \hline
      GJ 699
      & Na & $-$0.61 &     2 &  0.20 &  0.18 &  0.01 &  0.00 &  0.07 &  0.03 &  0.27 \\
      & Mg & $-$0.52 &     1 &  0.28 &  0.07 &  0.02 &  0.01 &  0.19 &  0.04 &  0.35 \\
      & Ca & $-$0.68 &     9 &  0.05 &  0.07 &  0.01 &  0.00 &  0.09 &  0.02 &  0.13 \\
      & Ti & $-$0.56 &    21 &  0.02 &  0.22 &  0.02 &  0.01 &  0.23 &  0.01 &  0.32 \\
      & Cr & $-$0.70 &     8 &  0.04 &  0.06 &  0.00 &  0.00 &  0.12 &  0.03 &  0.14 \\
      & Mn & $-$0.86 &     2 &  0.20 &  0.13 &  0.01 &  0.01 &  0.16 &  0.04 &  0.29 \\
      & Fe & $-$0.60 &     7 &  0.11 &  0.06 &  0.00 &  0.01 &  0.12 &  0.03 &  0.18 \\
      & Sr & $-$0.66 &     1 &  0.28 &  0.01 &  0.00 &  0.01 &  0.16 &  0.08 &  0.33 \\
      \hline
      LSPM J1306+3050
      & Na & $-$0.24 &     3 &  0.18 &  0.17 &  0.01 &  0.00 &  0.10 &  0.02 &  0.27 \\
      & Mg & $-$0.26 &     1 &  0.32 &  0.14 &  0.03 &  0.00 &  0.19 &  0.04 &  0.40 \\
      & Ca & $-$0.18 &     3 &  0.18 &  0.06 &  0.01 &  0.00 &  0.12 &  0.03 &  0.23 \\
      & Ti & $-$0.09 &    20 &  0.03 &  0.30 &  0.04 &  0.01 &  0.30 &  0.01 &  0.43 \\
      & Cr & $-$0.11 &     8 &  0.07 &  0.11 &  0.01 &  0.00 &  0.13 &  0.02 &  0.18 \\
      & Mn & $-$0.26 &     2 &  0.23 &  0.23 &  0.02 &  0.00 &  0.17 &  0.03 &  0.37 \\
      & Fe & $-$0.09 &     6 &  0.13 &  0.13 &  0.01 &  0.00 &  0.13 &  0.03 &  0.23 \\
      & Sr & $-$0.17 &     1 &  0.32 &  0.06 &  0.00 &  0.00 &  0.17 &  0.08 &  0.37 \\
      \hline
      LSPM J1523+1727
      & Na & $-$0.33 &     3 &  0.13 &  0.20 &  0.01 &  0.00 &  0.08 &  0.02 &  0.25 \\
      & Mg & $-$0.21 &     1 &  0.22 &  0.12 &  0.02 &  0.00 &  0.19 &  0.04 &  0.32 \\
      & Ca & $-$0.29 &     3 &  0.13 &  0.10 &  0.01 &  0.00 &  0.12 &  0.03 &  0.20 \\
      & Ti & $-$0.14 &    20 &  0.03 &  0.28 &  0.03 &  0.01 &  0.27 &  0.01 &  0.39 \\
      & Cr & $-$0.22 &     8 &  0.04 &  0.11 &  0.01 &  0.00 &  0.13 &  0.03 &  0.17 \\
      & Mn & $-$0.40 &     2 &  0.16 &  0.20 &  0.02 &  0.01 &  0.17 &  0.03 &  0.31 \\
      & Fe & $-$0.20 &     6 &  0.09 &  0.12 &  0.01 &  0.01 &  0.13 &  0.03 &  0.20 \\
      & Sr & $-$0.24 &     1 &  0.22 &  0.02 &  0.01 &  0.01 &  0.18 &  0.08 &  0.30 \\
      \hline
      LSPM J1652+6304
      & Na & $-$0.04 &     3 &  0.17 &  0.26 &  0.01 &  0.00 &  0.13 &  0.02 &  0.34 \\
      & Mg &  0.07 &     1 &  0.29 &  0.17 &  0.01 &  0.00 &  0.18 &  0.04 &  0.39 \\
      & Ca &  0.04 &     3 &  0.17 &  0.15 &  0.00 &  0.00 &  0.13 &  0.03 &  0.26 \\
      & Ti &  0.30 &    20 &  0.03 &  0.36 &  0.02 &  0.02 &  0.32 &  0.01 &  0.48 \\
      & Cr &  0.09 &     8 &  0.04 &  0.16 &  0.00 &  0.01 &  0.14 &  0.03 &  0.22 \\
      & Mn &  0.03 &     2 &  0.21 &  0.29 &  0.01 &  0.01 &  0.19 &  0.03 &  0.41 \\
      & Fe &  0.08 &     6 &  0.12 &  0.17 &  0.00 &  0.01 &  0.14 &  0.03 &  0.25 \\
      & Sr &  0.12 &     1 &  0.29 &  0.05 &  0.01 &  0.01 &  0.16 &  0.09 &  0.35 \\
      \hline
      LSPM J1703+5124
      & Na &  0.18 &     3 &  0.12 &  0.30 &  0.03 &  0.02 &  0.15 &  0.02 &  0.36 \\
      & Mg &  0.18 &     1 &  0.21 &  0.20 &  0.03 &  0.02 &  0.18 &  0.04 &  0.34 \\
      & Ca &  0.18 &     3 &  0.12 &  0.18 &  0.02 &  0.02 &  0.14 &  0.03 &  0.26 \\
      & Ti &  0.54 &    20 &  0.03 &  0.42 &  0.05 &  0.05 &  0.33 &  0.01 &  0.54 \\
      & Cr &  0.20 &     8 &  0.04 &  0.19 &  0.02 &  0.02 &  0.14 &  0.03 &  0.24 \\
      & Mn &  0.20 &     2 &  0.15 &  0.34 &  0.03 &  0.03 &  0.20 &  0.04 &  0.42 \\
      & Fe &  0.23 &     6 &  0.08 &  0.20 &  0.02 &  0.03 &  0.14 &  0.03 &  0.26 \\
      & Sr &  0.23 &     1 &  0.21 &  0.05 &  0.01 &  0.03 &  0.15 &  0.09 &  0.28 \\
      \hline
      LSPM J1802+3731
      & Na & $-$0.21 &     3 &  0.16 &  0.18 &  0.01 &  0.00 &  0.11 &  0.02 &  0.26 \\
      & Mg & $-$0.15 &     1 &  0.28 &  0.13 &  0.03 &  0.00 &  0.19 &  0.04 &  0.36 \\
      & Ca & $-$0.11 &     3 &  0.16 &  0.07 &  0.01 &  0.00 &  0.12 &  0.03 &  0.21 \\
      & Ti & $-$0.07 &    20 &  0.04 &  0.30 &  0.04 &  0.01 &  0.31 &  0.01 &  0.43 \\
      & Cr & $-$0.08 &     8 &  0.06 &  0.11 &  0.01 &  0.00 &  0.13 &  0.02 &  0.18 \\
      & Mn & $-$0.27 &     2 &  0.20 &  0.23 &  0.02 &  0.00 &  0.17 &  0.03 &  0.35 \\
      & Fe & $-$0.05 &     6 &  0.11 &  0.13 &  0.01 &  0.00 &  0.14 &  0.03 &  0.22 \\
      & Sr & $-$0.05 &     1 &  0.28 &  0.05 &  0.00 &  0.01 &  0.17 &  0.08 &  0.34 \\
      \hline
      LSPM J1816+0452
      & Na &  0.45 &     3 &  0.15 &  0.38 &  0.04 &  0.02 &  0.17 &  0.02 &  0.44 \\
      & Mg &  0.33 &     1 &  0.25 &  0.27 &  0.04 &  0.02 &  0.18 &  0.04 &  0.42 \\
      & Ca &  0.28 &     3 &  0.15 &  0.22 &  0.03 &  0.01 &  0.14 &  0.03 &  0.31 \\
      & Ti &  0.88 &    20 &  0.02 &  0.54 &  0.07 &  0.05 &  0.34 &  0.01 &  0.65 \\
      & Cr &  0.36 &     8 &  0.04 &  0.25 &  0.03 &  0.02 &  0.14 &  0.03 &  0.30 \\
      & Mn &  0.47 &     2 &  0.18 &  0.43 &  0.05 &  0.03 &  0.21 &  0.04 &  0.51 \\
      & Fe &  0.37 &     6 &  0.10 &  0.26 &  0.03 &  0.03 &  0.14 &  0.04 &  0.32 \\
      & Sr &  0.41 &     1 &  0.25 &  0.11 &  0.02 &  0.03 &  0.15 &  0.10 &  0.33 \\
      \hline
      LSPM J1909+1740
      & Na &  0.12 &     3 &  0.13 &  0.28 &  0.03 &  0.01 &  0.14 &  0.02 &  0.35 \\
      & Mg &  0.15 &     1 &  0.23 &  0.17 &  0.03 &  0.01 &  0.18 &  0.04 &  0.34 \\
      & Ca &  0.11 &     3 &  0.13 &  0.16 &  0.02 &  0.01 &  0.14 &  0.03 &  0.25 \\
      & Ti &  0.48 &    20 &  0.02 &  0.38 &  0.05 &  0.04 &  0.32 &  0.01 &  0.50 \\
      & Cr &  0.16 &     8 &  0.04 &  0.17 &  0.02 &  0.02 &  0.14 &  0.03 &  0.23 \\
      & Mn &  0.16 &     2 &  0.16 &  0.31 &  0.03 &  0.02 &  0.20 &  0.04 &  0.40 \\
      & Fe &  0.18 &     6 &  0.09 &  0.18 &  0.02 &  0.02 &  0.14 &  0.03 &  0.25 \\
      & Sr &  0.16 &     1 &  0.23 &  0.05 &  0.01 &  0.02 &  0.15 &  0.09 &  0.29 \\
      \hline
      LSPM J2026+5834
      & Na &  0.15 &     3 &  0.16 &  0.31 &  0.03 &  0.01 &  0.15 &  0.02 &  0.38 \\
      & Mg &  0.05 &     1 &  0.27 &  0.23 &  0.04 &  0.01 &  0.19 &  0.04 &  0.41 \\
      & Ca &  0.05 &     3 &  0.16 &  0.16 &  0.02 &  0.01 &  0.13 &  0.03 &  0.26 \\
      & Ti &  0.36 &    20 &  0.03 &  0.47 &  0.06 &  0.02 &  0.34 &  0.01 &  0.58 \\
      & Cr &  0.11 &     8 &  0.06 &  0.20 &  0.02 &  0.01 &  0.14 &  0.03 &  0.25 \\
      & Mn &  0.12 &     2 &  0.19 &  0.37 &  0.04 &  0.01 &  0.20 &  0.03 &  0.47 \\
      & Fe &  0.16 &     6 &  0.11 &  0.22 &  0.02 &  0.01 &  0.15 &  0.03 &  0.29 \\
      & Sr &  0.17 &     1 &  0.27 &  0.05 &  0.01 &  0.01 &  0.15 &  0.09 &  0.33 \\
      \hline
      LSPM J2043+0445
      & Na &  0.41 &     3 &  0.19 &  0.42 &  0.05 &  0.01 &  0.17 &  0.02 &  0.49 \\
      & Mg &  0.27 &     1 &  0.32 &  0.34 &  0.06 &  0.01 &  0.19 &  0.04 &  0.51 \\
      & Ca &  0.29 &     3 &  0.19 &  0.26 &  0.04 &  0.01 &  0.14 &  0.03 &  0.35 \\
      & Ti &  0.75 &    20 &  0.03 &  0.66 &  0.10 &  0.04 &  0.36 &  0.01 &  0.76 \\
      & Cr &  0.33 &     8 &  0.06 &  0.29 &  0.04 &  0.01 &  0.15 &  0.03 &  0.34 \\
      & Mn &  0.41 &     2 &  0.23 &  0.50 &  0.07 &  0.02 &  0.21 &  0.03 &  0.59 \\
      & Fe &  0.39 &     6 &  0.13 &  0.32 &  0.04 &  0.02 &  0.15 &  0.03 &  0.38 \\
      & Sr &  0.42 &     1 &  0.32 &  0.12 &  0.03 &  0.02 &  0.14 &  0.09 &  0.39 \\
      \hline
      LSPM J2053+1037
      & Na & $-$0.13 &     3 &  0.11 &  0.22 &  0.01 &  0.00 &  0.11 &  0.02 &  0.26 \\
      & Mg & $-$0.08 &     1 &  0.19 &  0.12 &  0.02 &  0.01 &  0.19 &  0.04 &  0.29 \\
      & Ca & $-$0.12 &     3 &  0.11 &  0.11 &  0.01 &  0.01 &  0.13 &  0.03 &  0.20 \\
      & Ti &  0.17 &    20 &  0.02 &  0.29 &  0.03 &  0.02 &  0.29 &  0.01 &  0.41 \\
      & Cr & $-$0.09 &     8 &  0.04 &  0.11 &  0.01 &  0.01 &  0.13 &  0.03 &  0.18 \\
      & Mn & $-$0.10 &     2 &  0.13 &  0.23 &  0.02 &  0.01 &  0.18 &  0.03 &  0.32 \\
      & Fe & $-$0.05 &     6 &  0.08 &  0.12 &  0.01 &  0.01 &  0.14 &  0.03 &  0.20 \\
      & Sr &  0.01 &     1 &  0.19 &  0.03 &  0.01 &  0.02 &  0.17 &  0.09 &  0.27 \\
      \hline
      LSPM J2343+3632
      & Na & $-$0.04 &     3 &  0.12 &  0.23 &  0.01 &  0.00 &  0.12 &  0.02 &  0.29 \\
      & Mg &  0.04 &     1 &  0.21 &  0.12 &  0.02 &  0.01 &  0.18 &  0.04 &  0.31 \\
      & Ca & $-$0.06 &     3 &  0.12 &  0.12 &  0.01 &  0.01 &  0.13 &  0.03 &  0.22 \\
      & Ti &  0.31 &    20 &  0.02 &  0.29 &  0.02 &  0.03 &  0.30 &  0.01 &  0.42 \\
      & Cr & $-$0.06 &     8 &  0.03 &  0.12 &  0.01 &  0.01 &  0.13 &  0.03 &  0.18 \\
      & Mn &  0.05 &     2 &  0.15 &  0.24 &  0.01 &  0.02 &  0.19 &  0.04 &  0.34 \\
      & Fe & $-$0.00 &     6 &  0.09 &  0.13 &  0.01 &  0.02 &  0.14 &  0.03 &  0.21 \\
      & Sr &  0.13 &     1 &  0.21 &  0.04 &  0.01 &  0.02 &  0.17 &  0.09 &  0.29 \\
      \hline
    \enddata
  \tablecomments{
    The calculation of $\sigma_\mathrm{SEM}$ is depicted in (1) of Section \ref{sec:IRD_analysis}. Those of $\sigma_{T_{\mathrm{eff}}}$,  $\sigma_{\log{g}}$, and $\sigma_{\xi}$ are described in (2). Those of $\sigma_\mathrm{OE}$  and  $\sigma_\mathrm{cont}$ are described in (3) and (4), respectively.
    $\sigma_\mathrm{Total}$ denotes the quadrature sum of all the errors. %
  }
\end{deluxetable*}
\begin{figure*}
  \includegraphics[width=77mm]{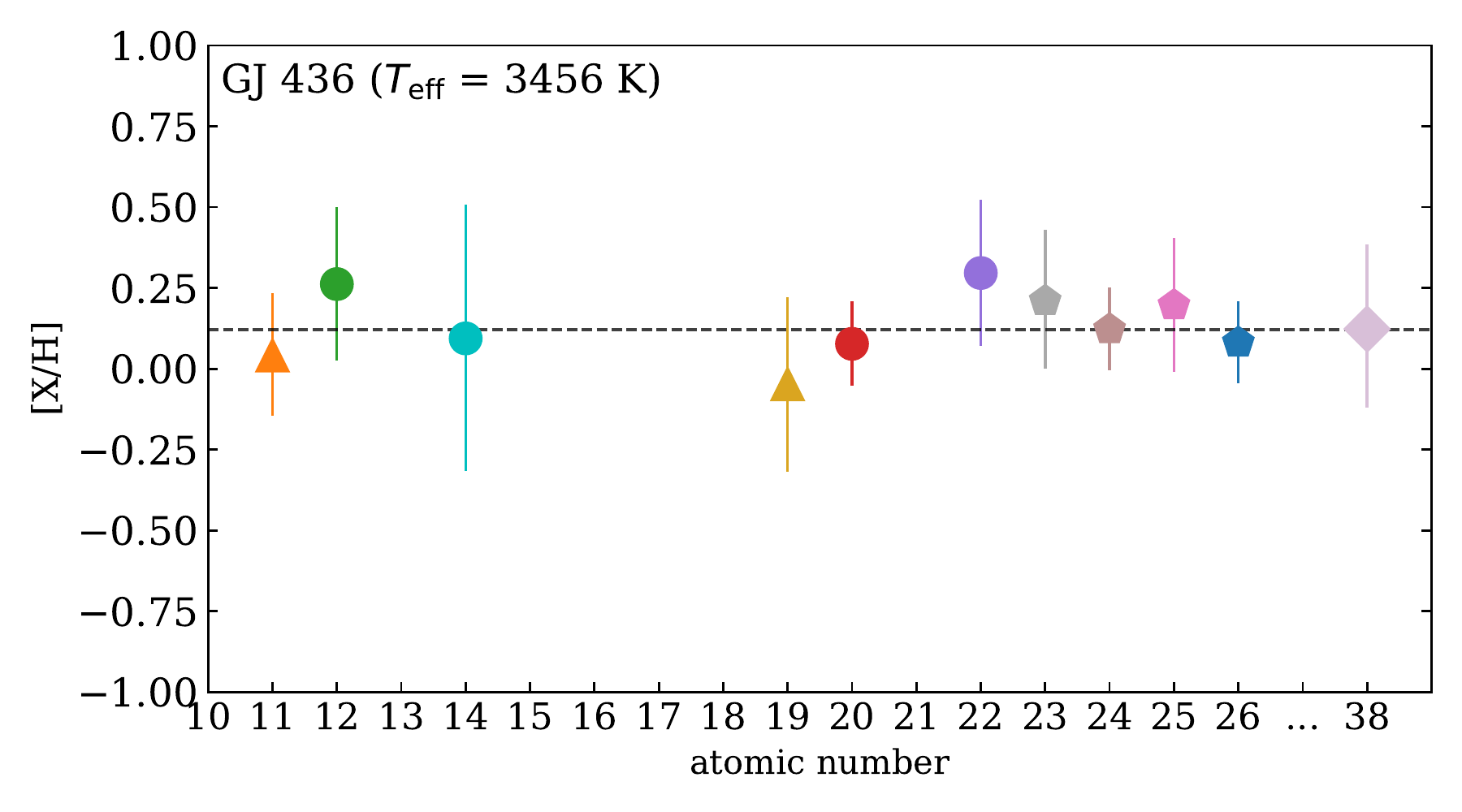}%
  \includegraphics[width=77mm]{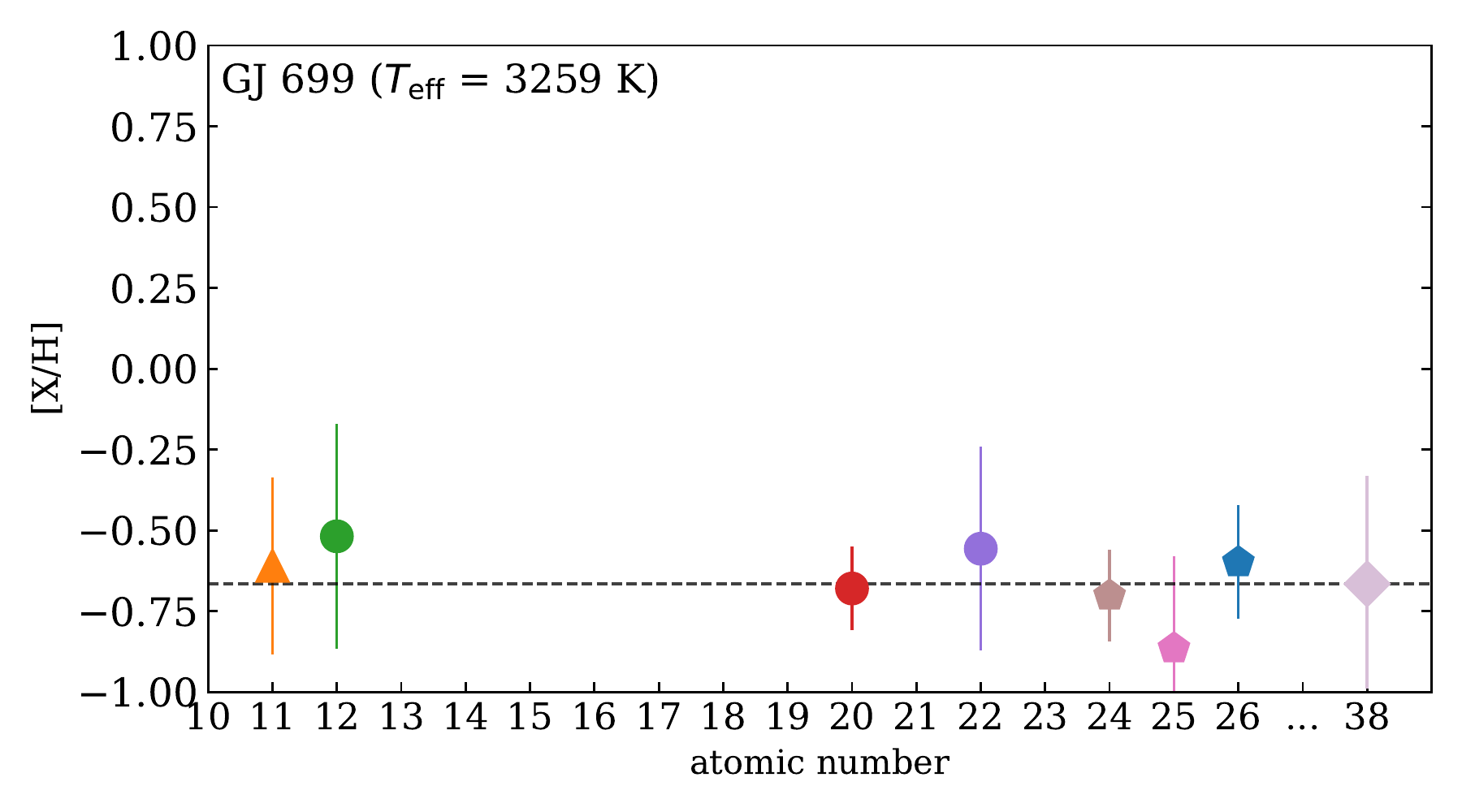}%
  \\
  \includegraphics[width=77mm]{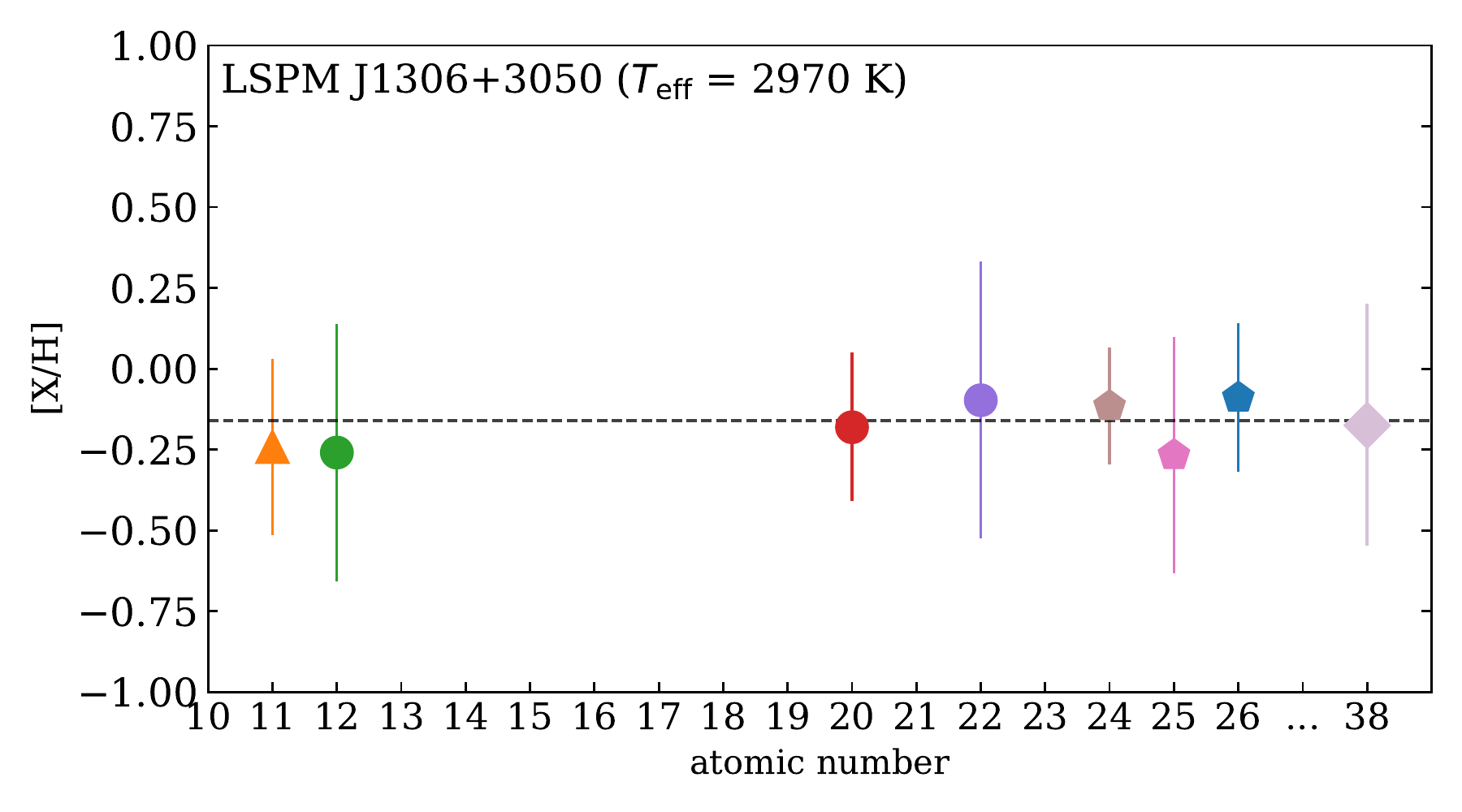}%
  \includegraphics[width=77mm]{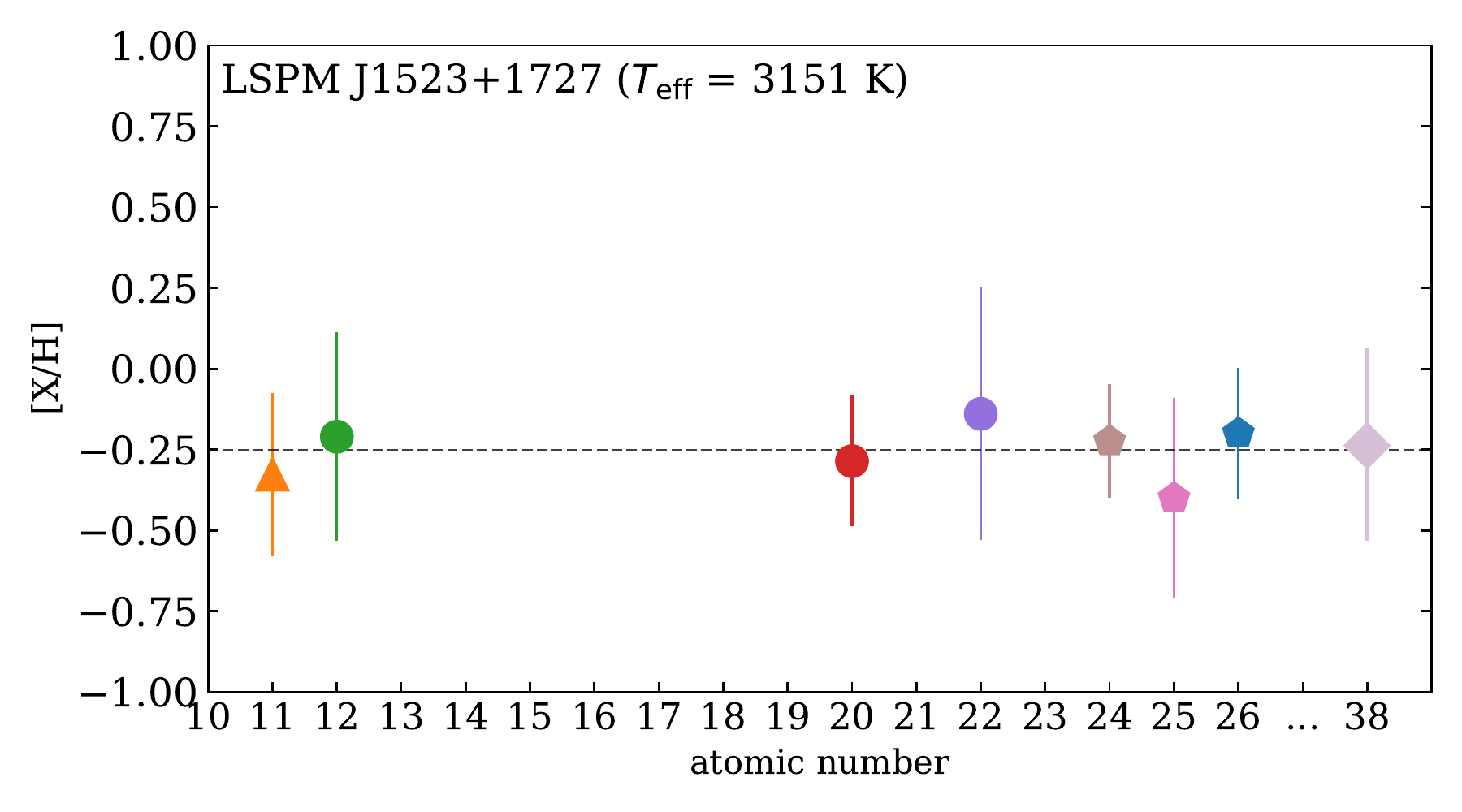}
  \\
  \includegraphics[width=77mm]{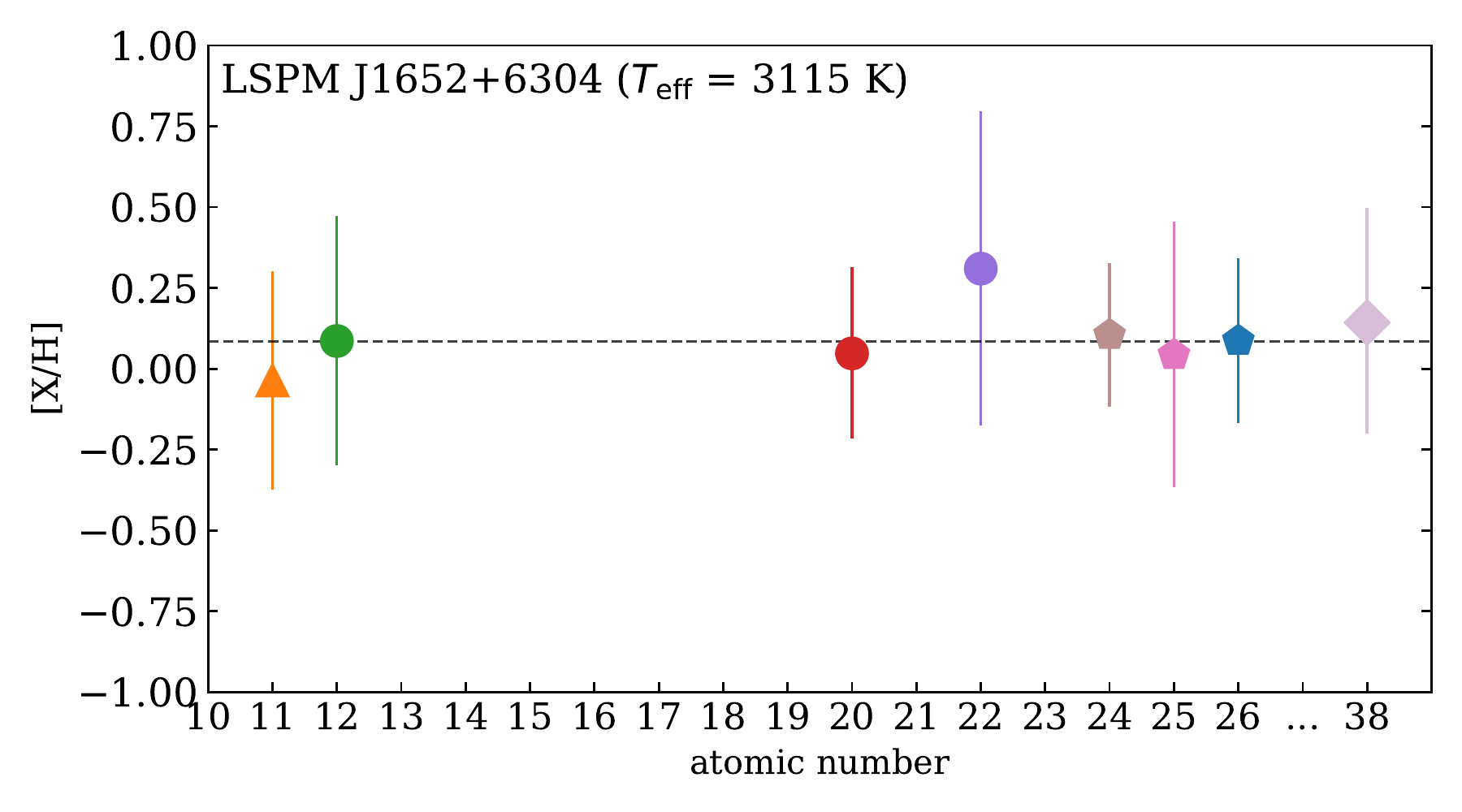}%
  \includegraphics[width=77mm]{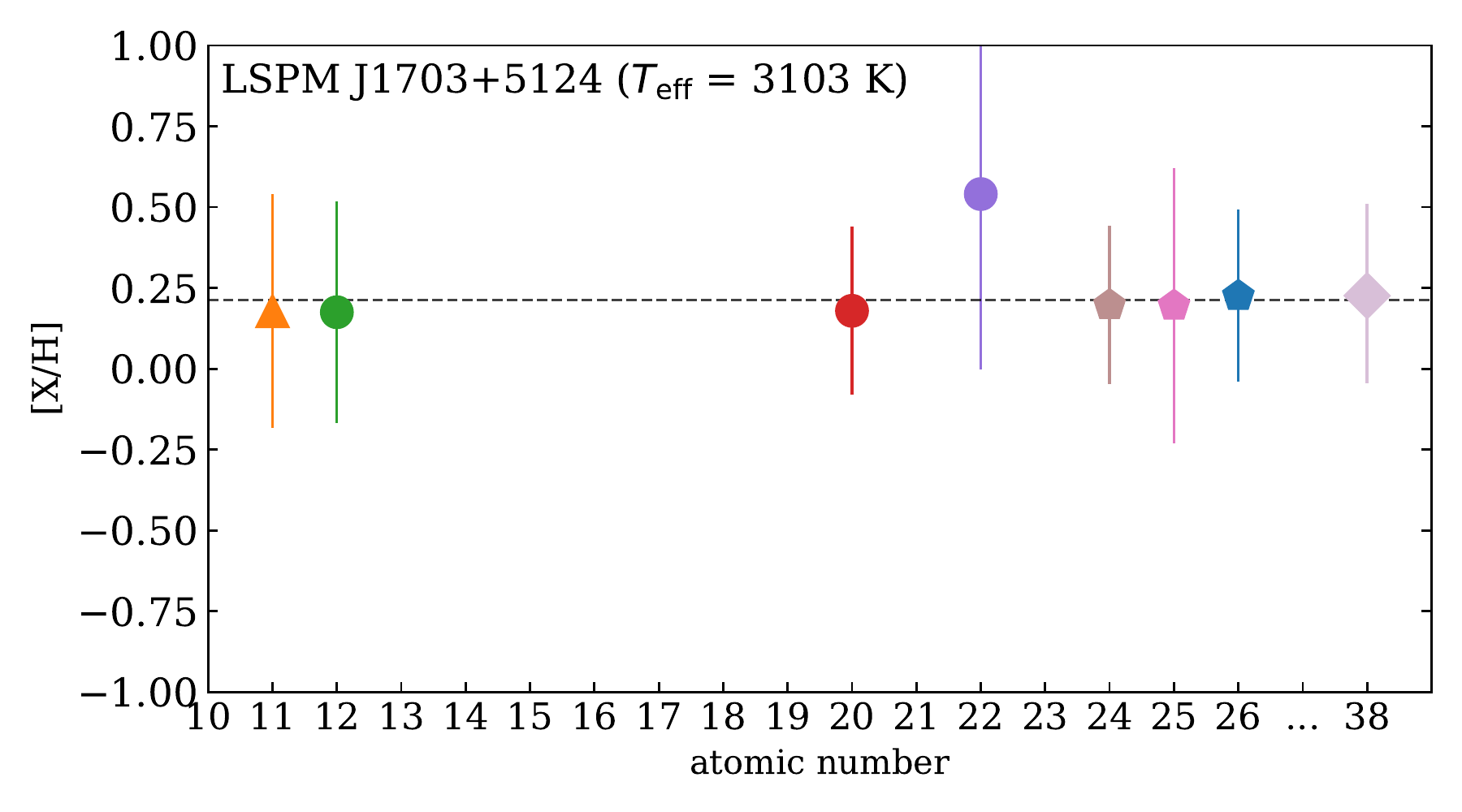}
  \\
  \includegraphics[width=77mm]{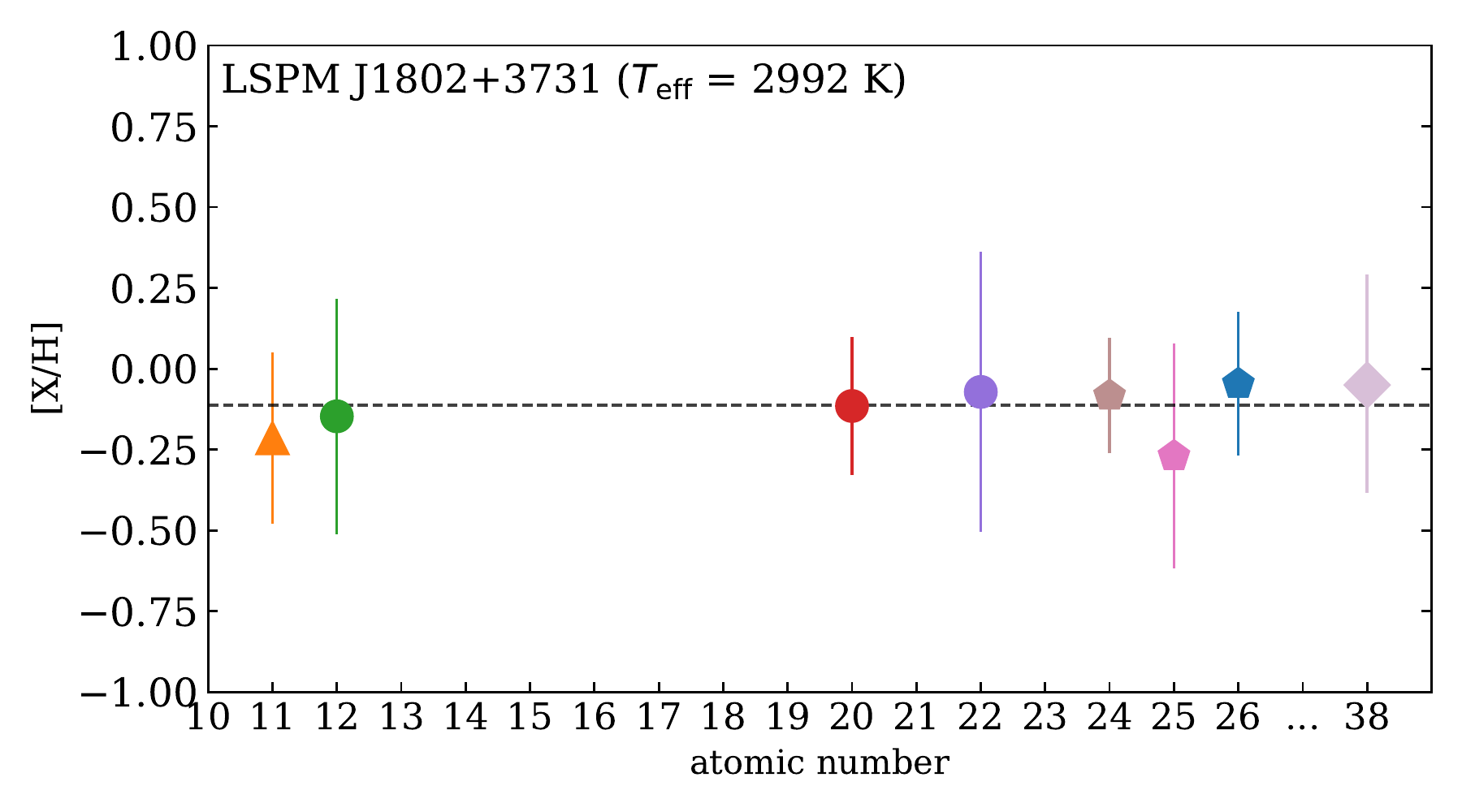}%
  \includegraphics[width=77mm]{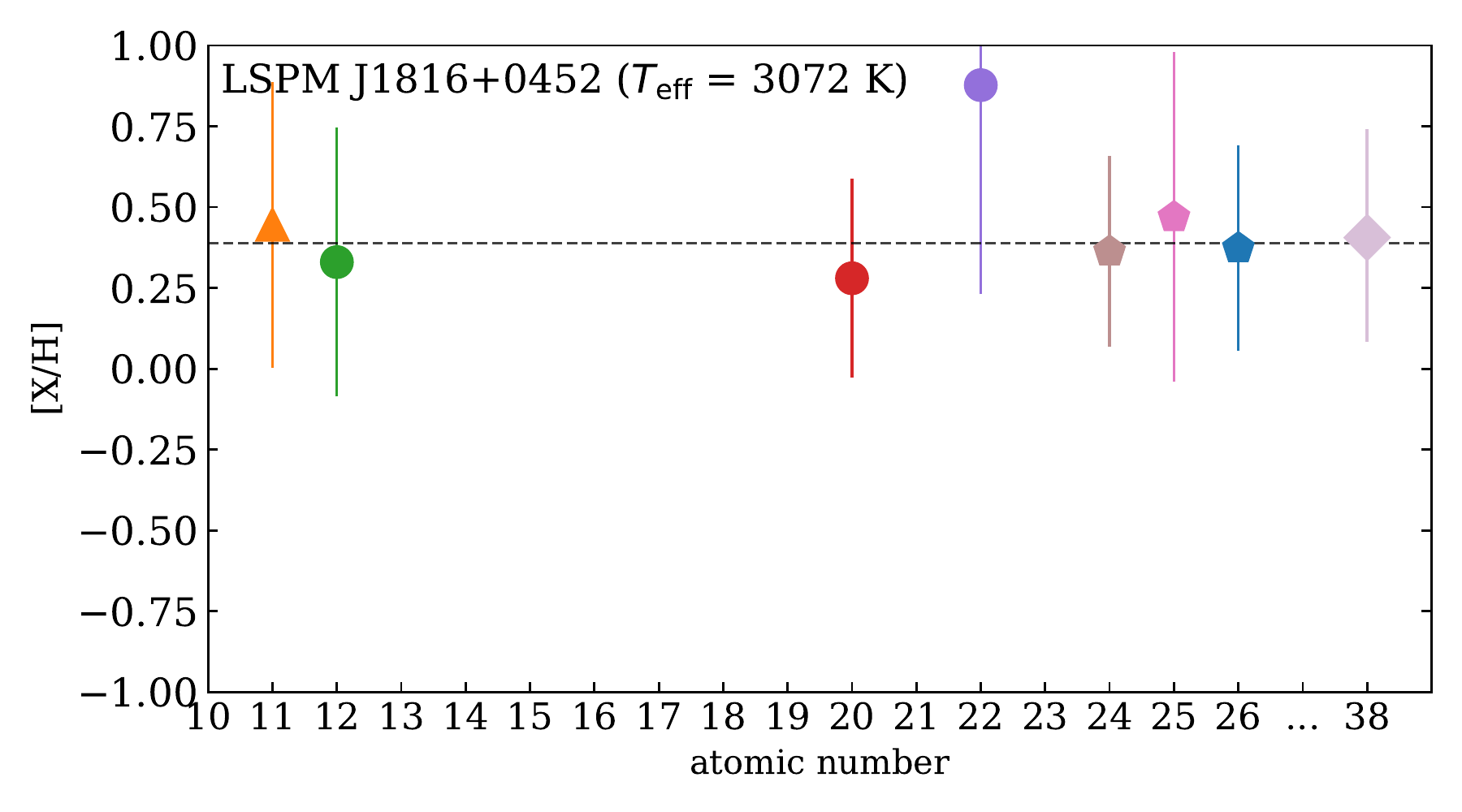}
  \\
  \includegraphics[width=77mm]{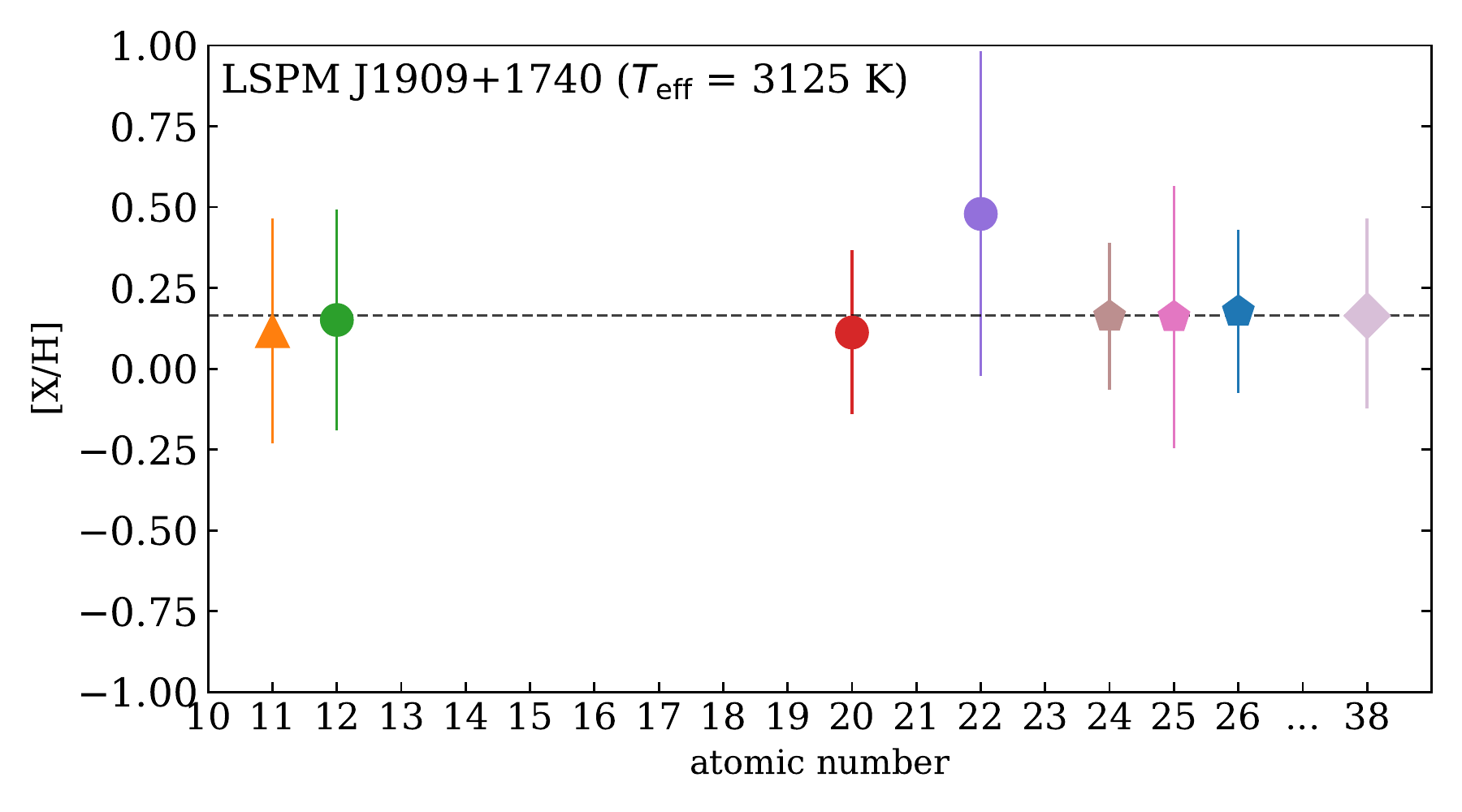}%
  \includegraphics[width=77mm]{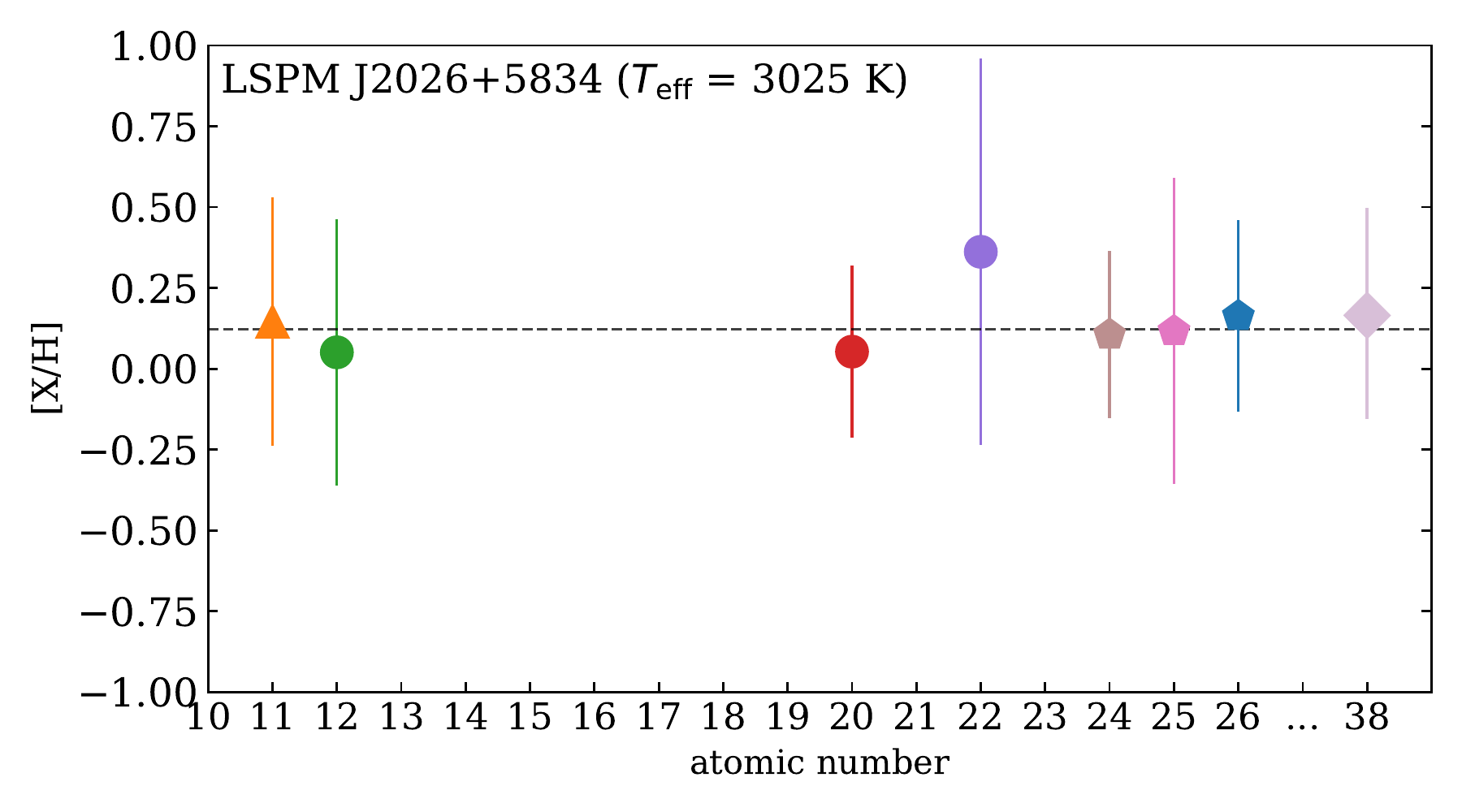}
  \caption{Abundance ratios [X/H] of individual elements for each M dwarf as a function of atomic number.
  \added{%
  The different symbols represent different families of elements: triangles for light odd-Z elements (Na and K), circles for alpha elements, pentagons for iron-peak elements, and diamonds for the neutron-capture element Sr.
  }%
  The error sizes are the $\sigma_\mathrm{Total}$ estimated in Section \ref{sec:IRD_analysis}. %
  The horizontal dashed line indicates the average of all the elements [M$_\mathrm{ave}$/H], weighted by the inverse of the square of the errors. %
  }\label{fig:res_IRD_SSP_targets}
\end{figure*}
\addtocounter{figure}{-1}
\begin{figure*}
  \includegraphics[width=77mm]{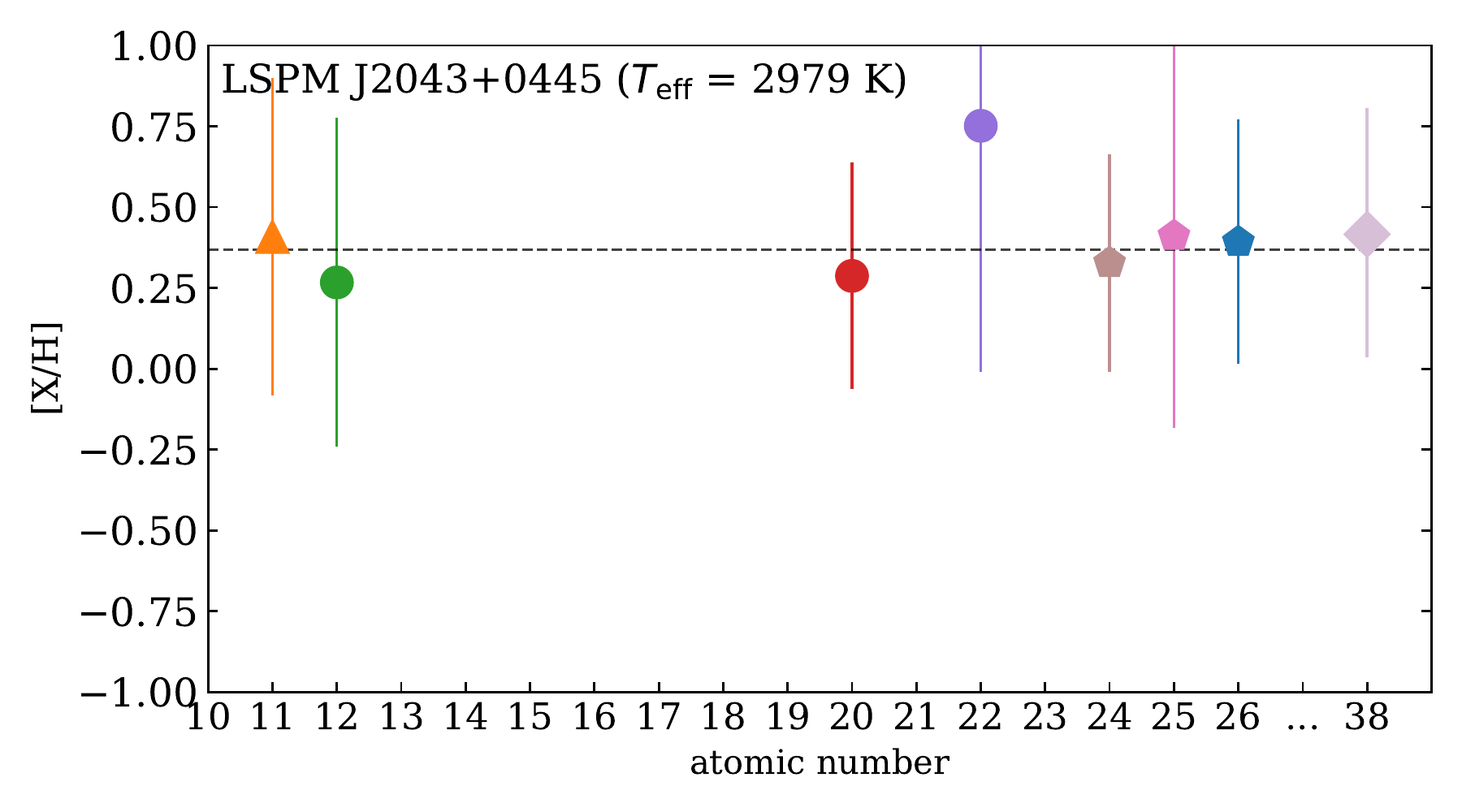}%
  \includegraphics[width=77mm]{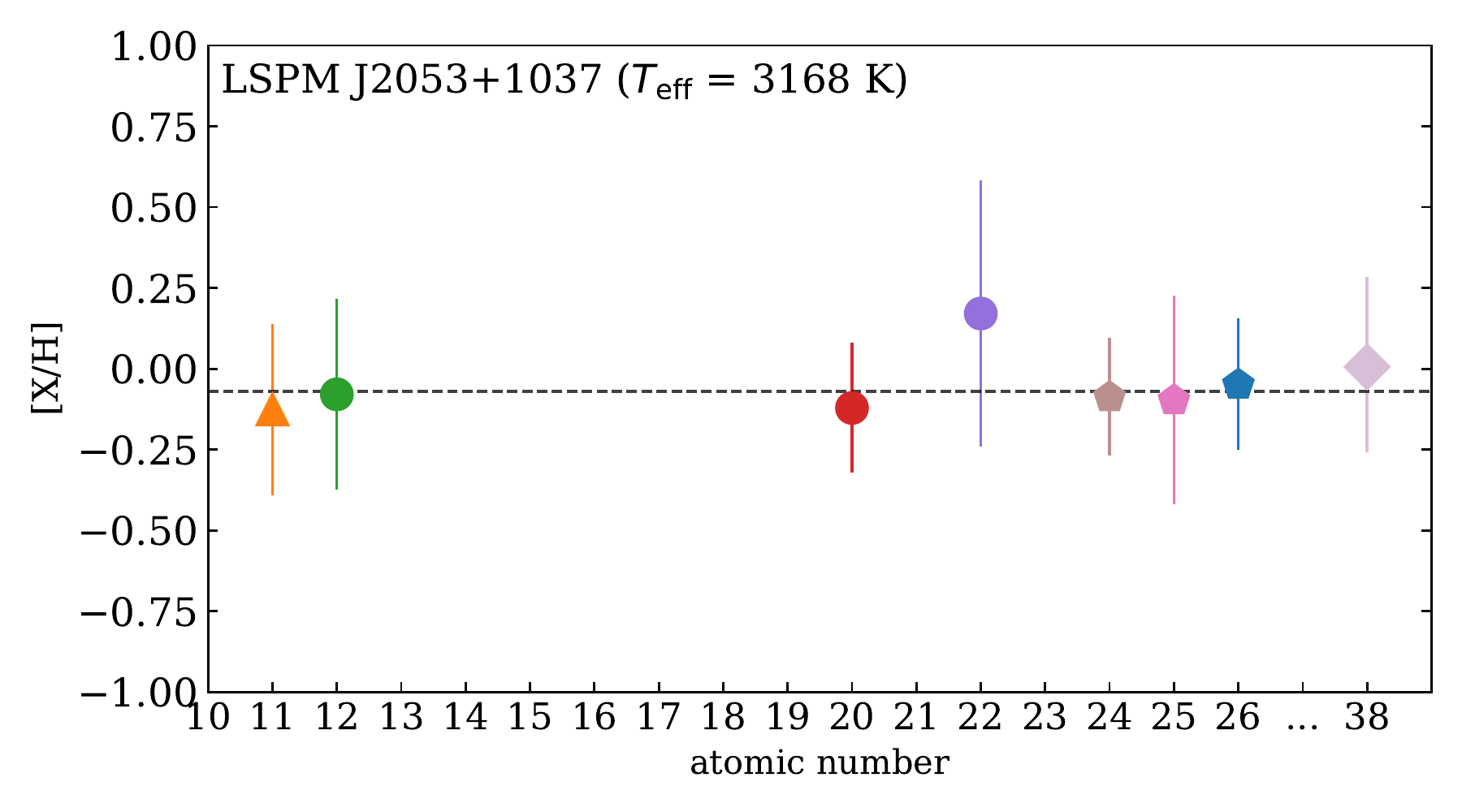}\\
  \includegraphics[width=77mm]{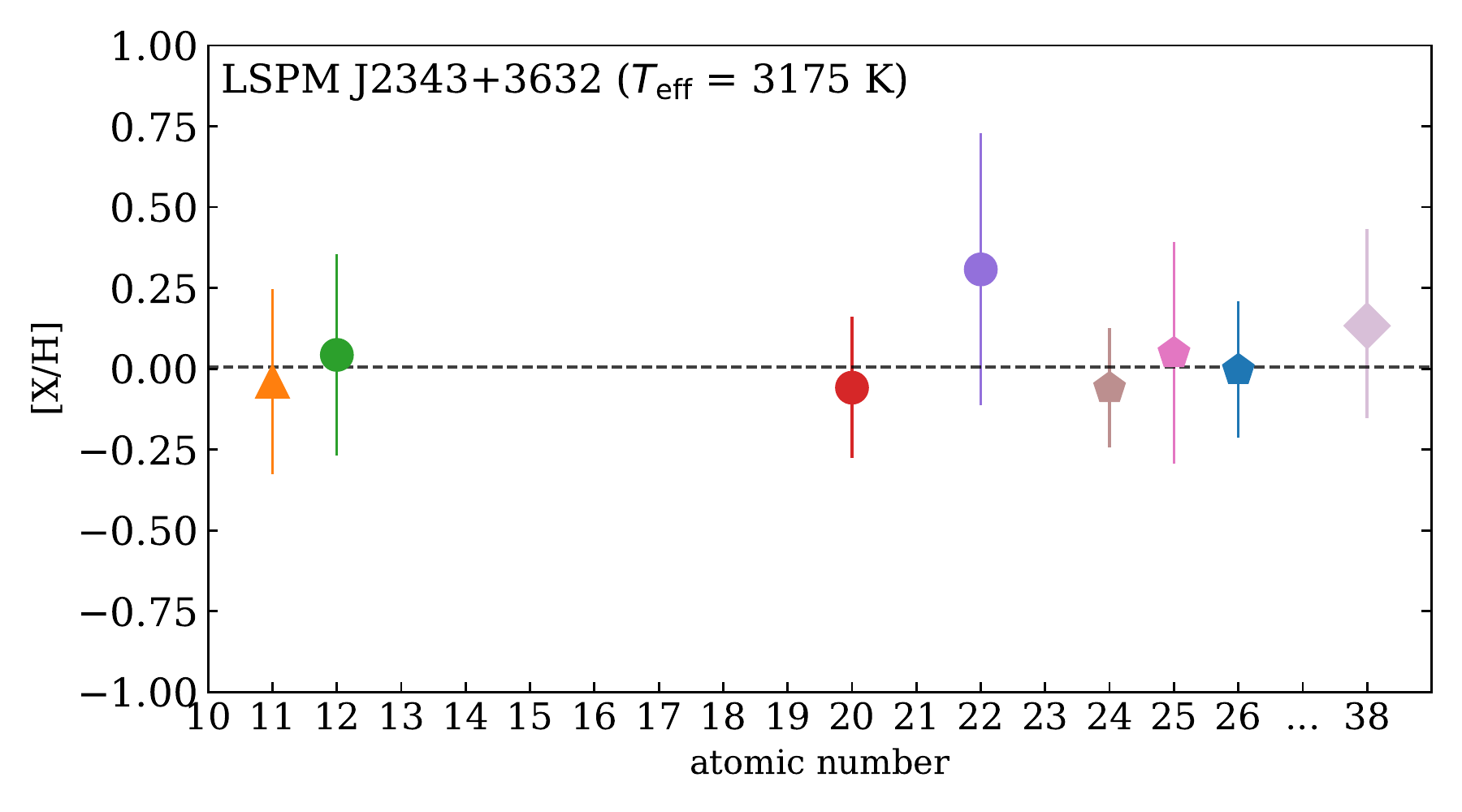}
  \caption{(Continued)}
\end{figure*}

Following the caveats presented by Ish20, we determined the elemental abundances so that the assumed abundance of each element is consistent with the resulting abundance. %
The similarity of the abundance patterns to the Sun derived by the analysis for individual elements indicates that scaled-solar abundance ratios could be a good approximation at least for these targets. %

\begin{figure}\begin{center}
  \plotone{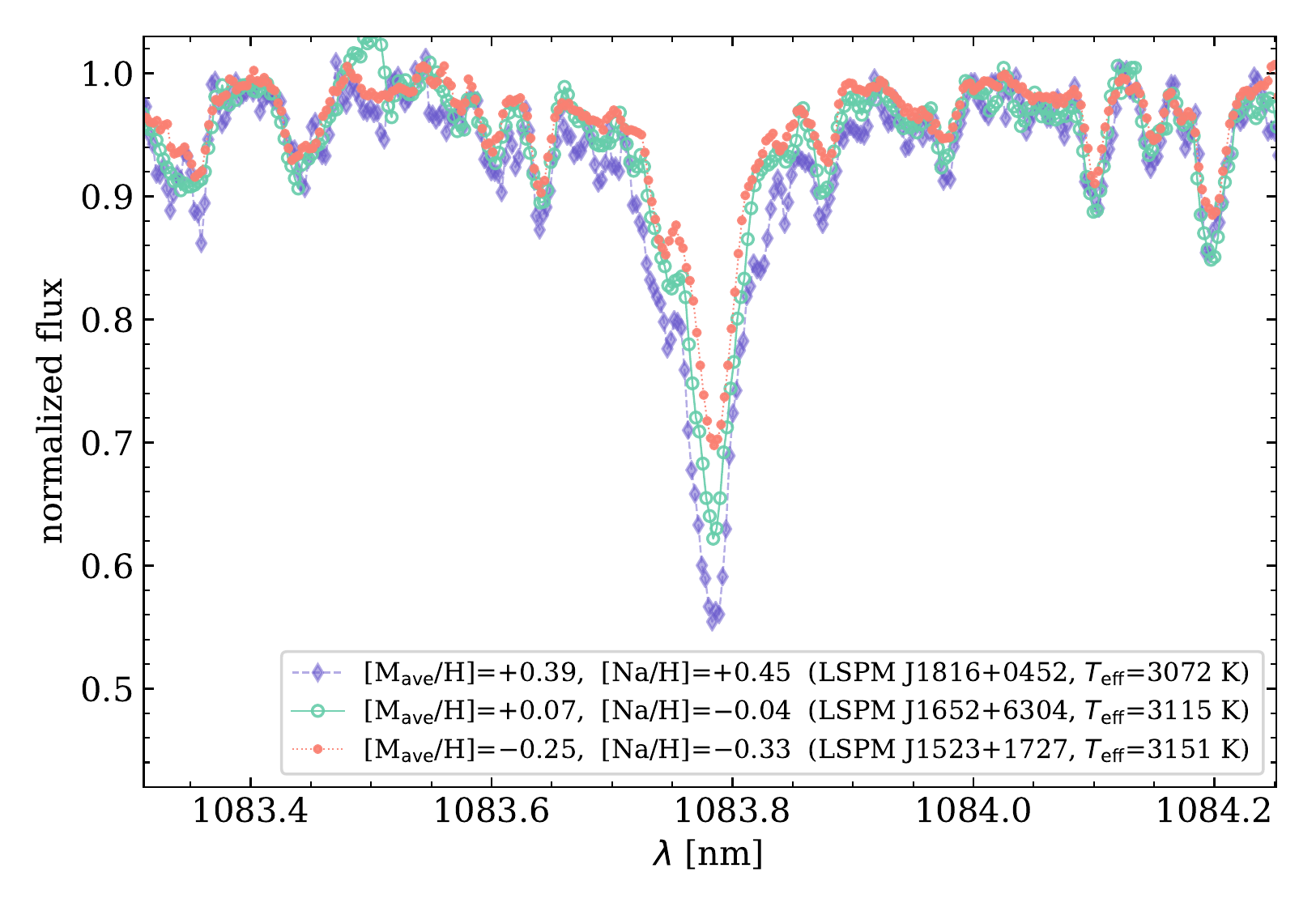} %
  \caption{The IRD spectra around a Na I line (1083.78 nm) of M dwarfs at $T_{\mathrm{eff}} \sim$ 3100 K with the low metallicity, high metallicity, and in-between.
  } \label{fig:Metallicity-dependence_111_10837}
\end{center}\end{figure}

The [M$_\mathrm{ave}$/H] ranges from $\sim-$0.6 to $\sim+$0.4, centered around the solar value.
These differences in metallicity can be visually confirmed by the difference in depth of a Na I line as shown in Figure \ref{fig:Metallicity-dependence_111_10837}.
Note that we show a Na I line here because Na I lines are the most insensitive to the abundances of other elements among the measured species %
due to the fact that the abundance of Na controls the continuum opacity dominantly as reported by Ish20. %

\begin{figure}\begin{center}
  \plotone{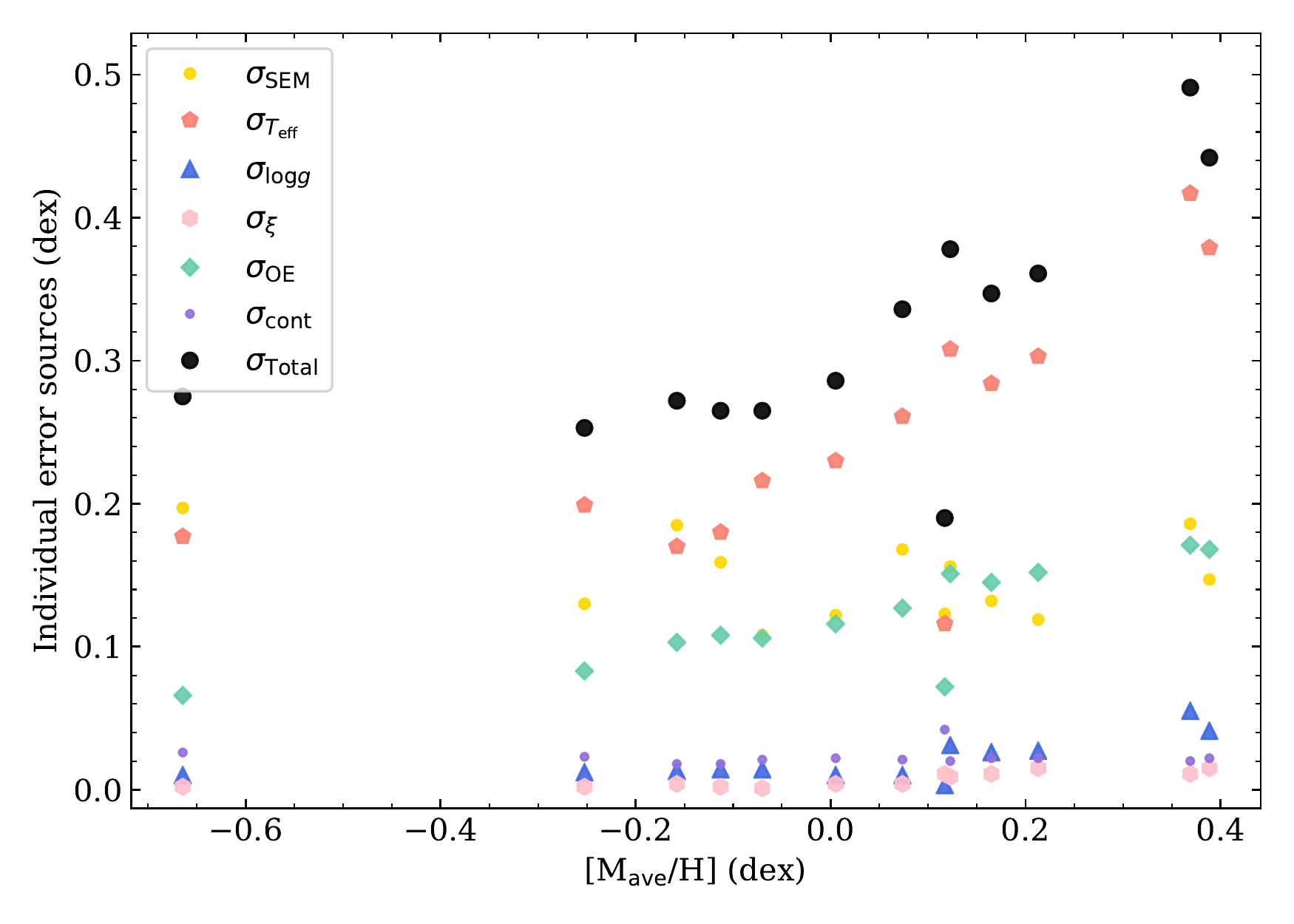} %
  \caption{Size of errors associated with the resulting [Na/H] caused by individual error sources for each target %
  as a function of [M$_\mathrm{ave}$/H] of the objects.
  The $\sigma_\mathrm{SEM}$ is a statistical error. %
  The $\sigma_{T_{\mathrm{eff}}}$,  $\sigma_{\log{g}}$, and $\sigma_{\xi}$ are errors propagated from the uncertainties of corresponding stellar parameters. %
  The $\sigma_\mathrm{OE}$ and  $\sigma_\mathrm{cont}$ are the errors due to uncertainties of other elements and errors associated with the continuum level uncertainty, respectively. %
  The $\sigma_\mathrm{Total}$ is the quadrature sum of all the above errors.
  The outlier with the smallest error ($\sigma_\mathrm{Total} \sim 0.2$ dex) is GJ 436, which is due to its higher temperature ($T_{\mathrm{eff}} \sim 3456$ K) than the other objects.
  } \label{fig:metal-sigma_total_111}
\end{center}\end{figure}

The precision of the [Na/H] determination is critical since it strongly affects the $\sigma_\mathrm{OE}$ of other elements through the continuum opacity.
We found that the precision of [Na/H] is degraded with increasing [M$_\mathrm{ave}$/H] values among the objects with similar temperatures. %
Figure \ref{fig:metal-sigma_total_111} presents a comparison of the amount of errors in [Na/H] caused by individual error sources for different targets. %
The error source $\sigma_{T_{\mathrm{eff}}}$ mainly contributes to the trend from lower left to upper right, and $\sigma_\mathrm{OE}$ and $\sigma_{\log{g}}$ also show gradual upward trends.
These trends are also observed in other elements, which may be attributed to the fact that the stronger the absorption lines, the more insensitive they become to the elemental abundance, due to effects such as the saturation of the line core. %
Figure \ref{fig:metal-sigma_total_111} also shows that the error source dominating total errors in most objects is the uncertainty of $T_{\mathrm{eff}}$, not the data quality or other parameters.
Future improvements of $T_{\mathrm{eff}}$ estimates by, for instance, an increase of interferometric measurements, will enable one to determine elemental abundances with significantly smaller errors. %

For LSPM J2043+0445, one of the most metal-rich objects, we found an RV variation that can be caused by a stellar companion (IRD-SSP team, in prep), consistent with its large value of {\tt\string ruwe}$\sim$10.2 \citep[the renormalized unit weight error for the astrometric solution introduced by][]{LL:LL-124} in Gaia EDR3. %
All the Na I lines used in this study become stronger at lower temperatures in the temperature range of interest. %
If the EWs of Na I lines are overestimated due to the contamination of a cooler companion star, the [Na/H] would be overestimated.
This may lead to an overall overestimation of the metal abundances of LSPM J2043+0445.
Note that the contamination of a cooler companion star should also make the $T_{\mathrm{eff\mathchar`-TIC}}$ underestimated, and as a result, [Na/H] is expected to be underestimated.
How much the two opposite effects of increasing the observed EW and decreasing the estimated $T_{\mathrm{eff\mathchar`-TIC}}$ affect the abundance results needs to be further investigated quantitatively.

Although Mg is an important element to constrain the nature of planets around the stars, the errors are relatively large because only one broad absorption line is available for the analysis.
We have identified several Mg I lines in the $H$-band, but they are all heavily contaminated by unidentified absorption lines and cannot be used to derive reliable abundances.
In the future, when the database of molecular absorption lines in this band is improved and the features around the Mg I lines can be reproduced well with synthetic spectra, those Mg lines could be used to further constrain [Mg/H].

\begin{figure*}\begin{center}
  \includegraphics[width=78mm]{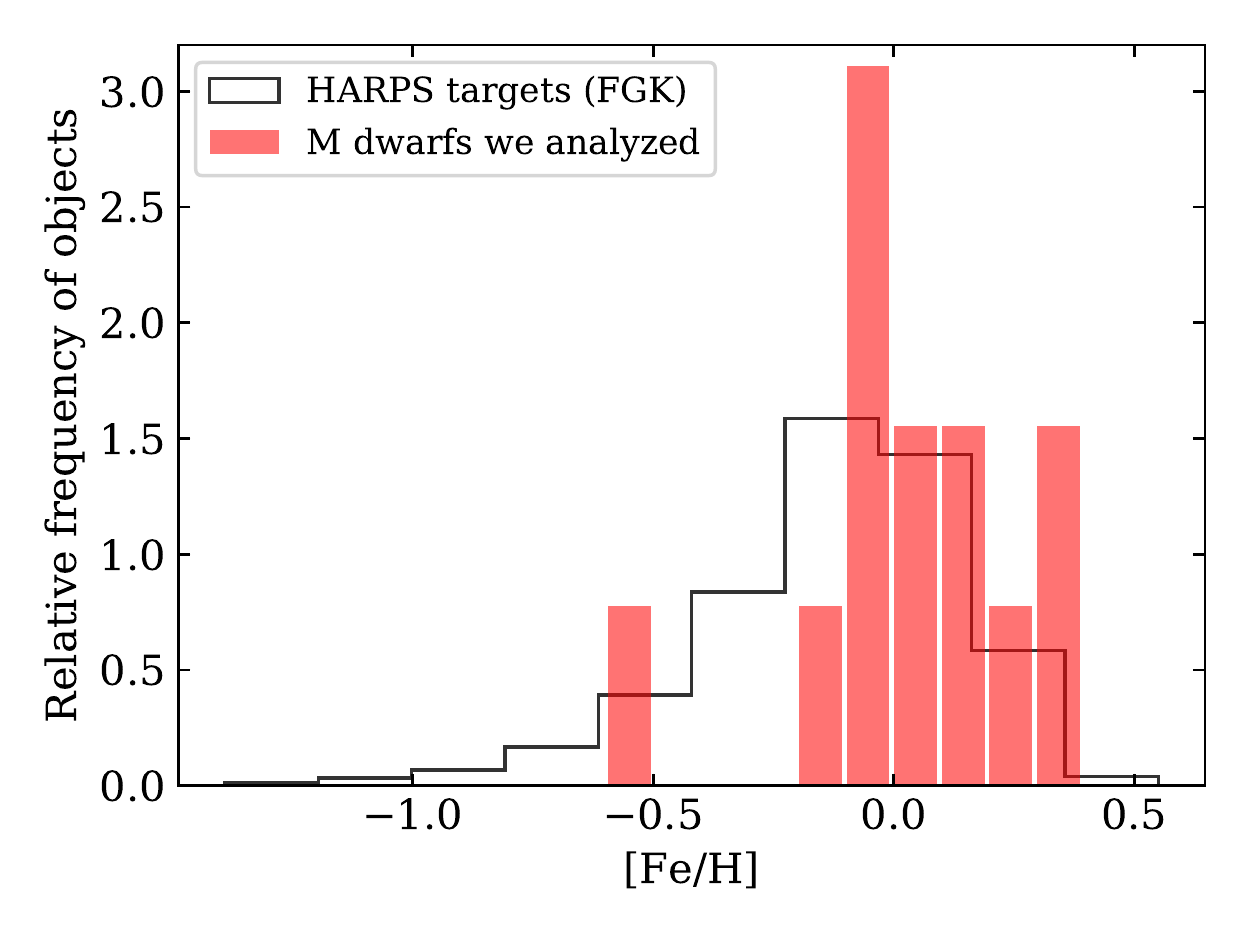} %
  \includegraphics[width=78mm]{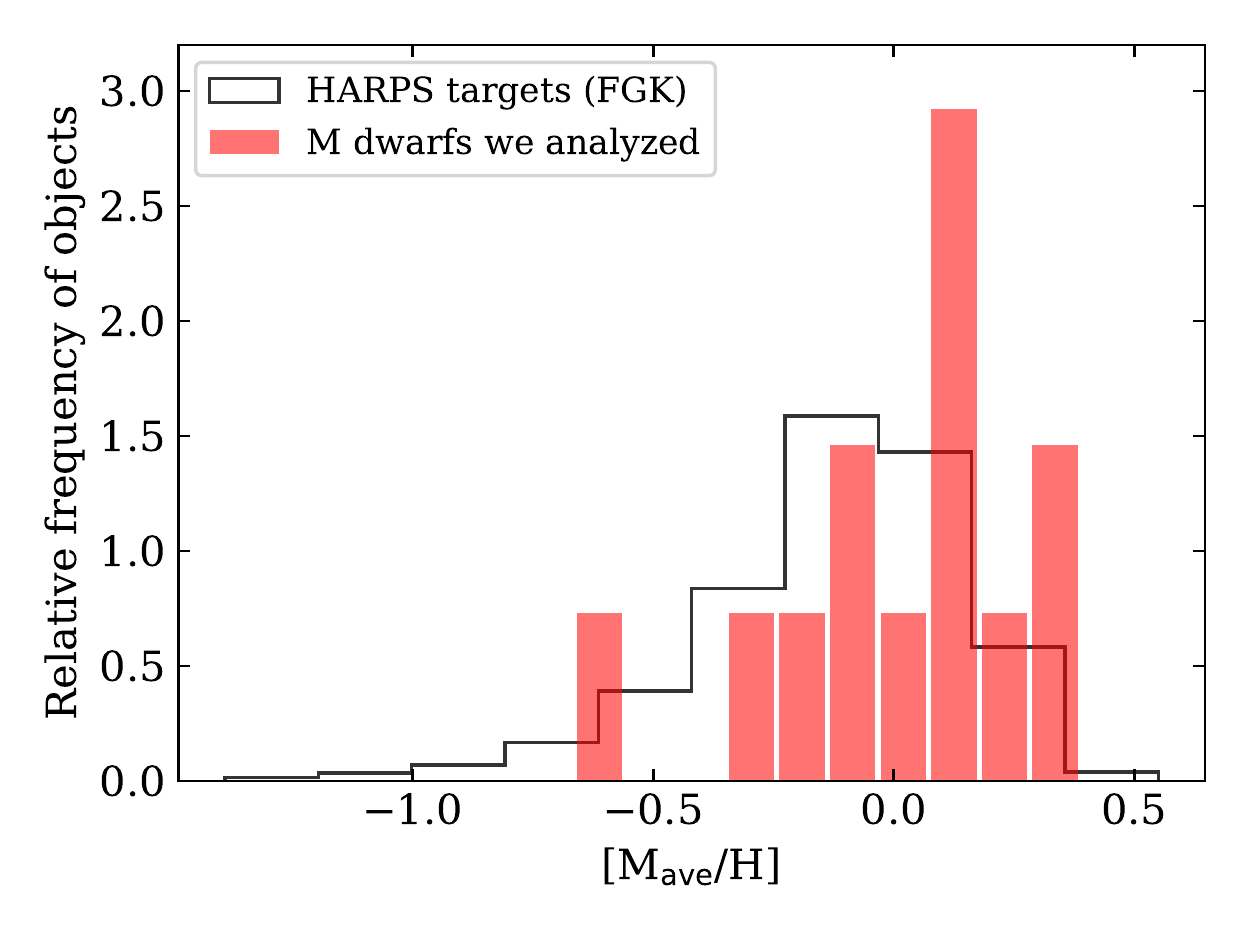} %
  \caption{Distribution of the 13 M dwarfs in the iron abundance (red histogram in the right panel) and in the error-weighted average of abundances of all the elements (that in the left panel). %
  The black histograms in both panels are the distribution of [Fe/H] values of 1111 FGK dwarfs reported in \citet{2012A&A...545A..32A}, where the [Fe/H] is treated as a proxy of overall metallicity.
  } \label{fig:hist_iron}
\end{center}\end{figure*}

Figure \ref{fig:hist_iron} shows the distribution of the 13 M dwarfs with the horizontal axis as iron abundance [Fe/H] in the left panel and averaged metallicity [M$_\mathrm{ave}$/H] in the right panel. %
The [Fe/H] of 1111 FGK stars determined by \citet{2012A&A...545A..32A} (hereafter Adi12) for the sample of the HARPS GTO planet search program~\citep{2003Msngr.114...20M} %
are also shown as a reference.
We note that [Fe/H] in the literature is treated as an indicator of overall metallicity and then [M/H] is not given separately. %
The FGK stars were selected for the purpose of an RV survey %
with similar criteria to that of the IRD-SSP targets, such as slow rotation and low level of chromospheric activity.
The [Fe/H] distributions of our M dwarfs are comparable to those of FGK stars and centered around the solar value or slightly higher.
Many previous studies on the metallicity distribution of M dwarfs in the solar neighborhood have also indicated the mean metallicity around the solar value.~\citep[e.g.,][]{2014MNRAS.443.2561G, 2015ApJS..220...16T, 2020AJ....159...30H, 2020MNRAS.494.2718W}. %

To compare the [Fe/H] or [M$_\mathrm{ave}$/H] of the M dwarfs with the [Fe/H] of the FGK stars, we apply the Mann-Whitney U-test, because the sample size is different and %
a normal distribution cannot be assumed according to the Shapiro-Wilk test.
The resulting p-value is 0.021 and 0.055 in the case of [Fe/H] and [M$_\mathrm{ave}$/H], respectively.
The null hypothesis of equality of [Fe/H] distribution between M and FGK stars is excluded at the significance level of 5\,\%, while it is not definitive. %
If the difference is real, some systematics due to analytical problems, such as the collection of the used lines or the adopted stellar parameters, may have led to an overestimation of the [Fe/H]. %
The possible systematic error in the adopted $T_{\mathrm{eff\mathchar`-TIC}}$ and its effect on the abundance results are discussed in more detail in Section \ref{sec:Teff_shift}.
Alternatively, differences of [Fe/H] distribution between M dwarfs and FGK stars might exist.
The Mann-Whitney U-test between the distribution of [Fe/H] and [M$_\mathrm{ave}$/H] (p $=$ 0.76) does not reject their equality and the difference between the averages for 13 objects of [Fe/H] ($\sim+$0.04 dex) and [M$_\mathrm{ave}$/H] ($\sim+$0.01 dex) is much less than the typical error.
We need a larger sample to discuss the possible difference of metallicity between M and FGK stars. %

Given the long life of the M dwarfs and the selection bias of the IRD-SSP targets toward magnetically inactive ones, it is more natural to think that they are biased toward lower metallicities. %
Whereas, if the initial mass function has a dependence on metallicity, it could make difference in metallicity distribution depending on the spectral type. %

Another interesting feature found in the FGK sample is
the extended tail of low-metallicity objects. %
Our sample is still insufficient to investigate this tail.
Future extension of the analysis to all IRD-SSP targets will corroborate these discussions. %

\subsection{Comparison with literature for the 11 objects with 2900 $< T_{\mathrm{eff}} <$ 3200} \label{sec:comparison_with_previous_studies} %

\begin{figure*}\begin{center}
  \includegraphics[width=150mm]{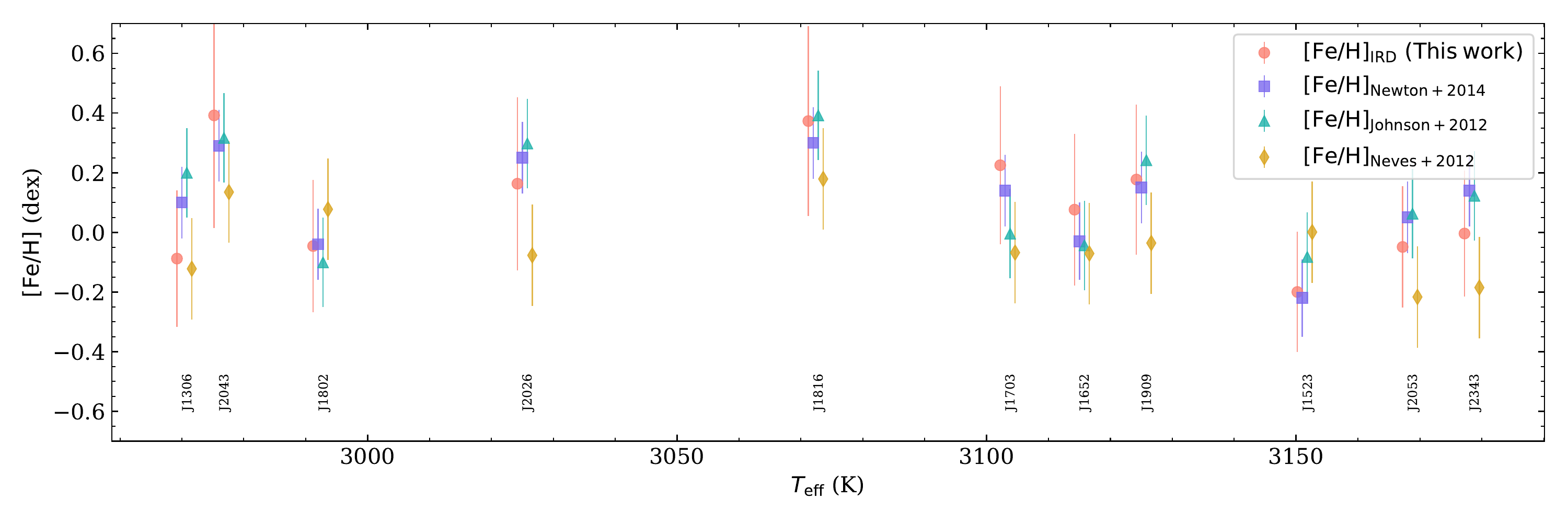} %
  \includegraphics[width=50mm]{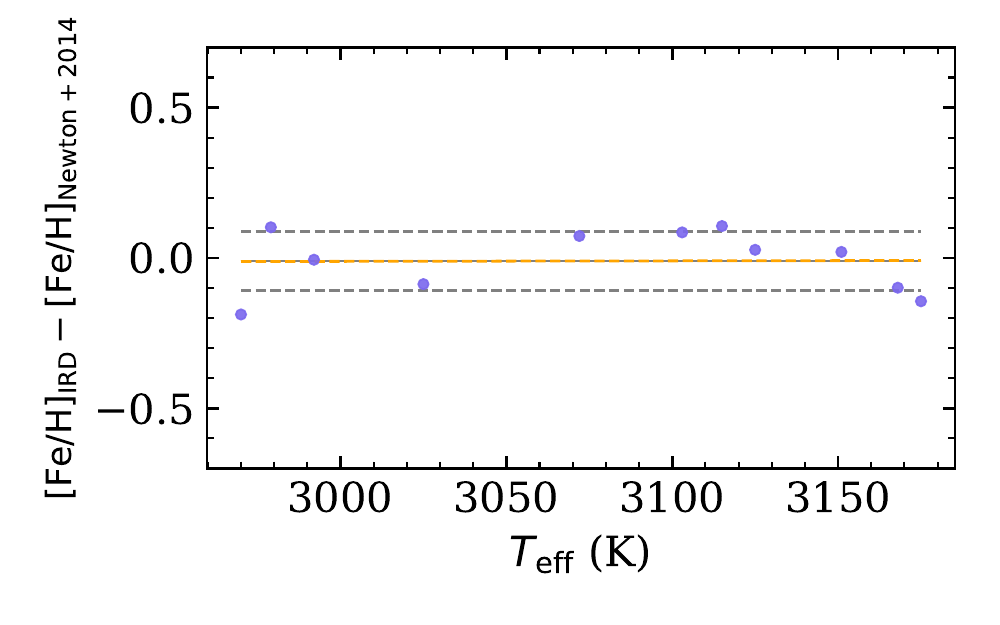}
  \includegraphics[width=50mm]{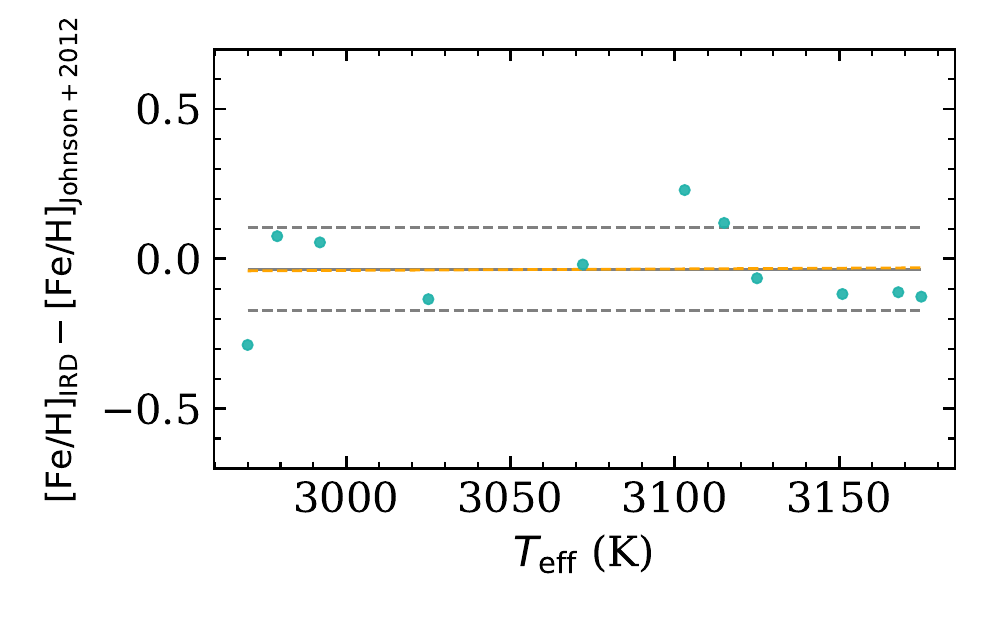}
  \includegraphics[width=50mm]{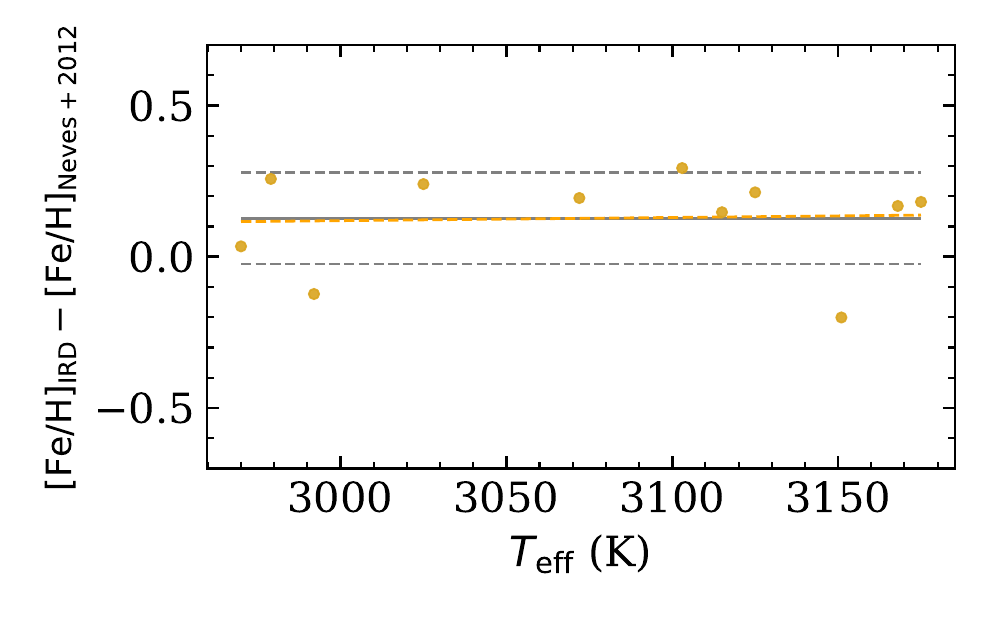}
  \includegraphics[width=50mm]{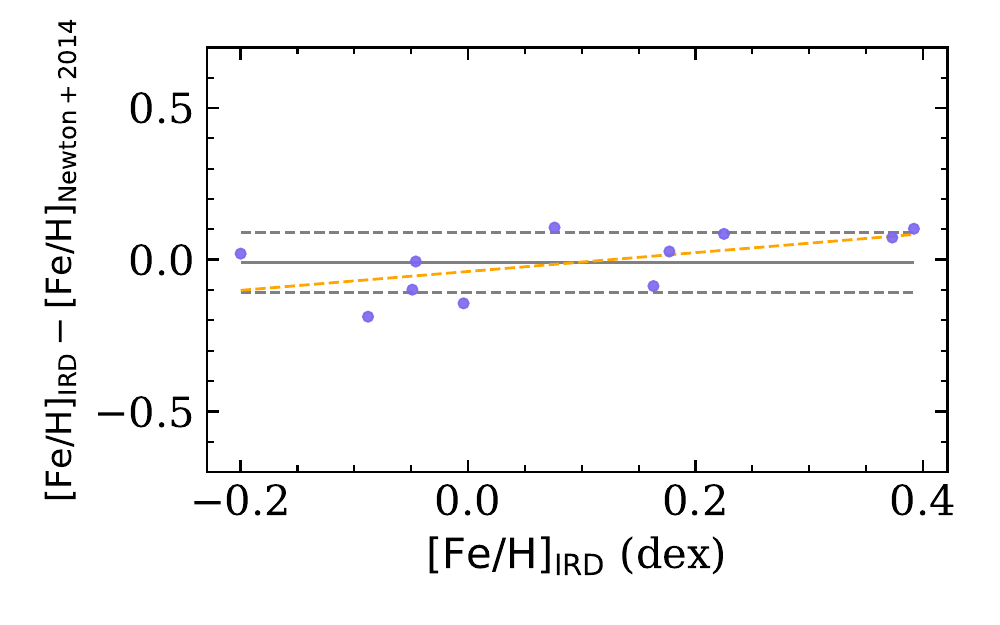}
  \includegraphics[width=50mm]{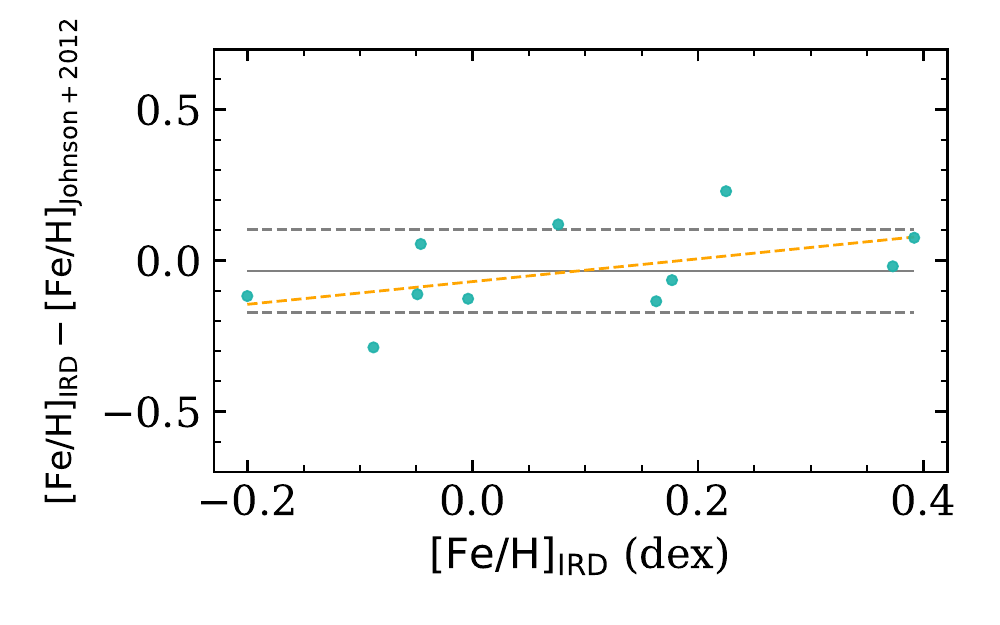}
  \includegraphics[width=50mm]{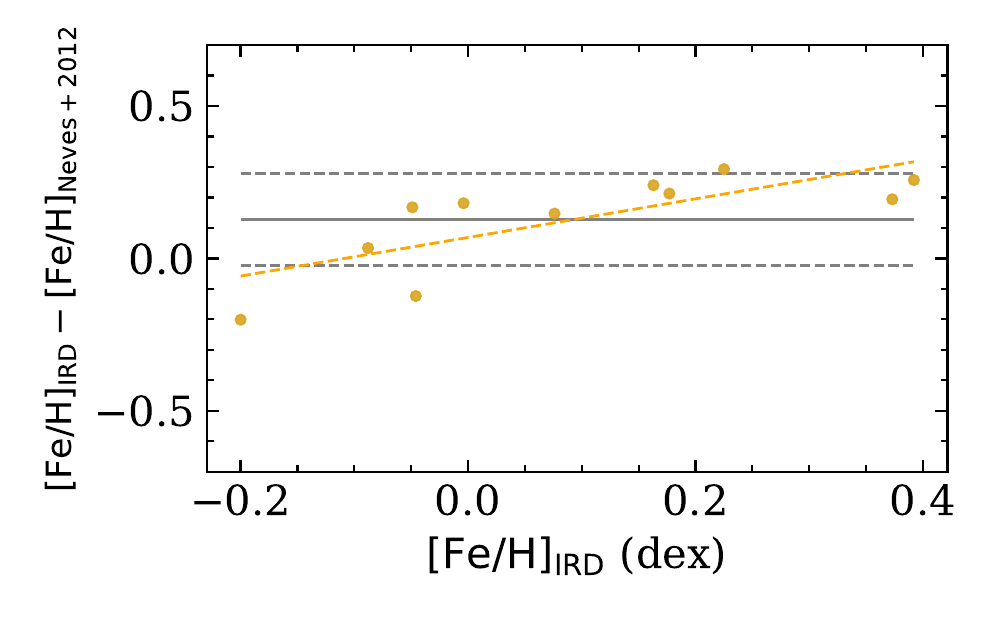}
  \caption{Comparison between our resulting [Fe/H] values and the previous estimates of the overall metallicity. %
  The red circles are our results.
  The purple squares are the estimates by \citet{2014AJ....147...20N} based on the EW of Na doublet at 2.2 $\mu$m.
  The green triangles and yellow diamonds are the estimates based on the photometric empirical formula of \citet{2012AJ....143..111J} and \citet{2012A&A...538A..25N}, respectively. %
  The lower six small panels show that the differences between our results and the other estimates as a function of $T_{\mathrm{eff}}$ or [Fe/H], where the gray solid lines, gray dashed lines, and orange dashed lines represent the mean, 1$\sigma$ scatter, and the fitted linear trend, respectively.
  } \label{fig:compare_mtlc}
\end{center}\end{figure*}

We compare the resulting iron abundances of eleven M dwarfs with $T_{\mathrm{eff}} < $ 3200 K with the metallicities estimated by previous studies (Figure \ref{fig:compare_mtlc}). %
The purple squares show the estimates by \citet{2014AJ....147...20N} (hereafter New14), who derived the [Fe/H] empirically from the EW of Na doublet at 2.2 $\mu$m ($K$-band) measured in medium-resolution spectra.
The green triangles and yellow diamonds are photometric estimates we obtain by using empirically calibrated formulae.
The former is based on an empirical relation between metallicity and the location in the ($J {-} K_s$)--($V {-} K_s$) color--color diagram calibrated by \citet{2012AJ....143..111J}.
The latter is based on an empirical relation between metallicity and the location in the ($V {-} K_\mathrm{s}$)--$M_{K_\mathrm{s}}$ color-magnitude diagram calibrated by \citet{2012A&A...538A..25N} ([Fe/H]$\mathrm{_{Nev12}}$). %
These metallicity estimates agree with our [Fe/H] values with a scatter of $\sim$0.1 dex, which is below the expected uncertainty, as shown in the top panel of Figure \ref{fig:compare_mtlc}.
The other smaller panels demonstrate that the differences do not have a significant systematic trend with $T_{\mathrm{eff}}$ or [Fe/H], with the exception of the bottom right panel, which implies that the higher the [Fe/H], the higher our result tends to be compared to [Fe/H]$\mathrm{_{Nev12}}$.

Note that \citet{2015ApJ...800...85N} point out that the method of New14 overestimates the metallicities of late-M dwarfs (M5V and later) compared with the more reliable estimation by \citet{2014AJ....147..160M} by up to $\sim$0.2 dex.
Possible reasons are that the estimation of New14 does not take the impact of $T_{\mathrm{eff}}$ into account, and their calibration sample is heavily biased toward mid-M dwarfs.
Their indication suggests that our [Fe/H] values also might be overestimated.

\subsection{Comparison with literature for Barnard's star} \label{sec:comparison_Barnard}

Our sample includes GJ 699 (Barnard's star). %
It is a well-studied bright and high proper-motion M dwarf and the closest single star to the Solar System.
It is also famous for the RV signal of a super-Earth candidate orbiting near its snow line by \citet{2018Natur.563..365R}, although \citet{2021AJ....162...61L} recently suggested the stellar activity origin of the signal rather than planetary origin.
Its metallicity has been estimated by many studies with a variety of methods, including Man15 and New14.

\begin{figure}\begin{center}
  \plotone{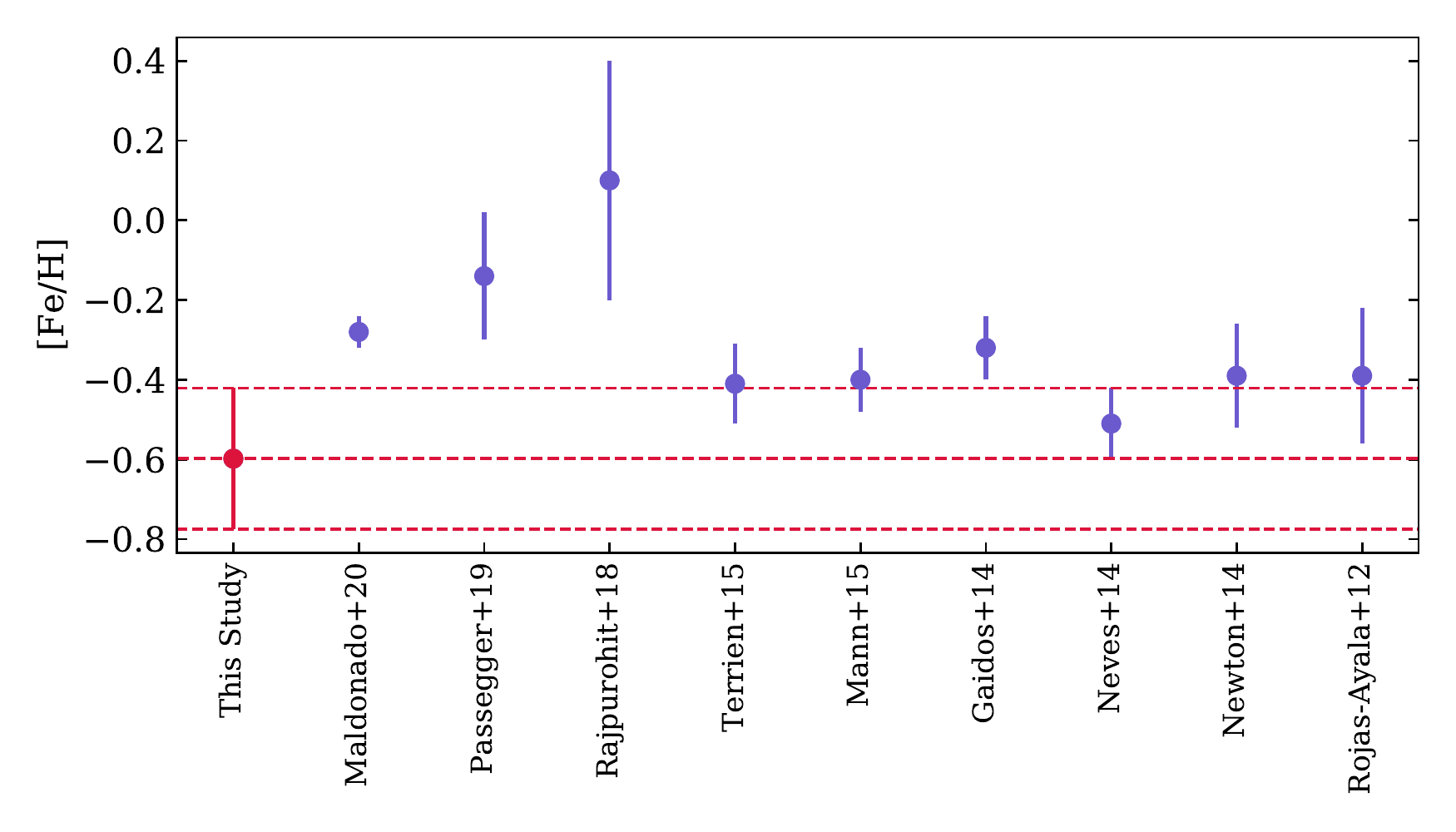} %
  \caption{Comparison of the [Fe/H] of GJ 699 from our analysis with those from previous studies.
  The values in the literature are not the iron abundances determined directly only from Fe lines.
  They refer to overall metallicities, except for that of \citet{2020A&A...644A..68M} who empirically determined abundances of iron using principal component analysis and sparse Bayesian method.
  The three horizontal dashed lines indicate the result of this study and its 1$\sigma_\mathrm{Total}$ range.
  } \label{fig:ComparePrevious_gj699}
\end{center}\end{figure}

We compare our [Fe/H] with estimates from the previous studies in Figure \ref{fig:ComparePrevious_gj699}.
The literature values range from $-$0.51 to $+$0.10 dex, whereas our [Fe/H] is $-0.60 \pm 0.18$ dex. %
Our result is consistent with several literature values within the error margin but is lower than those of any previous study.

The closest estimate to our result is that of \citet{2014A&A...568A.121N}.
Their method is a model-independent empirical analysis at optical wavelengths, which is very different from our method.
They empirically determined [Fe/H] from the pseudo-EW of 4104 absorption features in high-resolution optical spectra by the procedure calibrated by reference [Fe/H] values.
The reference values were based on an empirical relation between $M_{K_s}$, $V {-} K_s$, and [Fe/H] calibrated by \citet{2012A&A...538A..25N} with 23 FGK+M binaries. %

The results of other literature using medium-resolution $K$-band spectra, %
namely \citet{2015ApJS..220...16T}, Man15, \citet{2014MNRAS.443.2561G}, New14, and \citet{2012ApJ...748...93R}, consistently cluster around $-$0.4.
\citet{2015ApJS..220...16T} compared the results from a limited number of the features in each spectral band of $J$, $H$, and $K_s$ and found that the $K_s$-band estimates achieve the best precision.
The plot in Figure \ref{fig:ComparePrevious_gj699} is the result from $K_s$-band.
Man15 and \citet{2014MNRAS.443.2561G} determined [Fe/H] from the 22 metallicity-sensitive features on medium-resolution visible/near-infrared spectra that were statistically identified by the analysis of 112 FGK+M binary systems in \citet{2013AJ....145...52M}.
\citet{2012ApJ...748...93R} determined [Fe/H] from EWs of metallicity-sensitive Na I and Ca I lines and $T_{\mathrm{eff}}$-sensitive $\mathrm{H_{2}O}$ features in the medium-resolution $K$-band spectra.
The aforementioned overestimation of New14 is not seen because the $T_{\mathrm{eff}}$ of GJ 699 is higher than those of the objects in Section \ref{sec:comparison_with_previous_studies} and among the $T_{\mathrm{eff}}$ range where their method is well calibrated.

Recent estimates by synthetic spectral fitting on high-resolution spectra show even higher results.
\citet{2019A&A...627A.161P} and \citet{2018A&A...620A.180R} determined the metallicity and stellar parameters simultaneously by fitting the synthetic spectral grids of PHOENIX-SESAM and PHOENIX-BT-Settl, respectively, to the high-resolution optical and near-infrared spectra of CARMENES. %
They used spectral features of various elements all together to determine the overall metallicity. %
The features include a significant number of Ti I lines, which may have potentially degraded the abundance determination due to the anti-correlation of Ti I lines with the overall metallicity as is presented by Ish20.
In addition, the $T_{\mathrm{eff}}$ estimated through the analysis by \citet{2018A&A...620A.180R} (3400 K) is higher than that in other literature, including the $T_{\mathrm{eff\mathchar`-TIC}}$ (3259 K) we applied.
This estimate seems to be uncertain because the $\chi^2$ ($=$ 2.43) of their fitting is significantly larger than those of other objects in their sample. %
It is thus possible that their fitting is not working well at least for this object for some reason such as its low metallicity.

Among previous studies, only \citet{2020A&A...644A..68M} determined the abundances of individual elements.
They empirically determined the elemental abundances of GJ 699 with the principal component analysis and sparse Bayesian methods. %
Their abundance results are higher than ours by more than 1$\sigma$ for elements other than Na and Mg.
Note that their training dataset does not include M dwarfs cooler than 3600 K and with metallicity lower than $-$0.2 dex, so the reliability of the analysis on GJ 699 could not be assured.
Our analysis should provide the most robust abundances among all the attempts using high-resolution spectra
because we analyzed the spectral lines of individual elements separately and consistently.

\section{Discussion} \label{sec:IRD_discussion}

\subsection{Abundance distribution} \label{sec:Abundance_distribution_in_the_Galaxy}

Figures \ref{fig:Galactic_abundance_distribution} and \ref{fig:Galactic_abundance_distribution_Sr} show the distribution of the abundance ratios, relative to iron, of our M dwarfs compared with the FGK stars belonging to the individual kinematically separated stellar populations of the Galaxy.
In these figures, the error bars of [X/Fe] are calculated as the quadrature sum of both $\sigma_\mathrm{SEM}$ values of [X/H] and [Fe/H], taking account of the fact that the abundance changes due to changes in the adopted stellar parameters are mostly canceled between the [X/H] and [Fe/H]. %
$\sigma_\mathrm{Total}$ is adopted as the error bar of [Fe/H] on the horizontal axis.
The abundance ratios of FGK stars are adopted from Adi12 for Figure \ref{fig:Galactic_abundance_distribution} and from \citet{2019MNRAS.484.3846M} for Figure \ref{fig:Galactic_abundance_distribution_Sr}. %
\begin{figure*}\begin{center}
  \includegraphics[width=120mm]{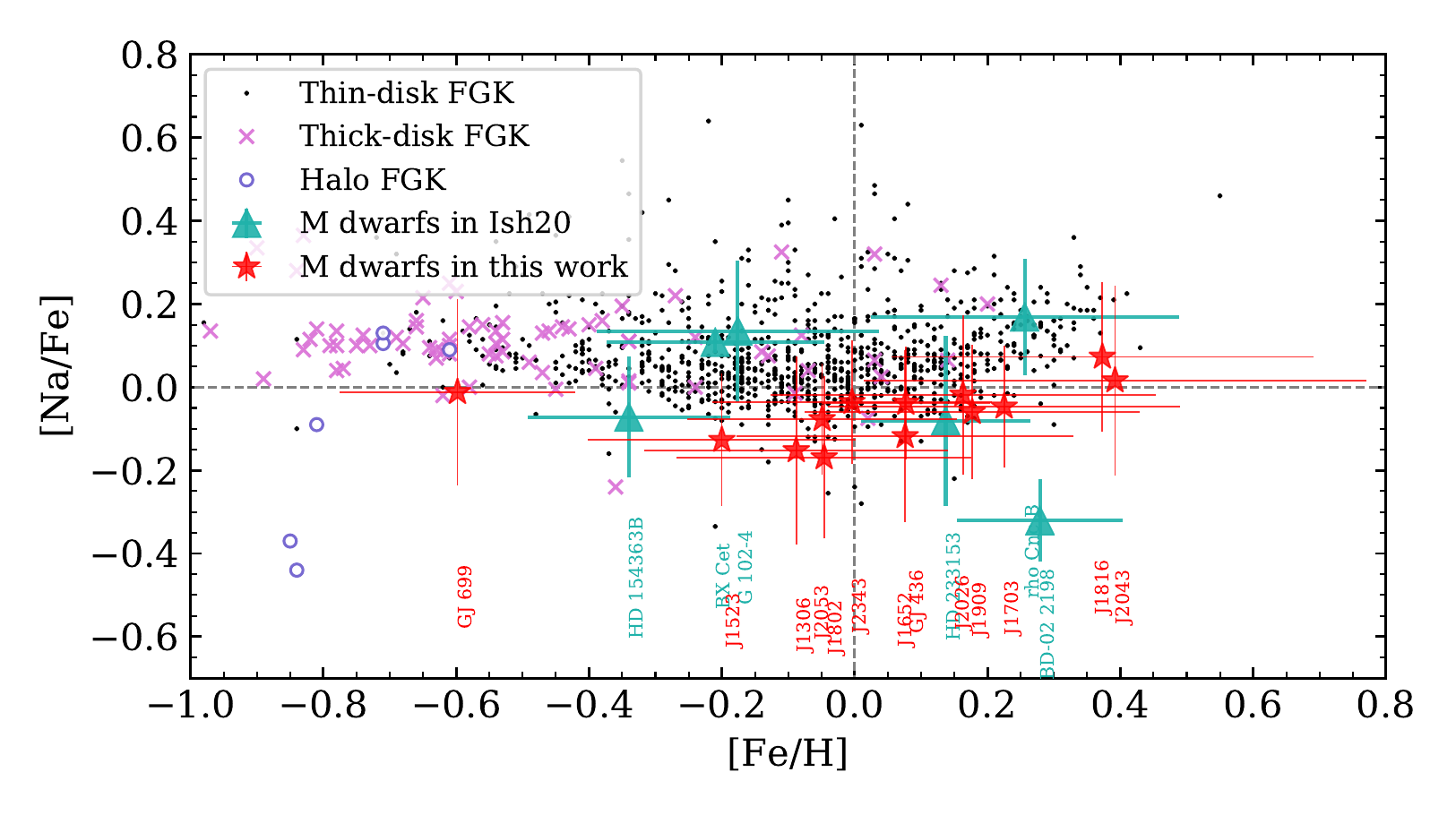}  %
  \includegraphics[width=120mm]{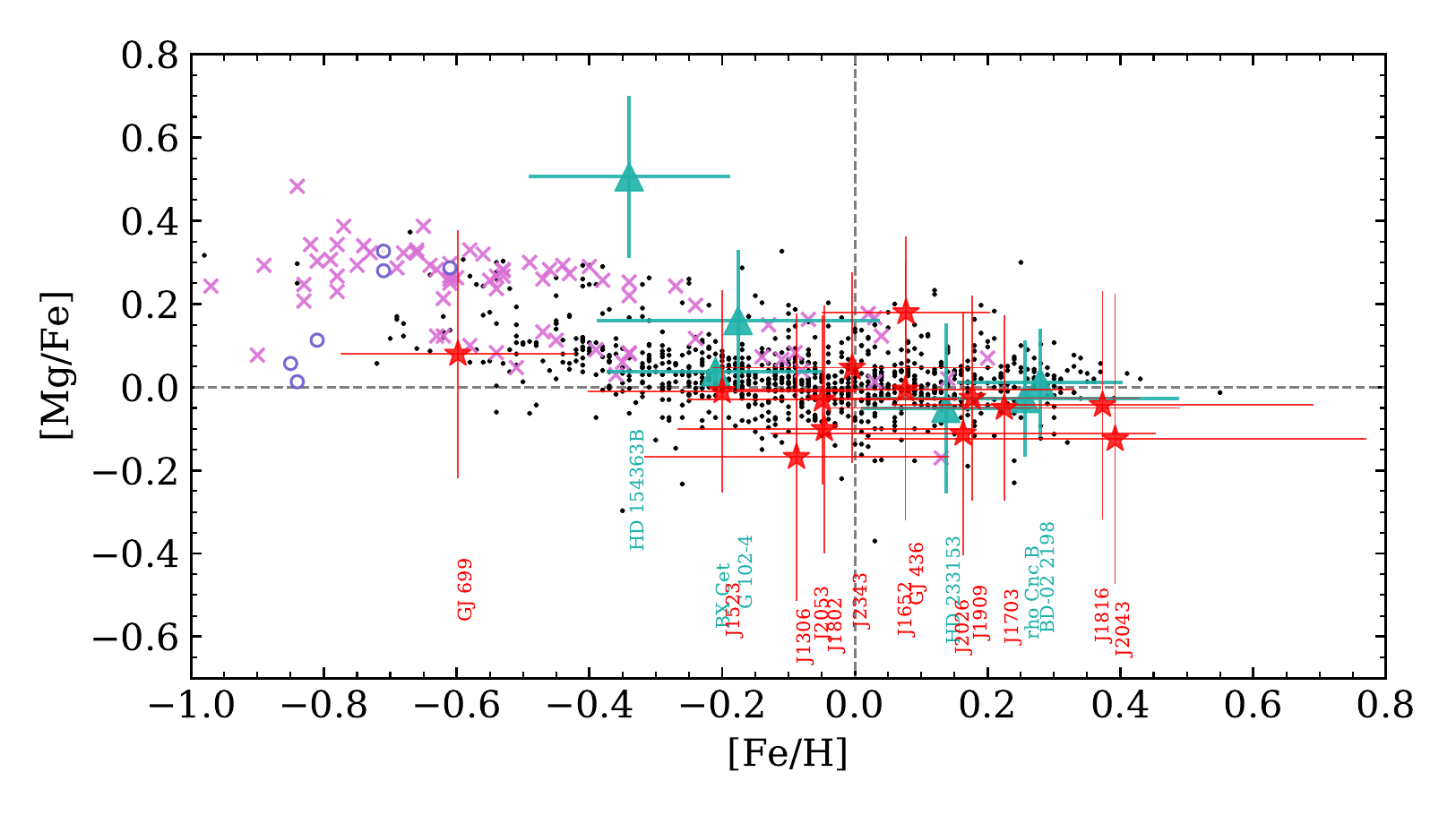}
  \includegraphics[width=120mm]{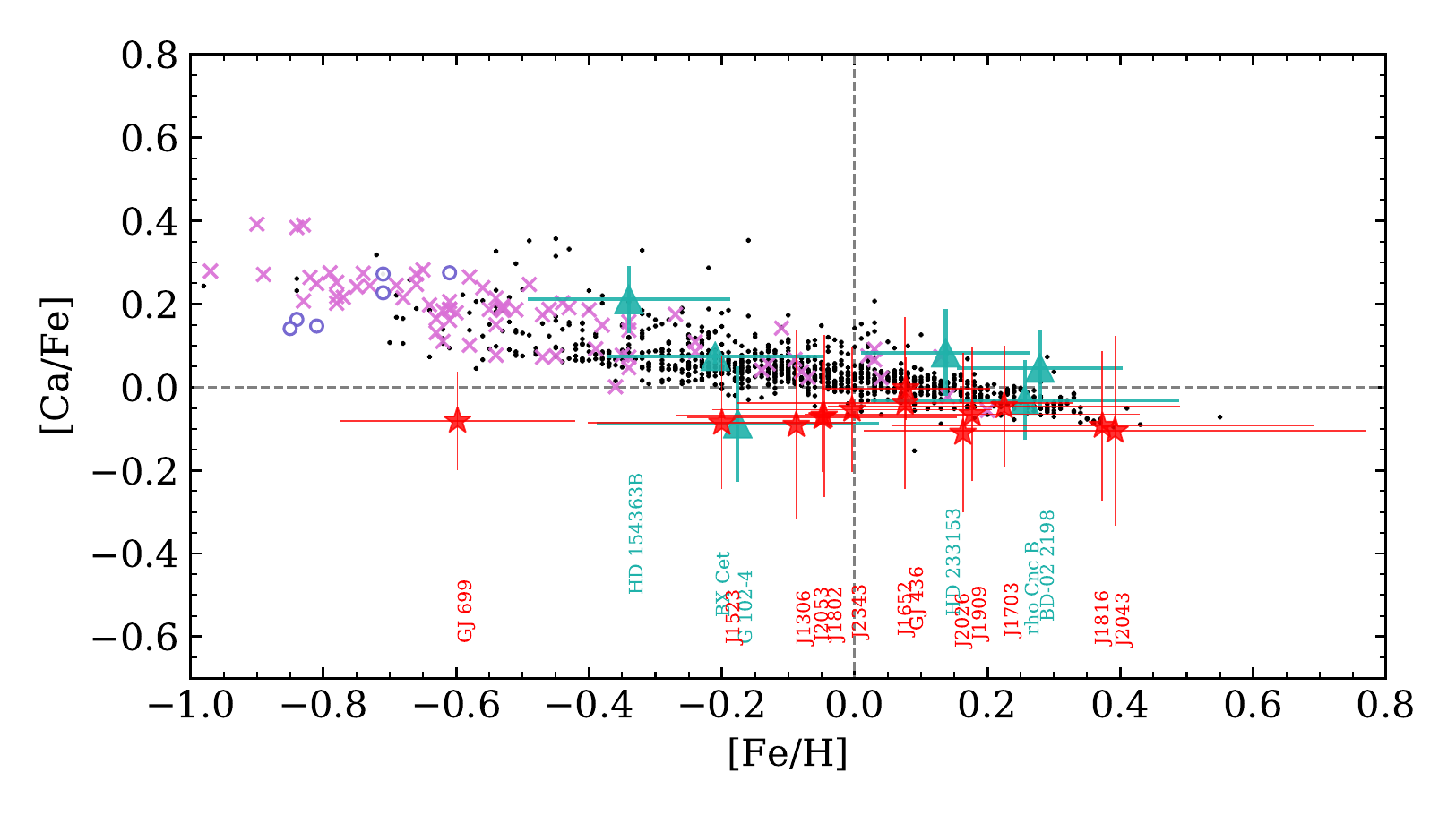}
  \caption{Iron-normalized abundances of six individual elements in the respective panels as a function of iron abundances. %
  The red stars and green triangles are the M dwarfs analyzed in this work and Ish20, respectively.
  FGK stars analyzed by Adi12 are displayed as references showing the abundance trends of individual Galactic subpopulations. %
  Among them, the gray dots, pink crosses, and purple circles are the thin disk stars, thick disk stars, and halo stars, respectively.
  } \label{fig:Galactic_abundance_distribution}
\end{center}\end{figure*}
\addtocounter{figure}{-1}
\begin{figure*}\begin{center}
  \includegraphics[width=120mm]{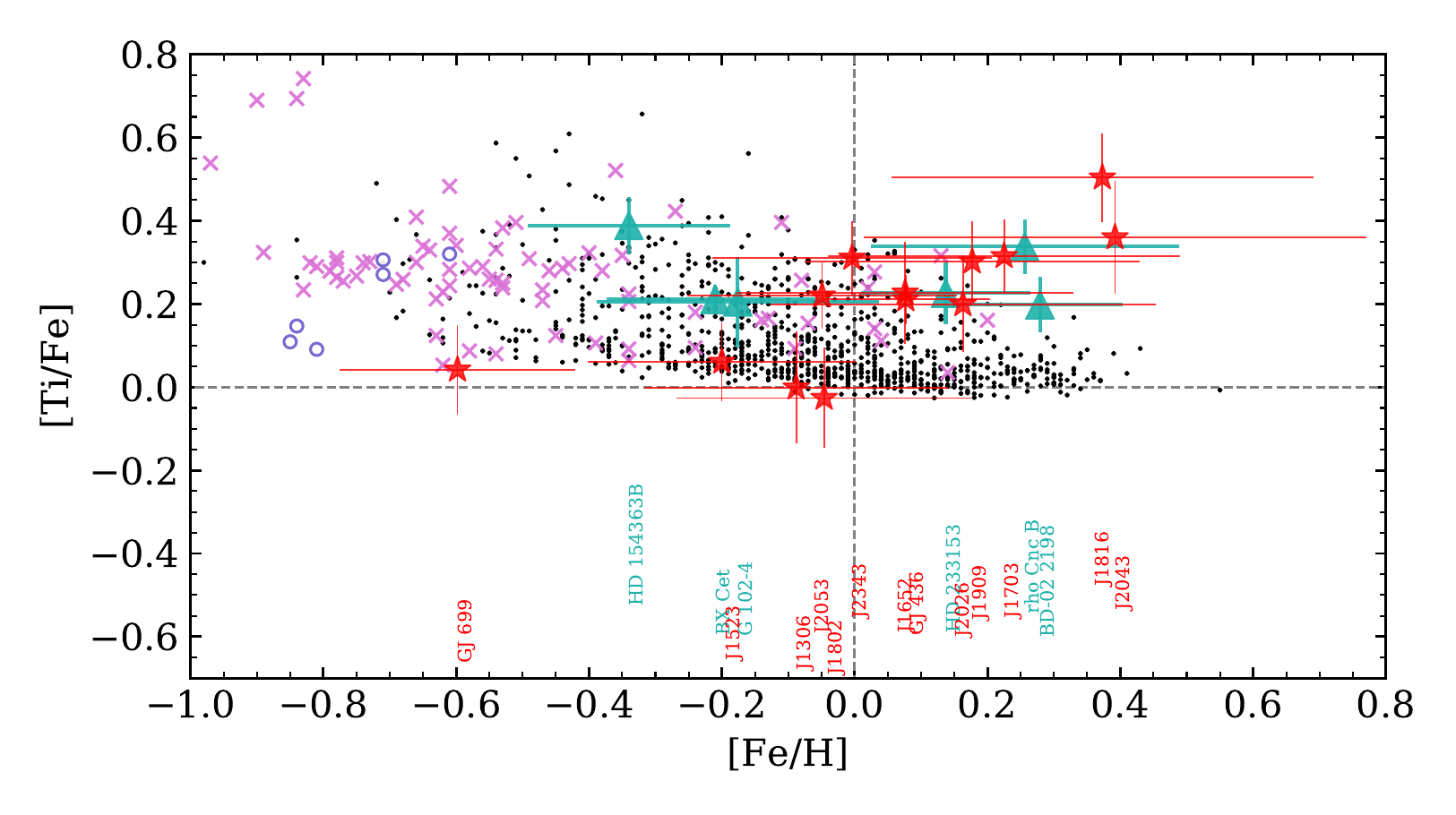}
  \includegraphics[width=120mm]{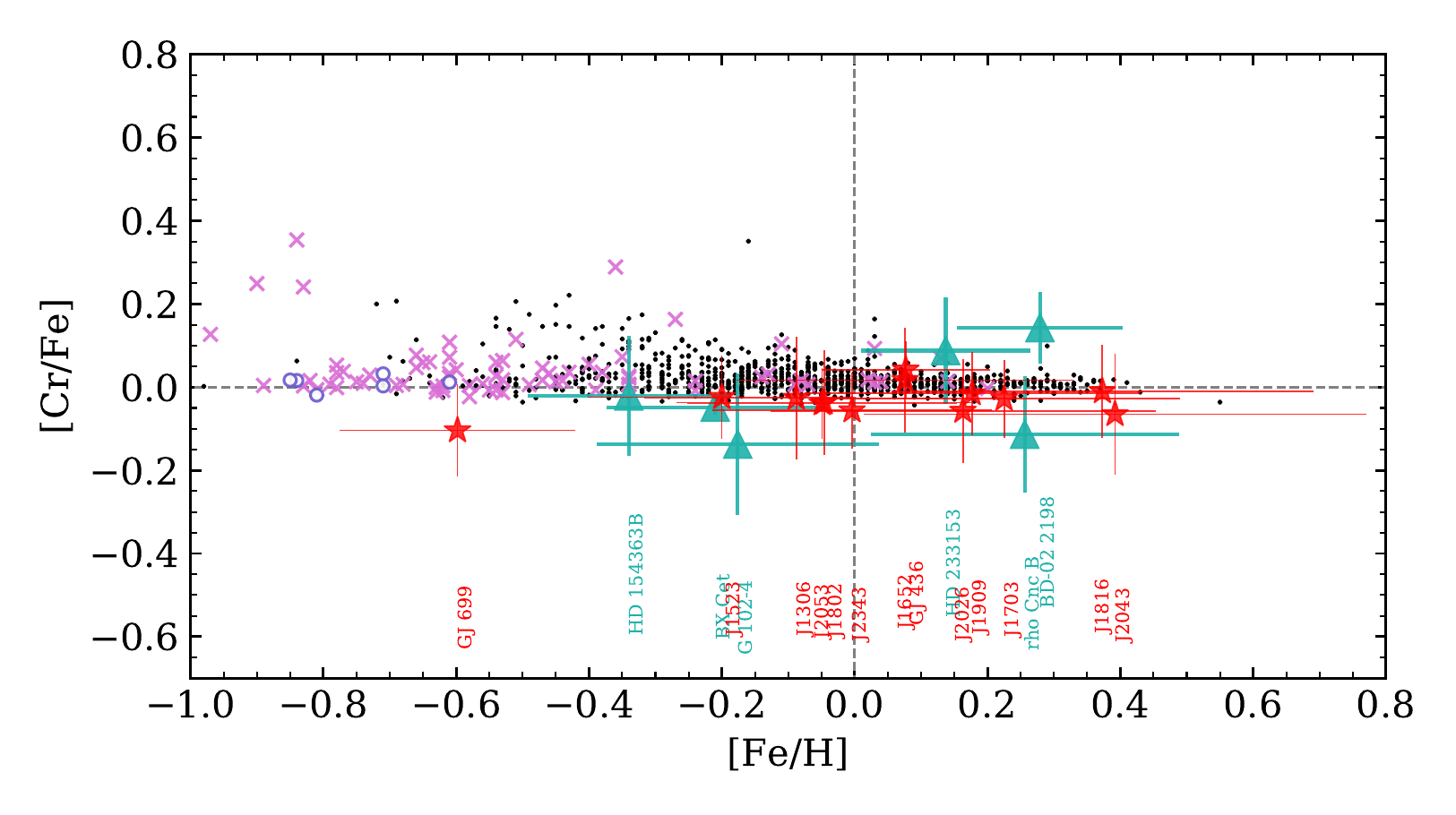}
  \includegraphics[width=120mm]{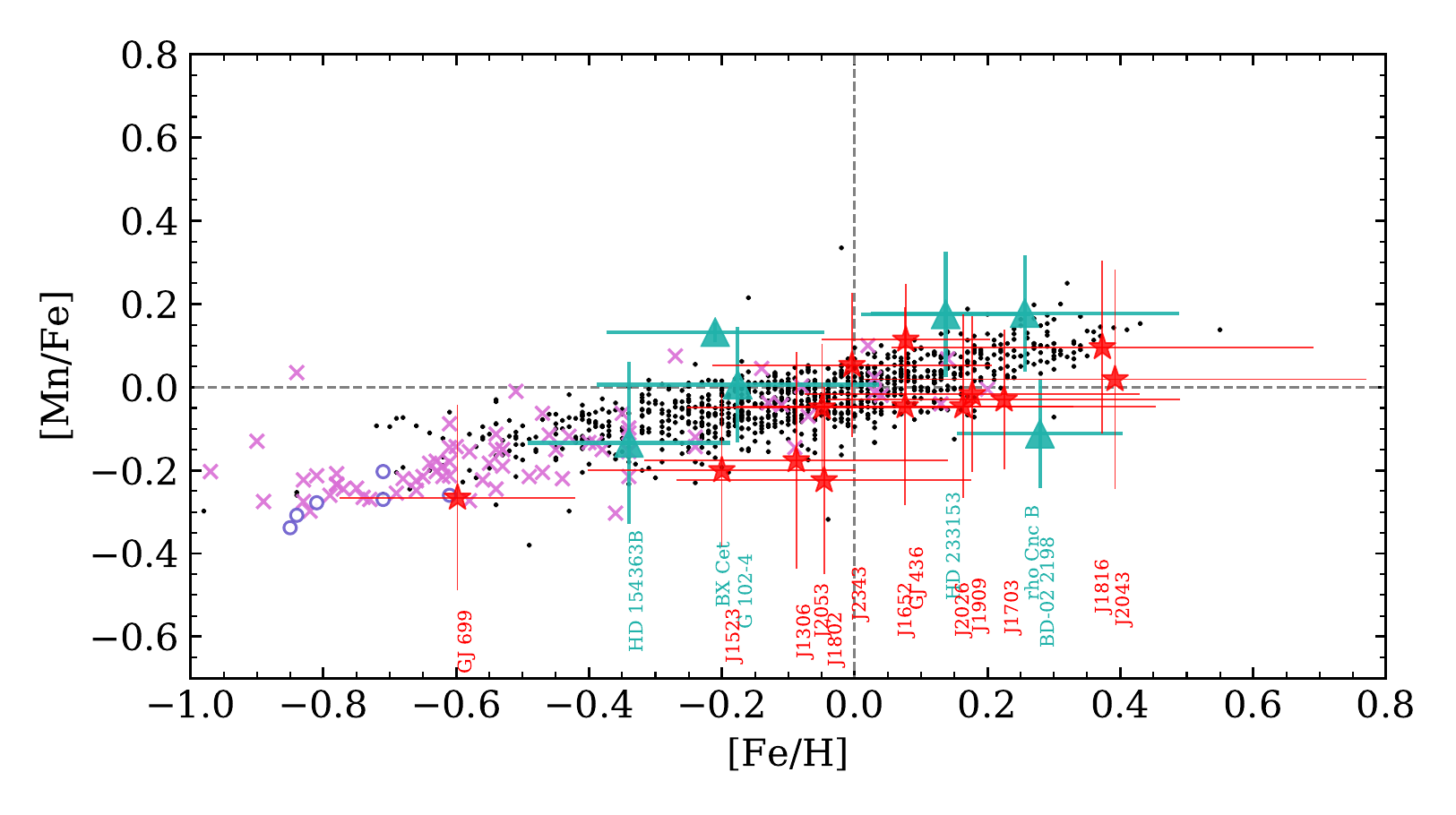}
  \begin{flushleft}
    \caption{(Continued)}
  \end{flushleft}
\end{center}\end{figure*}
\begin{figure*}\begin{center}
  \includegraphics[width=120mm]{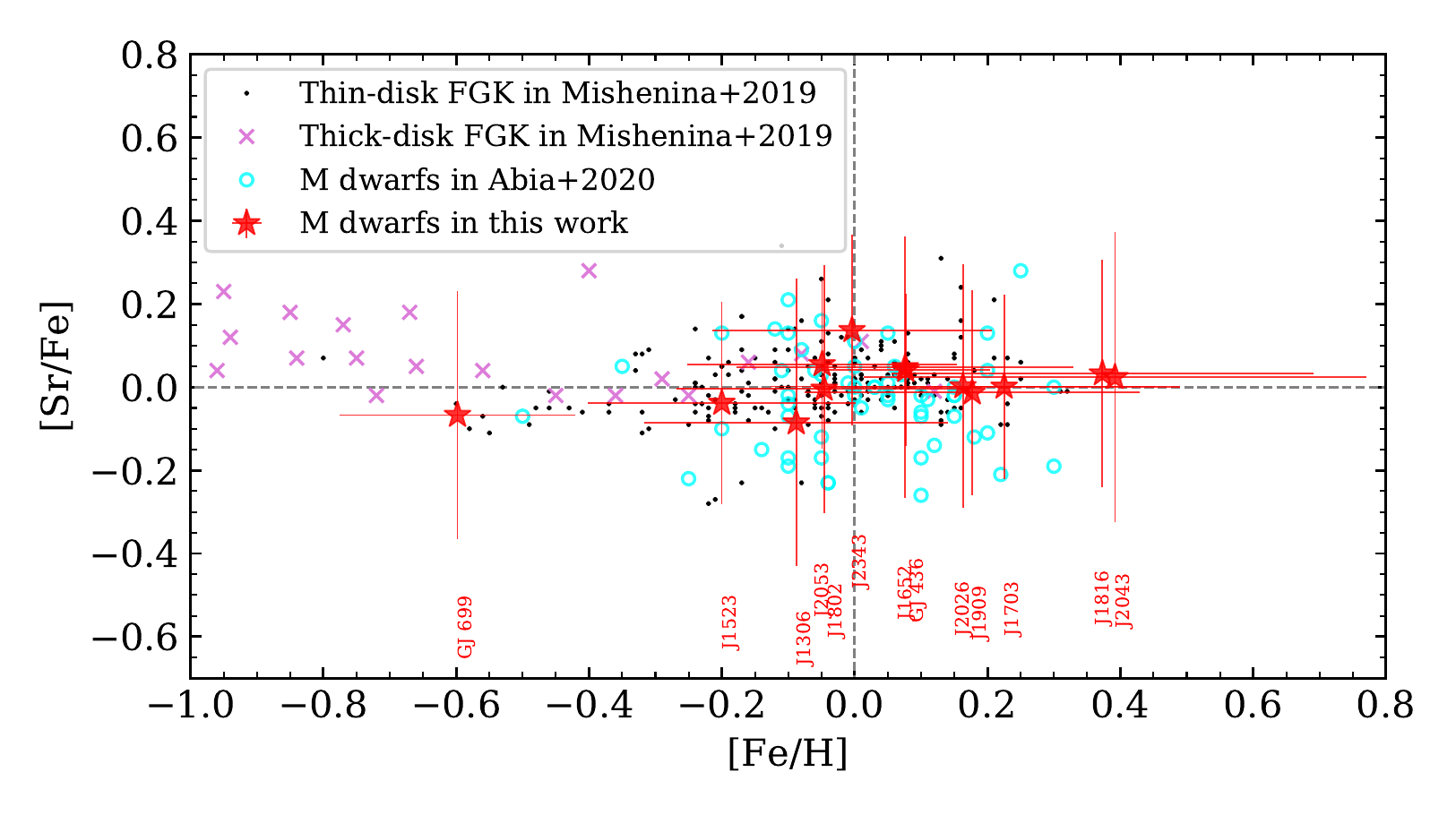}
  \caption{Iron-normalized abundances of Sr as a function of iron abundances.
  The red stars are the M dwarfs we analyzed in this work.
  FGK stars from \citet{2019MNRAS.484.3846M} are displayed as references showing the abundance trends of the thin-disk stars (gray dots) and the thick-disk stars (pink crosses).
  The cyan open circles represent the nearby early M dwarfs in \citet{2020A&A...642A.227A}.
  } \label{fig:Galactic_abundance_distribution_Sr}
\end{center}\end{figure*}

Our M dwarfs generally have a distribution similar to that of the FGK stars, most of which belong to the thin disk population. %
We suppose that most of the program stars belong to the Galactic thin disk with two exceptions, GJ 699 and HD$\:$154363$\:$B, which have [Fe/H] as low as those of the thick-disk FGK samples. %

Several previous studies have suggested that GJ 699 (Barnard's star) belongs to the older populations with an age of $\sim$7--12 Gyr based on observational properties such as a high space velocity, low metallicity, slow rotation ($P_\mathrm{rot} \sim $ 145 $\pm$ 15 days; \citealt{2019MNRAS.488.5145T}), non-detection of lithium~\citep{2004ARA&A..42..685Z}, low chromospheric activity~\citep{2006PASP..118..617R}, and low X-ray luminosity ($\log{L_\mathrm{X}} (\mathrm{erg \ s^{-1}}) = 26$; \citet{1999A&AS..135..319H}). %
Many literature reports have indicated that the star belongs to the thick disk population and some~\citep[e.g.,][]{2013ApJ...764..131C} suggested that it is even a halo star.
For such old stars, high [$\alpha$/Fe] is expected.
Nevertheless, [Mg/Fe] and [Ca/Fe] of GJ 699 are lower than the typical value of the thick-disk FGK samples.
There are, however, a fraction of thick disk stars that have [$\alpha$/Fe] values as low as those of thin disk stars.
Hence, the observed low values of [Mg/Fe] and [Ca/Fe] do not exclude GJ 699 from the thick disk population.
No excess of $\alpha$ elements in this object recalls the ``low-$\alpha$ stars'' in the halo reported by \citet{2010A&A...511L..10N}.
The low metallicity obtained by our analysis might also support the possibility that this star belongs to the halo structure, although a definitive classification of individual objects is not possible
with the current precision.
Note that, as mentioned in Section \ref{sec:comparison_Barnard}, [Fe/H] derived by this study is lower than most previous studies, and this is what highlights the peculiar and interesting position of this star in Figures \ref{fig:Galactic_abundance_distribution} and \ref{fig:Galactic_abundance_distribution_Sr}.
Even if the literature value of [Fe/H] ($\sim-$0.4) is adopted, this is the lowest metallicity in our sample, although it is not well distinguished from others.
In order to have more confidence in our analysis of this very metal-poor star, we need to extend the verification using binary systems of M dwarfs and warmer stars, as done by Ish20, to lower metallicity. %
Meanwhile,
the abundance pattern of HD$\:$154363$\:$B better agrees with that of thick-disk FGK samples.

For [X/Fe] of the other 17 M dwarfs (twelve from this work and five from Ish20), the distribution is indeed similar to the thin-disk stars.
However, for some elements,
the averages are lower than those of thin-disk FGK stars beyond the scatter of the data.
For example, the average of [Na/Fe] in our 17 objects is 0.12 dex lower than the average of FGK counterparts in Adi12, while the standard deviation in our sample is 0.12 dex.
In the case of [Ca/Fe], these discrepancy and standard deviation are 0.09 and 0.06 dex, respectively.
The discrepancies for other elements are smaller than the corresponding standard deviation.
However, the Mann-Whitney U-test suggests that the difference in the distributions of [X/Fe] is statistically significant for five elements other than Mn and Sr.
This indicates that there could be systematic errors in our measurements of abundance ratios with respect to Fe, or M dwarfs have small offsets in the abundance ratios from FGK stars.

In the following, we report what we find on the distribution of individual elements. %
First, [Na/Fe] is approximately constant, but shows an overall downward shift as mentioned above.
A closer look at the 13 objects analyzed in this work might be suggesting a trend that [Na/Fe] is higher at higher [Fe/H]. %
The [Mg/Fe] exhibits a similar trend to the FGK samples. %
The [Ca/Fe] values of M dwarfs are almost constant, with an average of $-$0.07.
This result does not follow the decreasing trend of [Ca/Fe] found for FGK stars and the values are lower on average.
The [Ti/Fe] shows a large scatter and does not follow the trend of FGK stars. %
We suppose that it is due to large uncertainties in the measurements. %
The error bars on the vertical axis %
include only the quadrature sum of both $\sigma_\mathrm{SEM}$ values of [Fe/H] and [Ti/H].
However, errors due to the uncertainties of other species are significant in the determination of Ti abundances as reported by Ish20. Hence, larger errors should be taken into consideration in the case of Ti.
The [Cr/Fe] is tightly aligned around the solar value as in the FGK samples.
This could be due to the relatively large number of the lines of Cr and Fe used in the analysis, and the similar ranges of excitation potentials (2--4 eV) or EWs (most lines $<$ 10 nm) of the lines of both elements. %
The [Mn/Fe] values of M dwarfs follow the increasing trend toward higher metallicities seen in the FGK samples. %
This trend is explained by the delayed contribution of Type Ia supernovae in chemical evolution~\citep[e.g.,][]{2013ARA&A..51..457N}. %
It is because the Mn yield relative to Fe is higher for Type Ia supernovae than for Type II supernovae/hypernovae.

As the reference for the [Sr/Fe], we adopt FGK stars in \citet{2019MNRAS.484.3846M} who recently reported Sr abundances separately for thin and thick disk stars,
instead of Adi12 who do not report Sr abundances.
We also plot the [Sr/Fe] results of \citet{2020A&A...642A.227A} for 57 nearby (within a few tens of parsecs) early-M dwarfs ($T_{\mathrm{eff}} >$ 3400 K) selected from the CARMENES-GTO targets.
Figure \ref{fig:Galactic_abundance_distribution_Sr} shows that the [Sr/Fe] ratios derived for our M dwarfs (red stars) agree well with those derived for the thin-disk FGK dwarfs (grey dots) and the early-M dwarfs (cyan open circles).
[Sr/Fe] values of thick and thin disk stars are not well separated in the FGK sample.
GJ 699 shows a [Sr/Fe] value that follows the trend found in FGK stars.

\subsection{Effect of $T_{\mathrm{eff}}$ shift} \label{sec:Teff_shift}
There may still be systematic errors in the $T_{\mathrm{eff\mathchar`-TIC}}$ values for cool stars ($<$ 3200 K) due to the scarcity of measurements of angular diameters as mentioned in Section \ref{sec:IRD_FeH_Teff}. %

Although the resulting metallicities from our analysis are similar to or even higher than the solar value,
in the HR diagram of Figure \ref{fig:HR-diagram},
all the targets are located between two isochrones of [Fe/H] = 0.0 and $-$0.5.
This suggests two possibilities: (1) the $T_{\mathrm{eff\mathchar`-TIC}}$ is systematically overestimated; (2) the $T_{\mathrm{eff}}$ of the PARSEC model is systematically underestimated. %
We discuss Case (1) in this section.
Case (2) cannot be ruled out, but the systematic errors embedded in the PARSEC model are beyond the scope of this paper.

If the $T_{\mathrm{eff}}$ is assumed to be lower than $T_{\mathrm{eff\mathchar`-TIC}}$ by 100K, data points in Figure \ref{fig:HR-diagram} move according to the direction and length of the red arrows. %
Note that the luminosity in TIC is calculated by the Stefan-Boltzmann law from the $T_{\mathrm{eff\mathchar`-TIC}}$ and radius estimated from the radius-$M_K$ relation in Man15, thus a downward adjustment of  $T_{\mathrm{eff\mathchar`-TIC}} $ also reduces the luminosities. %
The impact of the luminosity shift is, however, much smaller than that of $T_{\mathrm{eff}}$ in the figure. %

\begin{figure*}\begin{center}
  \includegraphics[width=78mm]{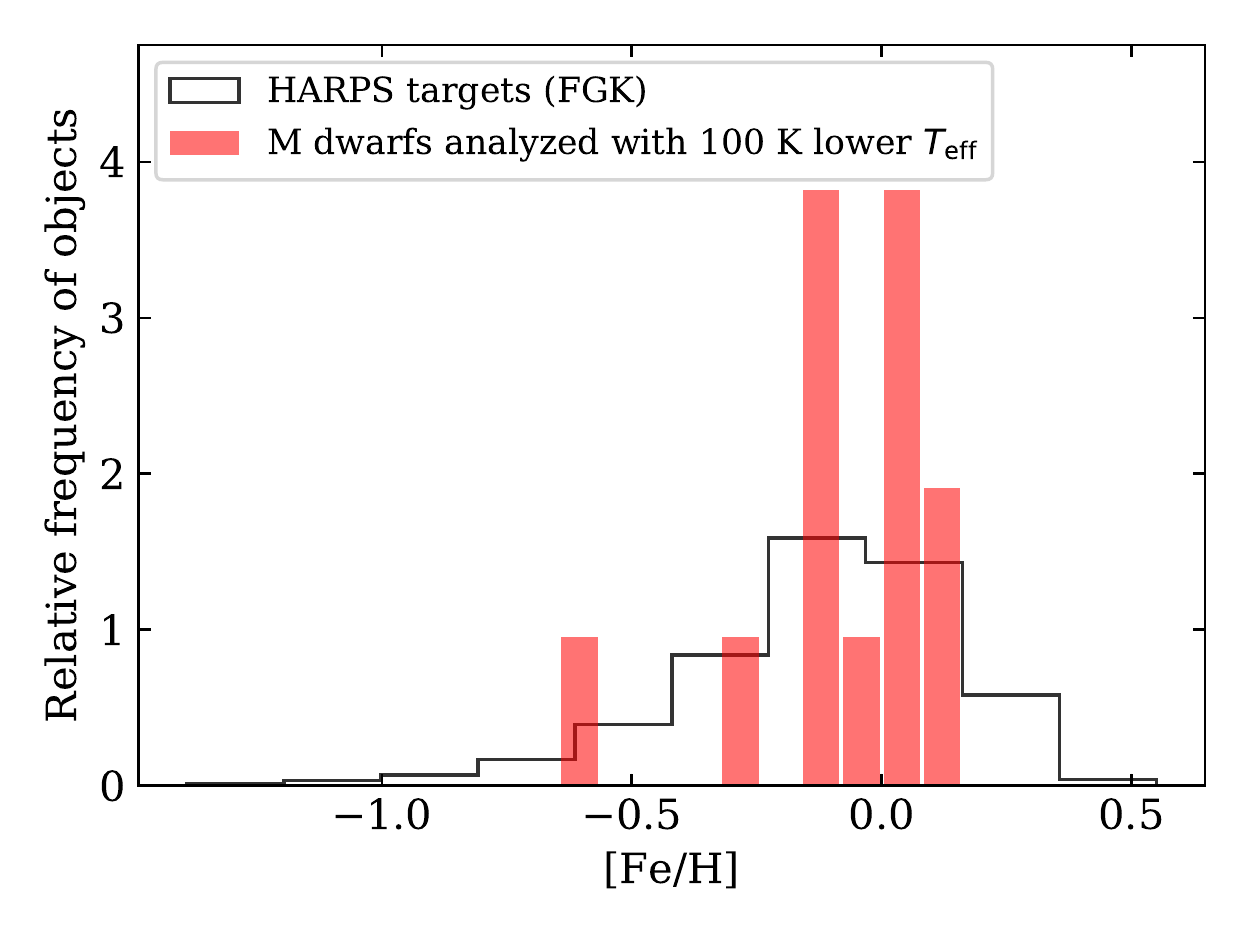}
  \includegraphics[width=78mm]{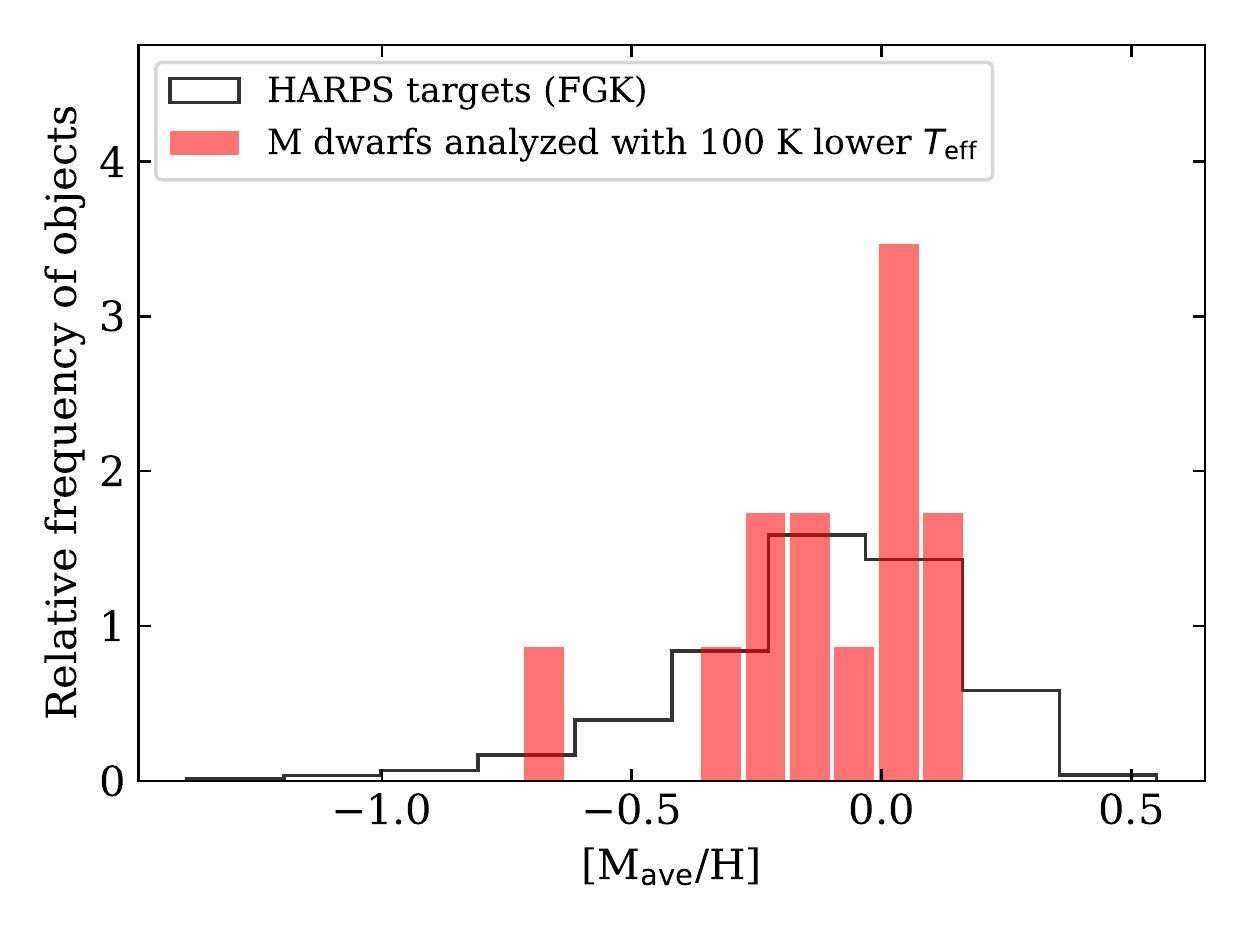}
  \caption{The same histograms as in Figure \ref{fig:hist_iron} but with 100K lower $T_{\mathrm{eff}}$ adopted. %
  } \label{fig:hist_iron_low}
\end{center}\end{figure*}

To investigate the effect of the possible overestimation of $T_{\mathrm{eff\mathchar`-TIC}}$ on the abundance analysis, we perform the same analysis as in Section \ref{sec:IRD_analysis} by employing $T_{\mathrm{eff}}$ that is 100 K lower than $T_{\mathrm{eff\mathchar`-TIC}}$.
The resulting metallicity distribution is shown in Figure \ref{fig:hist_iron_low}.
Both [Fe/H] and [M$_\mathrm{ave}$/H] shift to lower values by 0.11 dex on average from those shown in Figure \ref{fig:hist_iron}.
The distribution better agrees with those of FGK stars. %
The Mann-Whitney U-test could not reject the null hypothesis of no difference.

The abundances of other elements [X/H] also decrease by adopting lower $T_{\mathrm{eff}}$, as of Fe.
This is primarily due to the change of Na abundance.
The qualitative explanation is that if a lower $T_{\mathrm{eff}}$ is adopted in the spectral synthesis, the ionization degree of Na would decrease and make the EWs of Na I lines larger in the model, thus the result of [Na/H] reproducing the observed EW becomes lower.
In addition to the fact that the abundances of other elements also show similar behavior, the decrease in [Na/H] causes a general decrease in the abundances of all the other elements through the changes in H$^{-}$ opacity during the iterative process.

In contrast, the abundance ratios [X/Fe] of individual elements relative to Fe show different changes depending on the ionization state of the elements and the characteristics of the used lines.
Table \ref{tab:Teff_shift} tabulates the variation of each abundance ratio with a 100 K decrease in $T_{\mathrm{eff}}$.
The changes in [X/Fe] values are smaller than 0.05 dex in most cases.
The direction of the changes in Table \ref{tab:Teff_shift} do not lead to the agreement between the abundances of FGK dwarfs and M dwarfs, e.g., the [Na/Fe] distribution, which we find lower than those of FGK dwarfs (see Section \ref{sec:Abundance_distribution_in_the_Galaxy}), becomes even lower.
Thus, if lower $T_{\mathrm{eff}}$ is adopted in the analysis, %
the difference of [Fe/H] distribution between our M dwarfs and FGK stars mostly diminishes, while the small discrepancies in [X/Fe] trends remain.
Future $T_{\mathrm{eff}}$ estimation by reliable methods such as measurements of angular diameters is essential to reduce systematic errors.

\begin{deluxetable*}{cccccccc}
  \tablecaption{Changes in abundance ratios when the temperature is lowered by 100 K}
  \label{tab:Teff_shift}
    \tablehead{
      [Fe/H] & [Na/Fe] & [Mg/Fe] & [Ca/Fe] & [Ti/Fe] & [Cr/Fe] & [Mn/Fe] & [Sr/Fe] %
      }
    \decimals %
    \startdata
      $-$0.11 & $-$0.05 & $-$0.01 & $+$0.03 & $-$0.14 & $+$0.01 & $-$0.08 & $+$0.09 \\
    \enddata
  \tablecomments{ %
  These values are the average of the results for the 13 objects analyzed in this section.
  }
\end{deluxetable*}

\subsection{Kinematics} \label{sec:IRD_result_kinematics} %
\begin{figure}\begin{center}
  \plotone{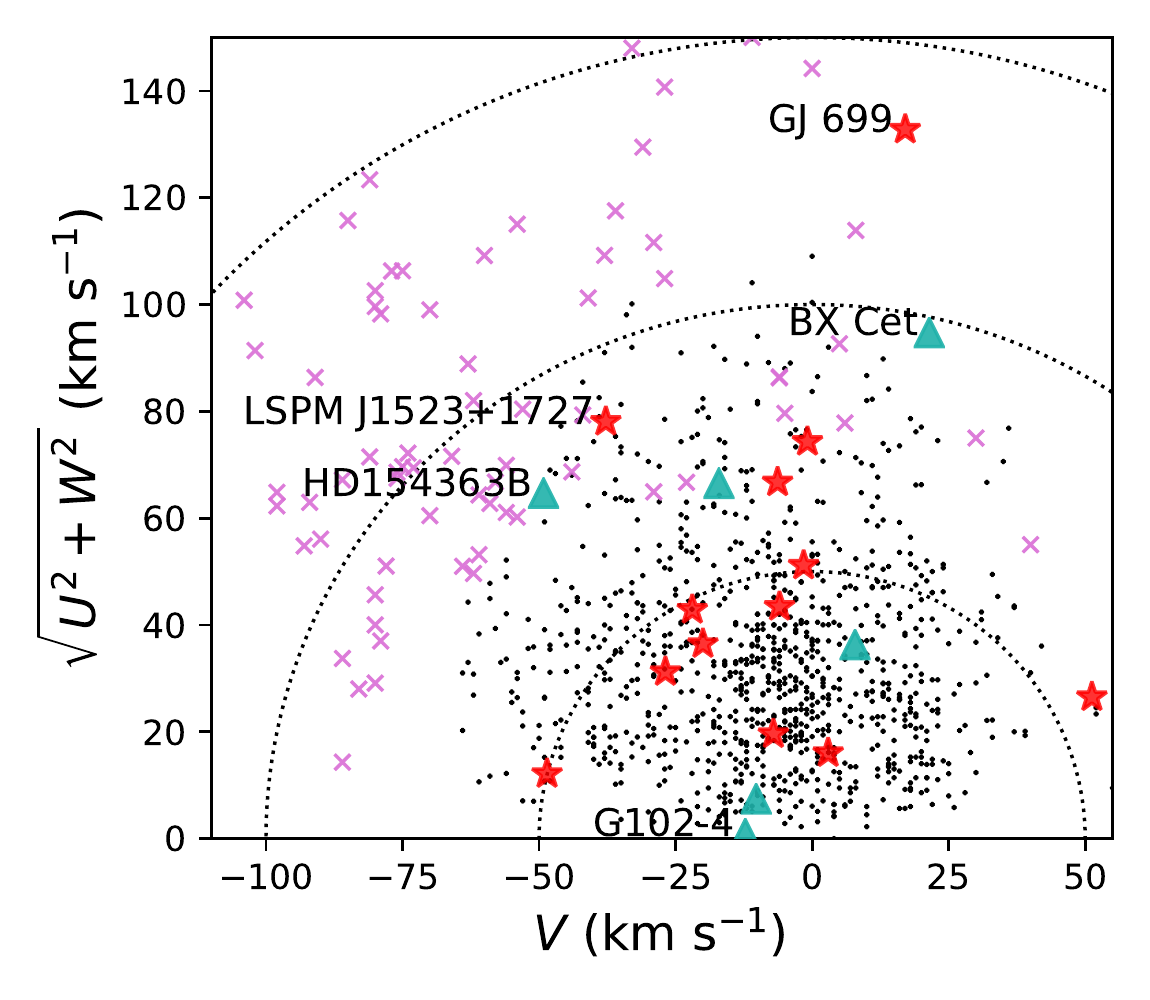} %
  \caption{Toomre diagram of our M dwarfs and the FGK stars in Adi12.
  The types of markers are the same as in Figure \ref{fig:Galactic_abundance_distribution}.
  The Galactocentric space velocities $U$, $V$, and $W$ are with respect to the LSR.
  Dotted lines represent constant values of the total space velocity $v_\mathrm{tot} = (U^2 + V^2 + W^2)^{1/2}$ in steps of 50 km s$^{-1}$.
  The five objects with the lowest [Fe/H] were labeled with names.
  } \label{fig:Toomre}
\end{center}\end{figure}

In Figure \ref{fig:Toomre}, we plot the Galactocentric space velocities of our 19 M dwarfs calculated in Section \ref{sec:IRD_kinematics} with those of FGK stars in Adi12 for comparison.
We find a rough trend that objects with lower [Fe/H] tend to have higher space velocities, which is consistent with the trend found for FGK stars.

We calculated the probability for each object belonging to either the thin disk ($D$), thick disk ($TD$), or the halo ($H$) based on the kinematical criteria of \citet{2014A&A...562A..71B} using the $UVW$.
The criteria are based on the assumption that the individual populations have Gaussian velocity distributions, different rotation velocities around the Galactic center, and occupy certain fractions of the stars in the solar neighborhood.
Then we classify stars into thin and thick disks as those with the thick-to-thin disk probability ratios $TD/D < 0.5$ and $TD/D > 2$, respectively.
As a result, GJ 699 is likely to belong to the thick disk, HD$\:$154363$\:$B and LSPM J1523+1727 are classified as ``in-between stars'' with equivalent chances of being the thin and thick disk, and the rest 16 objects belong to the thin disk.
There are no halo stars with $H/TD > 1$.
These are results expected from the fact that $\sim$90\,\%
of the FGK stars in the solar vicinity are thin-disk stars and halo stars account for less than 1\,\%~\citep[e.g.,][]{2017MNRAS.464.2610F, 2010ApJ...714..663D}.

\subsection{Indication to the chemical composition of planet-hosts} \label{sec:IRD_discussion_planet}

The sample of this study shows metallicity distribution in the range of about 1 dex.
The differences over the range could have an impact on planet formation as suggested by many previous studies~\citep[e.g.,][]{2019Geosc...9..105A}.

The abundance ratios are similar to solar values in most of our target M dwarfs.
Scaled solar abundances would be a useful approximation for them
in the current quality of the measurements, which needs to be improved by future studies for more detailed discussion on the impact of abundance ratios on the planet formation and structures.
Moreover, there could be thick disk stars with high $\alpha$/Fe ratios even in the solar neighborhood, which needs to be taken into account in the interpretation of the planets to be found by the IRD-SSP. %

This study also demonstrates that, for M dwarfs, elemental abundances and kinetics can be used complementarily to distinguish stellar populations in the Galactic context. %
Abundance patterns characteristic of older populations, such as low metallicity, coupled with spatial velocities that deviate from the LSR, provide evidence for classification as older populations.
Future extension of these analyses to larger samples will reveal the distribution of stellar populations in nearby M dwarfs.
Since it is difficult to determine the age of M dwarfs due to the slow changes of their properties (e.g., $T_{\mathrm{eff}}$ and luminosity) after reaching the main sequence, the chemical and kinematical information is a valuable clue to understanding the formation era and environment of M dwarfs. %
The formation history of M dwarfs directly connects to the formation and evolution of orbiting planets.

Combining the chemical composition of M dwarfs with the results of ongoing and future planet searches around M dwarfs will enable us to statistically explore trends such as the planet--metallicity correlation, which provides the constraints on planet formation scenarios around M dwarfs.
The difference in the occurrence rate of low-mass planets in different stellar populations indicated by \citet{2020A&A...643A.106B} can also be assessed. %

\section{Summary} \label{sec:IRD_summary}

We present an abundance determination of individual elements for the 13 nearby M dwarfs with 2900 $< T_{\mathrm{eff}} <$ 3500 K,
 which are selected from the IRD-SSP targets around which rocky planets might be found, %
 by the line-by-line EW analysis using the high-resolution ($\sim$70,000) near-infrared (970--1750 nm) spectra.
We used the high-S/N IP-deconvolved telluric-removed spectra produced for the RV measurements in the IRD-SSP project.
This is a pilot sample to consistently understand the chemical composition of a larger sample of nearby M dwarfs where the number of planet detection is increasing.

We determined the abundances of eight elements (Na, Mg, Ca, Ti, Cr, Mn, Fe, and Sr) for all 13 M dwarfs and those of three additional elements (Si, K, and V) for the hottest object GJ 436.
As indicated in Ish20, the consistent determination of the abundance of individual elements is important to determine the accurate chemical composition of M dwarfs because the results of abundance analysis of individual elements affect each other.
The error-weighted average metallicity [M$_\mathrm{ave}$/H] of the M dwarfs obtained from eight elements approximately range from $-$0.6 to $+$0.4 centered at around 0.0.
For all the objects, the abundance ratios of individual elements to Fe are generally aligned with the solar values within the measurement errors.
A notable object is the well-studied object GJ 699 (Barnard's star), whose abundances of all the measured elements were found to be low. %
This is consistent with the earlier suggestion of its old age ($\sim$7--12 Gyr), but besides, our [Fe/H] is even lower than most of the previous estimates. %

The abundance patterns of individual elements, i.e., the distribution of [X/Fe] as a function of [Fe/H], can now be compared with those of FGK stars in the solar neighborhood, most of which belong to the Galactic thin disk population.
Some of the trends known for FGK stars could also be suggested for M dwarfs in this work. %
Besides, the wide distribution of metallicity of our M-dwarf samples suggests that a few of them could be thick disk stars. %

We also measured the RVs from the high-resolution spectra of all our M dwarfs and those analyzed in Ish20.
They are combined with the astrometric measurements from Gaia EDR3 to calculate the space velocities $UVW$ with respect to the LSR.
The Kinematics based on the $UVW$ also suggests that a few of our program M dwarfs show similar features to FGK stars that are classified into the thick disk. %

The wide distribution of metallicity could have an impact on planet formation around M dwarfs.
Given that the existence of thick disk stars is suggested by our results even among the 13 nearby objects, different abundance ratios may be found in a larger sample.
Thus, abundance analyses of individual elements of the host stars are crucial to characterize planets to be found around the nearby M dwarfs by the IRD-SSP. %

The quality of the abundance determination is primarily limited by the uncertainty of $T_{\mathrm{eff}}$. %
\added{%
Whereas $T_{\mathrm{eff}}$ given in the TIC is adopted in the present work for abundance analyses, we also estimated
}%
$T_{\mathrm{eff}}$ using the line strength of FeH molecules.
The $T_{\mathrm{eff\mathchar`-FeH}}$ results in systematically higher values than the other empirical estimates based on photometry, especially at lower temperatures.
Meanwhile, the $T_{\mathrm{eff\mathchar`-FeH}}$ shows a good correlation with the $T_{\mathrm{eff\mathchar`-TIC}}$, which is empirically calculated from $G_\mathrm{BP} - G_\mathrm{RP}$ color.
Further interferometric measurements of angular diameters of such low-temperature M dwarfs are required to determine more reliable $T_{\mathrm{eff}}$.

This work provides the first reliable elemental abundances of nearby 13 M dwarfs including the objects with $T_{\mathrm{eff}}$ less than 3200 K that have not been previously investigated.
However, there is still a possibility of systematic errors that need to be addressed, as suggested for example by the lower distribution of [Na/Fe] and [Ca/Fe] than nearby FGK stars.
The verification in this $T_{\mathrm{eff}}$ range using visual binaries is required in future work. %
It also shows the possibility of locating M dwarfs on the Galactic chemical evolution based on the elemental abundance ratios and kinematics.
They will be useful to characterize the planet-hosting M dwarfs and their planets identified by the current and future planet search projects. %

The methods of this study will be utilized to characterize the primary stars of future planets discovered by IRD-SSP.
The chemical analysis of all the targets of IRD-SSP will shed light on the distribution of elemental abundances in very low mass stars in the solar neighborhood. %
Combined with the completion of the IRD-SSP planet search, this will lead to an understanding of the relation between the chemical composition of M dwarfs and the occurrence and properties of the orbiting planets, and hence, constraints on planet formation theories.

\begin{acknowledgments}
  This research is based on data collected at Subaru Telescope,
  which is operated by the National Astronomical Observatory of Japan.
  We are honored and grateful for the opportunity of observing the
  Universe from Maunakea, which has the cultural, historical, and natural
  significance in Hawaii.
  We gratefully acknowledge the staff at Subaru Telescope for their dedication and support of our observations.
  This work has made use of data from the European Space Agency (ESA) mission
  {\it Gaia} (\url{https://www.cosmos.esa.int/gaia}), processed by the {\it Gaia}
  Data Processing and Analysis Consortium (DPAC,
  \url{https://www.cosmos.esa.int/web/gaia/dpac/consortium}). Funding for the DPAC
  has been provided by national institutions, in particular the institutions
  participating in the {\it Gaia} Multilateral Agreement.
  This work made use of the VALD database, operated at Uppsala University, the Institute of Astronomy RAS in Moscow, and the University of Vienna. %
  This work has also made use of the SIMBAD database and the VizieR catalog access tool, both operated at the CDS, Strasbourg, France, and of NASA’s Astrophysics Data System Abstract Service.
  This research made use of Astropy, %
  a community-developed core Python package for Astronomy (Astropy Collaboration 2013, 2018). \nocite{2013A&A...558A..33A, 2018AJ....156..123A} %
  This work made use of PyRAF, a product of the Space Telescope Science Institute, which is operated by AURA for NASA.
  Data analysis was in part carried out on the Multi-wavelength Data Analysis System operated by the Astronomy Data Center (ADC), National Astronomical Observatory of Japan. %
  This work is partly supported by JSPS KAKENHI Grant Numbers JP21K20388, JP18H05442, JP15H02063, JP22000005, JP19K14783, JP21H00035, JP18H05439, and JP21H00055. It is also supported by JST PRESTO Grant Number JPMJPR1775 and the Astrobiology Center of National Institutes of Natural Sciences (NINS) Grant Number AB031010. %
\end{acknowledgments}

\bibliography{library}
\end{document}